\documentclass{aastex63}
\usepackage{amsmath}
\usepackage{amsfonts}
\usepackage{amssymb}
\usepackage{graphicx}
\usepackage{enumerate}
\usepackage{comment}
\usepackage{soul}
\usepackage{array}
\usepackage{xcolor}  
\usepackage{bm}
\usepackage{physics}

\def \apjs {{\rm {ApJS}}}

\def \aap {{\rm {A\&A}}}

\def \apjl{\rm {ApJL}}

\def \mj{\,$M_{\rm Jup}$\,}

\def \msun{\,$M_\odot$\,}

\def \mas{\,mas}
\def \masyr{\,mas\,yr$^{-1}$}

\date{\today}
\received{xxx}
\revised{xxx}
\accepted{xxx}
\submitjournal{ApJS}

\shorttitle{Modeling and Calibration of Astrometric Data for the Detection of Dark Companions}
\shortauthors{Feng et al.}

\begin{document}
\title{Modeling and Calibration of Gaia, Hipparcos, and Tycho-2 astrometric
  data for the detection of dark companions}
\author[0000-0001-6039-0555]{Fabo Feng}
\affiliation{Tsung-Dao Lee Institute, Shanghai Jiao Tong University, Shengrong Road 520, Shanghai, 201210, People's Republic Of China}
\affiliation{School of Physics and Astronomy, Shanghai Jiao Tong University, 800 Dongchuan Road, Shanghai 200240, People's Republic of China}

\author{Yicheng Rui}
\affiliation{Tsung-Dao Lee Institute, Shanghai Jiao Tong University, Shengrong Road 520, Shanghai, 201210, People's Republic Of China}
\affiliation{School of Physics and Astronomy, Shanghai Jiao Tong University, 800 Dongchuan Road, Shanghai 200240, People's Republic of China}

\author{Yifan Xuan}
\affiliation{Tsung-Dao Lee Institute, Shanghai Jiao Tong University, Shengrong Road 520, Shanghai, 201210, People's Republic Of China}
\affiliation{School of Physics and Astronomy, Shanghai Jiao Tong University, 800 Dongchuan Road, Shanghai 200240, People's Republic of China}

\author{Hugh Jones}
\affiliation{Centre for Astrophysics Research, University of Hertfordshire, College Lane, AL10 9AB, Hatfield, UK}

\correspondingauthor{Fabo Feng}
\email{ffeng@sjtu.edu.cn}

\begin{abstract}
Hidden within the Gaia satellite's multiple data releases lies a
valuable cache of dark companions. To facilitate the efficient and
reliable detection of these companions via combined analyses involving
Gaia, Hipparcos, and Tycho-2 catalogs, we introduce an astrometric modeling framework. This method incorporates analytical
least square minimization and nonlinear parameter optimization
techniques to a set of common calibration sources across the different
space-based astrometric catalogues. This enables us to discern the
error inflation, astrometric jitter, differential parallax zero-point,
and frame rotation of various catalogues relative to Gaia DR3. Our
findings yield the most precise Gaia DR2 calibration parameters to
date, revealing notable dependencies on magnitude and
color. Intriguingly, we identify sub-mas frame rotation between Gaia
DR1 and DR3, along with an estimated astrometric jitter of 2.16\,mas
for the revised Hipparcos catalog. In a thorough comparative analysis
with previous studies, we offer recommendations on calibrating and
utilizing different catalogs for companion detection. Furthermore, we provide a user-friendly pipeline\footnote{\url{https://github.com/ruiyicheng/Download_HIP_Gaia_GOST}} for catalog download and bias correction, enhancing accessibility and usability within the scientific community.
\end{abstract}

\section{Introduction}\label{sec:intro}
While our solar system boasts multiple rocky planets positioned closer
to the Sun and several giant planets situated on wider orbits, the
more than 4000 known planetary systems exhibit a variety of different configurations. The pursuit of Earth-like planets, particularly those resembling Earth in size and located within the habitable zone of Sun-like stars, is often regarded as the "holy grail" of exoplanet science. Additionally, giant planets play a crucial role in shaping planetary system architectures (e.g., \citealt{gomes05,tsiganis05}) and influencing the potential habitability of Earth-like worlds (e.g., \citealt{lunine01,horner10,horner20}). Detecting Earth-like planets poses challenges due to their faint signals (e.g., \citealt{hall18} and \citealt{ge22}), while our current capabilities enable relatively straightforward detection of cold giants \citep{wittenmyer20,feng22,laliotis23}.

The emergence of the Gaia satellite \citep{gaia16a,gaia16,gaia18,gaia21,gaia23} marks a golden age for the detection of cold giants. Leveraging Gaia's astrometric data, combined with other astrometric and radial velocity data, provides a powerful tool for this purpose \citep{snellen18,kervella19,feng19b,brandt19}. Specialized tools such as \small{orvara} \citep{brandt21} and \small{BINARYS} \citep{leclerc23} have been developed to conduct combined analyses. Typically, calibration of astrometric catalogs is performed beforehand to rectify biases like frame rotation and zero-point parallax in these catalogs \citep{brandt18,kervella19,brandt23}. However, in certain cases, calibration is integrated into the analysis itself, addressing potential biases alongside companion signals \citep{feng21}.

The techniques developed for exoplanet detection can also be applied
in the search for dark companions such as black holes (e.g.,
\citealt{elbadry23}), neutron stars (e.g., \citealt{shahaf23}), white dwarfs (e.g., \citealt{ganguly23}), and brown dwarfs (e.g.,
\citealt{feng22}). Detecting these dark companions opens a new window
to understand the formation and evolution of massive stars and binaries \citep{heger03} and to test whether there is a $2-5$\msun mass gap \citep{kreidberg12,lam22,ye22} between neutron stars and black holes.

In the absence of Gaia epoch data, the proper motion anomaly derived from Gaia and Hipparcos proper motions and positions is commonly used to constrain stellar reflex motion with decade-long orbital periods \citep[e.g.][]{kervella19,brandt21}. However, this method faces
  significant information loss in transforming astrometry into proper
  motion anomaly, making it inadequate for constraining the mass of
  Jupiter analogs\footnote{A Jupiter analog is an exoplanet of Jupiter’s mass (0.5 to 2\mj) orbiting a FGK star in a configuration akin to Jupiter, with a semi-major axis in the range 2.5 to 10\,au.} and distinguishing between retrograde and prograde
  orbits \citep{li21}. To address this limitation, a method
proposed by \cite{feng23} utilizes Gaia DR2 and DR3 five-parameter
astrometry, along with Hipparcos intermediate astrometric data (IAD),
combined with radial velocity data. This approach successfully
constrains the orbit and mass of nearby Jupiter analogs, such as
$\epsilon$ Eridani b, with a mass of $0.76_{-0.11}^{+0.14}$\mj and an
orbital period of $7.36_{-0.05}^{+0.04}$\,yr. Therefore, detecting
nearby cold Jupiters is feasible, considering the present precision of
 Gaia DR2 and DR3 of about 0.1\,mas for bright stars \citep{gaia23}, and the 24-year observational baseline established by Hipparcos and Gaia. This feasibility is contingent upon effectively mitigating biases and systematics in the astrometric catalogs to a comparable precision through calibration, either a priori or a posteriori.

Despite the vast dataset of one billion stars observed by Gaia, only a
small fraction (fewer than 10,000) have been studied with
high-precision spectrographs. To detect exoplanets and other dark companions
solely through astrometric data, we have developed an efficient
algorithm based on analytical $\chi^2$
minimization. This method considers various combinations of
astrometric catalogs, exploring combinations such as Gaia Data Release
1 (GDR1; \citealt{gaia16}), Gaia Data Release 2 (GDR2;
\citealt{gaia18}), Gaia Data Release 3 (GDR3; \citealt{gaia23}),
Tycho-2 (TYC; \citealt{hog00}), the original Hipparcos catalog (HIP1;
\citealt{perryman97}), and the revised Hipparcos catalog (HIP2;
\citealt{leeuwen07}).

We do not consider Gaia Early Data Release 3 (GEDR3; \citealt{gaia21})
since its astrometric data is the same as GDR3. Our Bayes factor (BF)
is derived from $\chi^2$ under the assumption of Laplace's
approximation \citep{schwarz78,kass95}, leading to the development of
a modeling framework for combined analyses of astrometric data. Additionally, we
employ orbital solutions from the GDR3 non-single-star (NSS;
\citealt{halbwachs23,holl23,arenou23}) catalog and stars with both
Gaia and Hipparcos data as test sets to estimate the frame rotations
of various catalogs relative to GDR3 and the error inflation within
the astrometric catalogs. 

This paper is organized as follows: Section \ref{sec:method} outlines
the formulae for the modeling framework applied to various combinations of astrometric
catalogs. Section \ref{sec:calibrate} introduces the calibration
procedure, while section \ref{sec:results} presents the calibration
outcomes. Finally, our findings and conclusions are summarized in
Section \ref{sec:conclusion}.

\section{Method}\label{sec:method}
In this section, we present the model framework designed for the
  integrated analysis of multiple astrometric catalogs. We introduce
  the circular reflex motion model and the simulation of Gaia epoch
  data using the Gaia Observation Forecast Tool (GOST) in subsection
  \ref{sec:astrometry}. The combined modeling of astrometric catalogs
  is categorized into three cases: GDR1+GDR2+GDR3, TYC+GDR2+GDR3, and
  HIP+GDR2+GDR3, which are detailed in subsections \ref{sec:gaia},
  \ref{sec:tyc}, and \ref{sec:hip}, respectively. For each case, we elucidate the astrometric model and provide the analytical solution for circular reflex motion. Subsequently, in subsection \ref{sec:ecc}, we augment the modeling framework with non-linear orbital parameters. The determination of the threshold for detecting a reflex motion signal induced by a companion is presented in subsection \ref{sec:threshold}.

\subsection{Astrometry model for circular orbits}\label{sec:astrometry}
In the case of unresolved binaries, Gaia observes the motion of the
system's photocenter around the barycenter, focusing on the
displacement of the combined center of light rather than tracking the
reflex motion of individual stars within the binary system. Given a
mass-luminosity function of $F(m)$, the angular semi-major axis of the photocentric motion is
\begin{equation}
a_p=\frac{a_b}{d}\frac{m_2}{m_1+m_2}\left[1-\frac{m_1F_2}{m_2F_1}\right]\left[1+\frac{F_2}{F_1}\right]^{-1},
\label{eq:ap}
\end{equation}
where $m_1$ and $m_2$ are respectively the masses of the primary and
the secondary, $F_1$ and $F_2$ are the observed fluxes of the primary and the secondary, $a_b$ is the binary semi-major axis, $d$ is the distance derived from parallax, $d=A_u/\varpi$ where $A_u\equiv1$\,au. For
dark companions, $F_2/F_1 \approx 0$, the photocentric motion is equivalent to
the reflex motion, and thus the angular semi-major axis of the reflex motion is $a_r=a_p=\frac{a_b}{d}\frac{m_2}{m_1+m_2}$. 

To derive analytical formulae for linear regression of astrometric model, we assume that
the orbit of a star is circular and could be described by
\begin{eqnarray}
  X_i&=&a_p\cos(\frac{2\pi}{P}\Delta t_i+\phi)~,\nonumber\\
  Y_i&=&a_p\sin(\frac{2\pi}{P}\Delta t_i+\phi)~,\\
  \label{eq:circular}
\end{eqnarray}
where $\Delta t_i=t_i-t_{\rm ref}$ is the time difference between
epoch $t_i$ and the reference epoch $t_{\rm ref}$ (GDR3 referernce epoch, J2016, by default),
$\phi$ is the orbital phase, and $P$ is the orbital period. The coordinates in the orbital plane are transformed into the sky plane using
\begin{eqnarray}
  \Delta\alpha_*^r&=&B'X_{i}+G'Y_{i}~,\\
  \Delta\delta^r&=&A'X_{i}+F'Y_{i}~,
\end{eqnarray}
where $A'$, $B'$, $F'$, $G'$ are scaled Thiele Innes constants, and are
functions of inclination $I$, argument of periastron $\omega_T$ of the
target star, longitude of ascending node $\Omega$. We multiply
$A',B',F',G'$ by $a_p$ to define the Thiele Innes constants $A, B, F,
G$. Here we assume that the reflex motion is equivalent to the
photocentric motion as long as the mass function is not derived from
the Thiele Innes constants, as in the case of GDR3 non-single-star
catalog (NSS). 

For circular orbit, $\omega_T$ can be arbitrarily chosen such that $\phi=0$. Hence we choose $\phi_j=0$
when $t_j=t_{\rm ref}$ and we have 
\begin{align}
  \Delta\alpha_{*i}^r&=BC^t_i+GS^t_i~,\\
  \Delta\delta_i^r&=AC^t_i+FS^t_i~,
\label{eq:adr}
\end{align}
where $C_i^t=\cos(\frac{2\pi}{P}\Delta t_i)$ and $S_i^t=\sin(\frac{2\pi}{P}\Delta t_i)$.

Relative to the reference epoch $t_j$, the motion of the target system
barycenter (TSB) viewed from the solar system barycenter is
\begin{eqnarray}
  \alpha_{*i}^b&=&\alpha_{*j}^b+\mu_{\alpha,j}\Delta t_{ij}~,\\
  \delta^b_{i}&=&\delta_j^b+\mu_{\delta,j}\Delta t_{ij}~,
\end{eqnarray}
where $\Delta t_{ij}=t_i-t_j$, $\alpha_{*j}^b\equiv \alpha_j^b/\cos\delta_j^b$, $\delta^b_j$, $\mu_{\alpha,j}^b$, $\mu_{\delta,j}^b$ are respectively R.A, Decl., and
proper motions in R.A. and Decl. at epoch $t_j$. Here
$\mu_\alpha\equiv \mu_{\alpha*}=\dot\alpha \cos\delta$. In this work, we only model the
deviation from the motion determined by the reference
astrometry. Hence the TSB relative motion is
\begin{eqnarray}
  \Delta
  \alpha_{*i}^b&=&(\alpha_{*j}^b-\alpha_*^{\rm{ref}})+(\mu_{\alpha,j}^b-\mu_{\alpha}^{\rm
                   ref})\Delta t_{ij}~,\\
  \Delta \delta_{i}^b&=&(\delta_j^b-\delta^{\rm{ref}})+(\mu_{\delta,j}^b-\mu_{\delta}^{\rm
                   ref})\Delta t_{ij}~,
\end{eqnarray}
where $(\alpha_{*j}^b,\delta_j^b,\mu_{\alpha,j}^b,\mu_{\delta,j}^b)$
represents the TSB astrometry at the reference
epoch $t_j$. By defining $\Delta \alpha_j^b=\alpha_{*j}^b-\alpha_*^{\rm ref}$,
$\Delta \delta_j^b=\delta_j^b-\delta^{\rm{ref}}$,
$\Delta\mu_{\alpha,j}=\mu^b_{\alpha,j}-\mu_{\alpha,j}^{\rm ref}$, and $\Delta\mu_{\delta,j}=\mu^b_{\delta,j}-\mu_{\delta,j}^{\rm ref}$,we have
\begin{eqnarray}
  \Delta \alpha_{i}^b&=&\Delta \alpha_j^b+\Delta\mu_{\alpha,j}^b\Delta t_{i}~,\\
  \Delta \delta_{i}^b&=&\Delta \delta_j^b+\Delta\mu_{\delta,j}^b\Delta t_{i}~.
\end{eqnarray}
The above propagation of the TSB does not account for second or higher
order effects such as perspective acceleration. We consider these
effects a priori by subtracting them from various catalogs. Because Gaia orbital solutions use the along-scan
(AL) coordinates of the star (i.e., abscissa or IAD), we project the combined motion of
the star onto the AL direction to simulate abscissa\footnote{In this
  work, we define abscissae as AL coordinates relative to the
  reference abscissae. Hence it is actually abscissa residual.}. With the Gaia
scan angle $\theta$ and AL parallax factor $f^{\rm AL}$ from GOST, the synthetic Gaia abscissae is
\begin{eqnarray}
  \xi_{i}&=&(\Delta \alpha_{i}^b+\Delta
              \alpha_{i}^r)\sin\theta_i+(\Delta \delta_{i}^b+\Delta
              \delta_{i}^r)\cos\theta_i+\Delta \varpi f^{\rm AL}_{i}\nonumber\\
  &=&(\Delta \alpha^b+\Delta\mu_{\alpha}^b\Delta
  t_{i}+BC^t_i+GS^t_i)S^p_i+(\Delta
  \delta^b+\Delta\mu_\delta^b\Delta
  t_{i}+AC^t_i+FS^t_i)C^p_i+\Delta \varpi^b f^{\rm AL}_i~,\label{eq:xi}
\end{eqnarray}
where $\Delta t_i=t_i-t_{\rm ref}$, $S^p_i\equiv \sin\theta_i$,
$C^p_i\equiv \cos\theta_i$. We define the parameters of TSB astrometry at the Gaia DR3 reference epoch as 
$\vec{\beta}^b\equiv(\Delta\alpha^b,\Delta\delta^b,\Delta\varpi^b,\Delta\mu_\alpha^b,\Delta\mu_\delta^b)^T$
and reflex-motion parameters as $\vec{\beta}^r\equiv({\rm A, B, F, G})^T$.
The coefficients of  $\vec\beta^b$ are defined as
$\bm{\kappa^r}\equiv(\overrightarrow{CC},\overrightarrow{CS},\overrightarrow{SC},\overrightarrow{SS})$,
where $\overrightarrow{CC}\equiv
(C_1^tC_1^p,C_2^tC_2^p,...,C_N^tC_N^p)^T$, $\overrightarrow{CS}\equiv
(C_1^tS_1^p,C_2^tS_2^p,...,C_N^tS_N^p)^T$, $\overrightarrow{SC}\equiv
(S_1^tC_1^p,S_2^tC_2^p,...,S_N^tC_N^p)^T$, $\overrightarrow{SS}\equiv
(S_1^tS_1^p,S_2^tS_2^p,...,S_N^tS_N^p)^T$ and $N$ is the number of
data points. The coefficient matrix of $\vec\beta^r$ is
$\bm{\kappa^b}=(\vec{S}^p,\vec{C}^p,\vec{f}^{\rm
  AL},\overrightarrow{TS},\overrightarrow{TC})$, where $\vec{S}^p$ and
$\vec{C}^p$ are one-column vectors with a length of $N$, 
$\overrightarrow{TS}=(\Delta t_1S_1^p,\Delta t_2S_2^p,...,\Delta
t_NS_N^p)$, and $\overrightarrow{TC}=(\Delta t_1C_1^p,\Delta
t_2C_2^p,...,\Delta t_NC_N^p)$. With these definitions, the synthetic abscissae could be written as $\vec\xi=\bm{\kappa^b}\vec\beta^b+\bm{\kappa^r}\vec\beta^r$.

There are five astrometric parameters for the TSB of targets in GDR2
and GDR3. For a three-catalog combination, the catalog cross-matched with GDR2 and GDR3 falls into one of the following categories:
\begin{itemize}
\item {\bf GDR1 five-parameter solutions (G1P5):}. Stars without TYC data in GDR1 only have R.A. and Decl. given. In this case, only position is fitted to the GOST synthetic data, denoted as $\hat{\vec{\beta}}^b_{\rm GDR1}=(\Delta \hat\alpha_{\rm GDR1}, \Delta \hat\delta_{\rm GDR1}, 0, 0, 0)^T$. The model is described in subsection \ref{sec:gaia}.
\item {\bf TYC:}  Given that TYC proper motions are derived through combined analyses of TYC and previous photographic catalogs \citep{hog00}, we use only the TYC R.A. and Decl. to constrain the orbital parameters. Although the abscissae of TYC are not provided, we model the position of TYC, and the analytical solution is described in subsection \ref{sec:tyc}.
\item {\bf Hipparcos:} For stars with Hipparcos data, GDR1 solutions are not independent of the Hipparcos data. We utilize the Hipparcos IAD in combination with GDR2 and GDR3 catalog data to constrain binary orbits. The detailed modeling of Hipparcos IAD is described in subsection \ref{sec:hip}.
\end{itemize}

\subsection{Modeling Gaia catalog data}\label{sec:gaia}
The five-parameter model for synthetic abscissae for the Gaia DR$j$ is
\begin{equation}
  \hat{\vec\xi}_j=\bm{\kappa^b_j}\hat{\vec\beta}_j~,
\end{equation}
where $\hat{\vec\beta}_j$ is the astrometry fitted for Gaia DR$j$. The
astrometry of Gaia DR$j$ relative to the reference astrometry (Gaia
DR3 by default) is $\vec\beta_j$. We define the synthetic abscissae
corresponding to the three Gaia DRs as
$\vec{\xi}=(\vec\xi_{\rm GDR1}^T,\vec\xi_{\rm GDR2}^T,\vec\xi_{\rm GDR3}^T)^T$. Then the residual is
\begin{eqnarray}
  \delta\vec{\xi}&=&\vec{\xi}-\hat{\vec{\xi}}\nonumber\\
  &=&\begin{bmatrix}
    \bm{\kappa^b_{\rm GDR1}}\\
    \bm{\kappa^b_{\rm GDR2}}\\
    \bm{\kappa^b_{\rm GDR3}}
  \end{bmatrix}\vec\beta^b+
  \begin{bmatrix}
    \bm{\kappa^r_{\rm GDR1}}\\
    \bm{\kappa^r_{\rm GDR2}}\\
    \bm{\kappa^r_{\rm GDR3}}
  \end{bmatrix}\vec\beta^r-
  \begin{bmatrix}
    \bm{\tilde{\kappa}^b_{\rm GDR1}}\hat{\vec\beta}^b_{\rm GDR1}\\
    \bm{\tilde{\kappa}^b_{\rm GDR2}}\hat{\vec\beta}^b_{\rm GDR2}\\
    \bm{\tilde{\kappa}^b_{\rm GDR3}}\hat{\vec\beta}^b_{\rm GDR3}
  \end{bmatrix}
  \nonumber\\
  &=&\bm{\mu^b}\vec\beta^b+\bm{\mu^r}\vec\beta^r-\bm{\mu}\hat{\vec\beta}^b~,
\end{eqnarray}
where $\tilde{\kappa}_{\rm GDRn}$ is similar to $\kappa_{\rm GDRn}$
but the reference time is now the reference epoch of GDRn, and
\[\bm{\mu^b}=\begin{bmatrix}
    \bm{\kappa^b_{\rm GDR1}}\\
    \bm{\kappa^b_{\rm GDR2}}\\
    \bm{\kappa^b_{\rm GDR3}}
  \end{bmatrix},~
\bm{\mu^r}=\begin{bmatrix}
    \bm{\kappa^r_{\rm GDR1}}\\
    \bm{\kappa^r_{\rm GDR2}}\\
    \bm{\kappa^r_{\rm GDR3}}
  \end{bmatrix},~
\bm{\mu}=\begin{bmatrix}
    \bm{\tilde{\kappa}_{\rm GDR1}^b}&&\\
    &\bm{\tilde{\kappa}_{\rm GDR2}^b}&\\
    &&\bm{\tilde{\kappa}_{\rm GDR3}^b}
  \end{bmatrix},~
  \hat{\vec\beta}^b=\begin{bmatrix}
    \hat{\vec\beta}^b_{\rm GDR1}\\
    \hat{\vec\beta}^b_{\rm GDR2}\\
    \hat{\vec\beta}^b_{\rm GDR3}
  \end{bmatrix}.
  \]

  For GDR1 targets without Tycho data, only the positions are
  measured. To be compatible with GDR2 and GDR3 solutions, the
  GDR1 proper motions and parallax are fixed at zero and the covariance matrix is
\[
 \bm{\Sigma_{\rm GDR1}}=\begin{bmatrix}
   \sigma_{\alpha*}^2&\rho\sigma_{\alpha*}\sigma_{\delta}&0&0&0\\
   \rho\sigma_{\alpha*}\sigma_{\delta}&\sigma_\delta^2&0&0&0\\
   0&0&\sigma_\varpi^2&0&0\\
   0&0&0&\sigma_{\mu_\alpha}^2&0\\
   0&0&0&0&\sigma_{\mu_\delta}^2
 \end{bmatrix},
\]
where $\rho$ is the the correlation between $\alpha*$ and
$\delta$ for GDR1, the parallax uncertainty $\sigma_\varpi$ is set
to 1000\mas, and the proper motion uncertainties,
$\sigma_{\mu_\alpha}$ and $\sigma_{\mu_\delta}$, are set to 1000\masyr.

The $\chi^2$ of the above model of abscissae is
$\chi^2_\xi=\delta\vec{\xi}^T\bm{\Sigma_\xi}^{-1}\delta\vec{\xi}$,
where $\bm{\Sigma_\xi}$ is the covariance of abscissae. Assuming
that the synthetic abscissae have constant measurement errors $\sigma_\xi$,
$\chi^2_\xi$ could be simplified as
$\chi^2_\xi=\delta\vec{\xi}^T\cdot\delta\vec{\xi}/\sigma^2$. The minimization
of $\chi^2_\xi$ leads to
\begin{equation}
  \pdv{\chi^2_\xi}{\hat{\vec\beta}^b}=-2(\bm{\mu^b}\vec\beta^b+\bm{\mu^r}\vec\beta^r-\bm{\mu}\hat{\vec\beta}^b)^T\bm{\mu}=0~.
\end{equation}
Transposing both sides of the above equation leads to 
\begin{equation}
  \bm{\mu}^T\bm{\mu}\hat{\vec\beta}^b=\bm{\mu}^T(\bm{\mu^b}\vec\beta^b+\bm{\mu^r}\vec\beta^r).
\end{equation}
and thus
\begin{equation}
  \hat{\vec\beta}^b=(\bm{\mu}^T\bm{\mu})^{-1}\bm{\mu}^T(\bm{\mu^b}\vec\beta^b+\bm{\mu^r}\vec\beta^r).
\end{equation}
The coefficient matrices of $\vec{\beta}^b$ and $\vec{\beta}^r$ are respectively
\begin{equation}
  \bm{\lambda}\equiv
  \begin{bmatrix}
    \bm{\lambda_{\rm GDR1}}\\
    \bm{\lambda_{\rm GDR2}}\\
    \bm{\lambda_{\rm GDR3}}\\
  \end{bmatrix}
  =
  \begin{bmatrix}
    [\bm{(\tilde{\kappa}_{\rm GDR1}^b)^T}\bm{\tilde{\kappa}_{\rm GDR1}^b}]^{-1}\bm{(\tilde{\kappa}_{\rm GDR1}^b)^T}\bm{\kappa_{\rm GDR1}^b}\\
   [\bm{(\tilde{\kappa}_{\rm GDR2}^b)^T}\bm{\tilde{\kappa}_{\rm GDR2}^b}]^{-1}\bm{(\tilde{\kappa}_{\rm GDR2}^b)^T}\bm{\kappa_{\rm GDR2}^b}\\
   [\bm{(\tilde{\kappa}_{\rm GDR3}^b)^T}\bm{\tilde{\kappa}_{\rm GDR3}^b}]^{-1}\bm{(\tilde{\kappa}_{\rm GDR3}^b)^T}\bm{\kappa_{\rm GDR3}^b}
 \end{bmatrix}
\label{eq:lambda}
\end{equation}
and
\begin{equation}
  \bm{\gamma}\equiv
  \begin{bmatrix}
    \bm{\gamma_{\rm GDR1}}\\
    \bm{\gamma_{\rm GDR2}}\\
    \bm{\gamma_{\rm GDR3}}\\
  \end{bmatrix}
  =
  \begin{bmatrix}
    [\bm{(\tilde{\kappa}_{\rm GDR1}^b)^T}\bm{\tilde{\kappa}_{\rm GDR1}^b}]^{-1}\bm{(\tilde{\kappa}_{\rm GDR1}^b)^T}\bm{\kappa_{\rm GDR1}^r}\\
   [\bm{(\tilde{\kappa}_{\rm GDR2}^b)^T}\bm{\tilde{\kappa}_{\rm GDR2}^b}]^{-1}\bm{(\tilde{\kappa}_{\rm GDR2}^b)^T}\bm{\kappa_{\rm GDR2}^r}\\
   [\bm{(\tilde{\kappa}_{\rm GDR3}^b)^T}\bm{\tilde{\kappa}_{\rm GDR3}^b}]^{-1}\bm{(\tilde{\kappa}_{\rm GDR3}^b)^T}\bm{\kappa_{\rm GDR3}^r}
  \end{bmatrix}
   \label{eq:gamma}
  ~.
\end{equation}
Hence 
\begin{equation}
  \hat{\vec\beta}^b=\bm{\lambda}\vec\beta^b+\bm{\gamma}\vec\beta^r.
\end{equation}

Because we do not consider perspective acceleration in the above linear modeling of synthetic GOST data, we propagate the
GDR3 astrometry to the reference epochs of GDR1 and GDR2, and
define the astrometric data as
$\vec{y}=(\vec\beta_{\rm GDR1},\vec\beta_{\rm GDR2},\vec\beta_{\rm GDR3})^T$. We
calculate the residual vector by subtracting $\hat{\vec\beta}^b$ from $\vec{y}$,
\begin{equation}
  \delta\vec{y}=\vec{y}-(\bm{\lambda}\vec\beta^b+\bm{\gamma}\vec\beta^r).
  \label{eq:beta}
\end{equation}
The $\chi^2$ for this model of catalog astrometry is
\begin{equation}
  \chi^2_{\rm
    GDR}=\delta\vec{y}^T\bm{\Sigma_{\beta}}^{-1}\delta\vec{y}~,
  \label{eq:sigma_beta}
\end{equation}
where
\[
  \bm{\Sigma_\beta}^{-1}=
  \begin{bmatrix}
    \bm{\Sigma_{\rm GDR1}}^{-1}&&\\
    &\bm{\Sigma_{\rm GDR2}}^{-1}&\\
    &&\bm{\Sigma_{\rm GDR3}}^{-1}\\
    \end{bmatrix}~,
\] $\bm{\Sigma_{\rm GDR1}}$, $\bm{\Sigma_{\rm GDR2}}$, and $\bm{\Sigma_{\rm GDR3}}$ are respectively the
covariances for Gaia DR1, DR2, and DR3. The minimization of $\chi^2_{\rm GDR}$ leads to 
\begin{align}\label{eq:dchi2}
\pdv{\chi^2_{\rm GDR}}{\vec\beta^b}&=2(\bm{\lambda}\vec\beta^b+\bm{\gamma}\vec{\beta}^r-\vec{y})^T\bm{\Sigma_\beta}^{-1}\bm{\lambda}=0~,\\
\pdv{\chi^2_{\rm GDR}}{\vec\beta^r}&=2(\bm{\lambda}\vec\beta^b+\bm{\gamma}\vec{\beta}^r-\vec{y})^T\bm{\Sigma_\beta}^{-1}\bm{\gamma}=0~.
\end{align}
The above equation array could be simplified as 
\begin{equation}
  \bm{\eta}\vec\beta^{br}=\vec{b}~,
  \label{eq:betabr}
\end{equation}
where
\[
  \bm{\eta}=
               \begin{bmatrix}
                 \bm{\lambda}^T\bm{\Sigma_\beta}^{-1}\bm{\lambda}&\bm{\lambda}^T\bm{\Sigma_\beta}^{-1}\bm{\gamma}\\
                 \bm{\gamma}^T\bm{\Sigma_\beta}^{-1}\bm{\lambda}&\bm{\gamma}^T\bm{\Sigma_\beta}^{-1}\bm{\gamma}
               \end{bmatrix},~
  \vec{\beta}^{br}=
  \begin{bmatrix}
    \vec{\beta}^b\\
    \vec{\beta}^r
  \end{bmatrix},~
  \vec{b}=
  \begin{bmatrix}
   \bm{\lambda}^T\bm{\Sigma_\beta}^{-1}\vec{y}\\
   \bm{\gamma}^T\bm{\Sigma_\beta}^{-1}\vec{y}\\
  \end{bmatrix}.
\]
Then the optimal parameters are
$\vec\beta^{br}=\bm{\eta}^{-1}\vec{b}$ and the corresponding
covariance is $\bm{C^{br}}=\bm{\eta}^{-1}$.
The baseline model is simply a covariance-weighted mean of the catalog astrometry,
\begin{equation}
\vec{\beta}_0^b=\bm{C_0^b}(\bm{\Sigma_{\rm GDR1}}^{-1}\vec\beta_{\rm GDR1}+\bm{\Sigma_{\rm GDR2}}^{-1}\vec\beta_{\rm GDR2}+\bm{\Sigma_{\rm GDR3}}^{-1}\vec\beta_{\rm GDR3}),
\end{equation}
where $\bm{C_0^b}=(\bm{\Sigma_{\rm GDR1}}^{-1}+\bm{\Sigma_{\rm GDR2}}^{-1}+\bm{\Sigma_{\rm GDR3}}^{-1})^{-1}$ measures the parameter covariance.

\subsection{Modeling TYC data}\label{sec:tyc}
For targets with TYC positional data, the target position relative to
the barycenter is
\begin{align}
  \Delta\hat\alpha&=\Delta\alpha+BC_i^t+GS_i^t~,\\
  \Delta\hat\delta&=\Delta\delta+AC_i^t+FS_i^t~.
\end{align}
The vectorized version is
\begin{equation}
  \hat{\vec{y}}_{\rm TYC}=\bm{\lambda}_{\rm TYC}\vec{\beta}^b+\bm{\gamma}_{\rm TYC}\vec{\beta}^r~,
\end{equation}
where
\begin{align}
  \hat{\vec{y}}_{\rm TYC}&=(\Delta\hat\alpha,\Delta\hat\delta)^T\\
  \bm{\lambda_{\rm TYC}}&=
                          \begin{bmatrix}
                            1&0&0&0&0\\
                            0&1&0&0&0\\
                          \end{bmatrix}\\
  \bm{\gamma_{\rm TYC}}&=
                          \begin{bmatrix}
                            0&C_i^t&0&S_i^t\\
                            C_i^t&0&S_i^t&0\\
                          \end{bmatrix}~.
  \label{eq:tyc}
\end{align}
Hence the residual is
\begin{align}
  \delta \vec{y}_{\rm TYC}=\vec{y}_{\rm TYC}-\hat{\vec{y}}_{\rm TYC}~,
\end{align}
where $\vec{y}_{\rm TYC}\equiv \vec{\beta}_{\rm TYC}=(\Delta
\alpha_{\rm TYC},\Delta \delta_{\rm TYC})$ is the TYC position
relative to the position propagated from Gaia DR3 to the TYC R.A. and
Decl. reference epochs. With the above definitions, $\bm{\lambda}$ and $\bm{\gamma}$ in eq. \ref{eq:lambda} become
\[
  \bm{\lambda}=
  \begin{bmatrix}
    \bm{\lambda_{\rm TYC}}\\
    \bm{\lambda_{\rm GDR2}}\\
    \bm{\lambda_{\rm GDR3}}
  \end{bmatrix},
  \bm{\gamma}= 
  \begin{bmatrix}
    \bm{\gamma_{\rm TYC}}\\
    \bm{\gamma_{\rm GDR2}}\\
    \bm{\gamma_{\rm GDR3}}\\
  \end{bmatrix},
  \bm{\Sigma_\beta}^{-1}=
  \begin{bmatrix}
    \bm{\Sigma_{\rm TYC}}^{-1}&&\\
    &\bm{\Sigma_{\rm GDR2}}^{-1}&\\
    &&\bm{\Sigma_{\rm GDR3}}^{-1}
  \end{bmatrix},
  \vec{y}=
                 \begin{bmatrix}
                   \vec\beta_{\rm TYC}\\
                   \vec\beta_{\rm GDR2}\\
                   \vec\beta_{\rm GDR3}
                 \end{bmatrix}.\\
\]
The optimal parameter vector for the corresponding baseline model is
\begin{equation}
  \vec\beta^b_0=\bm{C_0^b}[(\bm{\lambda_{\rm TYC}}^T\bm{\Sigma_{\rm TYC}}^{-1}\vec\beta_{\rm TYC}+\bm{\Sigma_{\rm GDR2}}^{-1}\vec\beta_{\rm GDR2}+\bm{\Sigma_{\rm GDR3}}^{-1}\vec\beta_{\rm GDR3}],
\end{equation}
where covariance $\bm{C_0^b}=[\bm{\lambda_{\rm TYC}}^T\bm{\Sigma_{\rm HIAD}}^{-1}\bm{\lambda_{\rm TYC}}+\bm{\Sigma_{\rm GDR2}}^{-1}+\bm{\Sigma_{\rm GDR3}}^{-1}]^{-1}$.

\subsection{Modeling Hipparcos intermediate data}\label{sec:hip}
Because the astrometric offset parameters shown in eq. \ref{eq:xi} are
defined with respect to GDR3, the Hipparcos raw IAD ($\vec\xi_{\rm
  raw}$) are transformed into relative Hipparcos IAD following
\begin{equation}
  \vec\xi_{\rm HIAD}=\vec\xi_{\rm raw}+\bm{\tilde{\kappa}_{\rm HIAD}^b}\vec{\beta}^b_{\rm HG}~,
\label{eq:hiad_correct}
\end{equation}
where $\vec{\beta}^b_{\rm HG}$ is the difference between
Hippacos astrometry and the astrometry propagated from the GDR3 epoch
to the Hipparcos epoch. After this correction, the Hipparcos abscissa
model is the same as that in eq. \ref{eq:xi}. The Gaia scan angle
$\theta$ is complementary to the Hipparcos scan angle $\psi$,
i.e. $\theta=\pi/2-\psi$. The residual is
\begin{equation}
  \delta\vec\xi_{\rm HIAD}=\vec\xi_{\rm HIAD}-(\bm{\kappa_{\rm HIAD}^b}\vec\beta^b+\bm{\kappa_{\rm HIAD}^r}\vec\beta^r)~.
\end{equation}
Here we use $\bm \kappa_{\rm HIAD}^b$ (with GDR3 as the reference epoch) instead of $\tilde{\bm \kappa}_{\rm HIAD}^b$ because the Hipparcos abscissae are already modified to be
compatible with the GDR3 solution in eq. \ref{eq:hiad_correct}.

The $\chi^2$ for $\delta\vec\xi_{\rm HIAD}$ is
\begin{equation}
\chi^2_{\rm HIAD}=\delta\vec\xi_{\rm HIAD}^T\bm{\Sigma_{\rm HIAD}}^{-1}\delta\vec\xi_{\rm HIAD}~,
\end{equation}
where $\bm{\Sigma_{\rm HIAD}}^{-1}$ is the covariance matrix of Hipparcos IAD. This covariance is
${\rm diag}(\sigma_1^2,\sigma_2^2,...,\sigma_{N_{\rm HIAD}}^2)$, where $N_{\rm HIAD}$ is the
number of Hipparcos epochs. The minimization of $\chi^2$ leads to
\begin{equation}
  \begin{bmatrix}
    (\bm{\kappa_{\rm HIAD}^b})^T\bm{\Sigma_{\rm HIAD}}^{-1}\bm{\kappa_{\rm HIAD}^b}&(\bm{\kappa_{\rm HIAD}^b})^T\bm{\Sigma_{\rm HIAD}}^{-1}\bm{\kappa_{\rm HIAD}^r}\\
    (\bm{\kappa_{\rm HIAD}^r})^T\bm{\Sigma_{\rm HIAD}}^{-1}\bm{\kappa_{\rm HIAD}^b}&(\bm{\kappa_{\rm HIAD}^r})^T\bm{\Sigma_{\rm HIAD}}^{-1}\bm{\kappa_{\rm HIAD}^r}
  \end{bmatrix}
  \cdot
    \vec{\beta}^{br}
  =
  \begin{bmatrix}
    (\bm{\kappa_{\rm HIAD}^b})^T\cdot
  \vec\xi_{\rm HIAD}\\
    (\bm{\kappa_{\rm HIAD}^r})^T\cdot
  \vec\xi_{\rm HIAD}
  \end{bmatrix}
  ~,
  \label{eq:hiad}
\end{equation}
where $\vec{\beta}^{br}=(\vec{\beta}^b,\vec{\beta}^r)^T$. To model
Hipparcos IAD together with Gaia DRs, we insert eq. \ref{eq:hiad} into
the definition of $\bm{\lambda}$ in eq. \ref{eq:lambda}, $\bm{\gamma}$
in eq. \ref{eq:gamma}, $\bm{\Sigma_\beta}^{-1}$ in eq. \ref{eq:sigma_beta} and $\vec{y}$ in
eq. \ref{eq:beta} as follows:
\[
  \bm{\lambda}=
  \begin{bmatrix}
    \bm{\kappa^b_{\rm HIAD}}\\
    \bm{\lambda_{\rm GDR2}}\\
    \bm{\lambda_{\rm GDR3}}\\
  \end{bmatrix},
  \bm{\gamma}= 
  \begin{bmatrix}
    \bm{\kappa^r_{\rm HIAD}}\\
    \bm{\gamma_{\rm GDR2}}\\
    \bm{\gamma_{\rm GDR3}}\\
  \end{bmatrix},
  \bm{\Sigma_\beta}^{-1}=
  \begin{bmatrix}
    \bm{\Sigma_{\rm HIAD}}^{-1}&&\\
    &\bm{\Sigma_{\rm GDR2}}^{-1}&\\
    &&\bm{\Sigma_{\rm GDR3}}^{-1}
  \end{bmatrix},
  \vec{y}=
                 \begin{bmatrix}
                   \vec\xi_{\rm HIAD}\\
                   \vec\beta_{\rm GDR2}\\
                   \vec\beta_{\rm GDR3}
                 \end{bmatrix}.\\
\]
With these new definitions, eq. \ref{eq:betabr} is still valid to find
solutions of $\vec\beta^b$ and $\vec\beta^r$. The optimal parameter vector
for the corresponding baseline model becomes
\begin{equation}
  \vec\beta^b_0=\bm{C_0^b}[(\bm{\kappa_{\rm HIAD}^b})^T\bm{\Sigma_{\rm HIAD}}^{-1}\vec\xi_{\rm HIAD}+\bm{\Sigma_{\rm GDR2}}^{-1}\vec\beta_{\rm GDR2}+\bm{\Sigma_{\rm GDR3}}^{-1}\vec\beta_{\rm GDR3}],
\end{equation}
where covariance $\bm{C_0^b}=[(\bm{\kappa_{\rm HIAD}^b})^T\bm{\Sigma_{\rm HIAD}}^{-1}\bm{\kappa_{\rm HIAD}^b}+\bm{\Sigma_{\rm GDR2}}^{-1}+\bm{\Sigma_{\rm GDR3}}^{-1}]^{-1}$.

\subsection{Modeling eccentric orbits}\label{sec:ecc}
By sampling orbital periods and minimizing $\chi^2$ for each period,
we find astrometric signals with circular orbits. This linear
least-square optimization provides initial parameters for more robust
non-linear regressions to find solutions for eccentric orbits. The Keplerian motion of system photocenter in the orbital plane is
\begin{eqnarray}
  X_i&=&a_p(\cos E_i-e)~,\nonumber\\
  Y_i&=&a_p(\sqrt{1-e^2}\sin E_i)~,\nonumber\\
  \dot{X}_i&=&-\frac{2\pi a_p}{P\varpi}\frac{\sin E_i}{1-e\cos E_i}~,\nonumber\\
  \dot{Y}_i&=&\frac{2\pi a_p}{P\varpi}\sqrt{1-e^2}\frac{\cos E_i}{1-e\cos E_i}~,
  \label{eq:keplerian}
\end{eqnarray}
where $P$ is orbital period, $e$ is eccentricity, $E_i$ is the
eccentric anomaly at time $t_i$ and can be determined by solving the Kepler's equation:
\begin{equation}
  M_i=E_i-e\sin E_i~,
\end{equation}
where $M_i$ is the mean anomaly at $t_i$. The mean anomaly is 
\begin{equation}
M_i=M_0+\frac{2\pi\Delta t_i}{P},
\end{equation}
where $M_0$ is the mean anomaly at the reference epoch. Compared with
circular orbits, a Keplerian orbit has two nonlinear parameters, $e$ and
$M_0$, in addition to orbital period $P$. Hence there
would be three nonlinear parameters for a Keplerian model. We define
this set of nonlinear Keplerian parameters as
$\vec\beta^n\equiv (P,e,M_0)^T$. Hence the parameter vector
for a full astrometry model is
\begin{equation}
  \vec\beta^{brn}=
  \begin{bmatrix}
    \vec\beta^b\\
    \vec\beta^r\\
    \vec\beta^n
  \end{bmatrix}.
\end{equation}

For a given $\vec\beta^n$, $C_i^t$ and $S_i^t$ in eq. \ref{eq:adr} are
respectively replaced by $C_i^{tr}\equiv \cos E_i-e$ and $S_i^{tr}\equiv
\sqrt{1-e^2}\sin E_i$. Following the procedures introduced in section \ref{sec:gaia}, \ref{sec:hip}, and
\ref{sec:ecc}, the linear parameters can be analytically optimized for
a given $\vec\beta^n$. Thus we only need to optimize $\vec\beta^n$
nonlinearly using algorithms such as the Levenberg-Marquardt optimization algorithm \citep{levenberg44,marquardt63}. The Thiele Innes elements $A$, $B$, $F$, $G$, can be converted to the
Campbell elements, $a_p$, $I$,$\omega_T$, and $\Omega$ although
$\omega_T$ and $\Omega$ are respectively degenerated with $\omega_T+\pi$ and
$\Omega+\pi$. 

\subsection{Detection threshold}\label{sec:threshold}
The preceding subsections outline the analytical formulae for computing optimal parameters through $\chi^2$ minimization. Assuming uniform priors for model parameters and employing Laplace's approximation \citep{schwarz78, kass95}, we transform the minimum $\chi^2$ of two models into the logarithmic Bayes Factor (lnBF) using the following equations:

\begin{align}
\ln{\rm BF}_{21} &= \ln\mathcal{L}_2 - \ln\mathcal{L}_1 - \frac{1}{2}\nu \ln{N}\\\nonumber
&= \frac{1}{2}(\chi^2_1 - \chi^2_2) - \frac{1}{2}\nu \ln{N}~,
\end{align}

Here, $\mathcal{L}_1$ and $\mathcal{L}_2$ represent the maximum
likelihood for models 1 and 2, while $\chi^2_1$ and $\chi^2_2$ denote
the respective minimum $\chi^2$ values for models 1 and 2. The
variable $\nu$ stands for the number of extra free parameters in model
2 compared to model 1, and $N$ represents the number of data
points. We consider lnBF$> 5$ as the threshold for identifying statistically significant signals. This is based on section 3.2 and table 2 of \cite{kass95}, where the criteria for decisive evidence in favor of a hypothesis are stated as 2lnBF $> 10$ or lnBF $> 5$ or BF $> 150$. We also note that \cite{feng16} shows that lnBF $> 5$ is suitable for identifying signals across various noise properties, particularly in the context of exoplanet detection. However, we note that our study, not yet having developed the modeling framework into a periodogram for signal detection, is not highly sensitive to the choice of threshold. We include a detection threshold primarily for the sake of completeness.

\section{Catalog calibration}\label{sec:calibrate}
\subsection{Calibration sources}
The models introduced in section \ref{sec:method} assume unbiased
catalog data, a presumption that might not hold true, especially in
combined analyses of different catalogs. While it's possible to model
bias concurrently with stellar reflex motion on a case-by-case basis,
as demonstrated in \cite{feng19b}, such bias models often involve
nonlinear parameters, impeding analytical optimization and diminishing
the efficiency of periodogram computation. Consequently, we
proactively address potential astrometric bias through calibration,
leveraging both GDR3 NSS and single-star (SS) catalogs.
While we opt for both NSS and SS as representations of distinct
astrometric solutions, it is important to note that our choice of calibration sources lacks comprehensiveness in terms of capturing various magnitudes, colors, sky positions, and other pertinent parameters. In subsection \ref{sec:ruwe}, we will evaluate the extent of contamination from undetected binaries in the SS catalogs by examining the correlation between calibration parameters and the renormalized unit weight error (RUWE; \citealt{lindegren18}).

In the calibration using the NSS catalog, our attention is directed
towards the 134,598 NSS targets within the GDR3
\verb|nss_two_body_orbit| table of the {\tt Orbital}
type\footnote{This catalog is for non-single-star orbital models for
  sources compatible with orbital model for an astrometric binary with
  Campbell orbital elements, or Vizier catlaog with a designation of
  I/357/tbooc.}, collectively referred to as ``G3NS''. We crossmatch
G3NS with GDR2 and catalog CAT1 to create a hybrid calibration
catalog. CAT1 encompasses two-parameter GDR1 (G1P2), five-parameter
GDR1 (G1P5), TYC, the original (HIP1), or the revised (HIP2) Hipparcos
catalogs, resulting in hybrid samples with 130,096, 116,271, 13,825,
14,947, 142, and 142 targets, respectively. The barycentric astrometry
for a G3NS target at the GDR3 reference epoch can be either the G3NS
catalog's provided barycentric value (G3NSB) or calculated by
subtracting the photocentric motion from the target's photocentric
astrometry (G3NST). This leads to designations such as NST, NSTP2,
NSTP5, NSTT, NSTH2, and NSTH1 for G3NST cross-matched with GDR2, G1P2,
G1P5, TYC, HIP1, and HIP2, respectively. Similary, we define NSB, NSBP2,
NSBP5, NSBT, NSBH2, and NSBH1 for G3NSB cross-matched with GDR2, G1P2,
G1P5, TYC, HIP1, and HIP2, respectively. 

Recognizing that Gaia systematics may vary with solution types
\citep{lindegren21a}, we cross-match the GDR3 single-star catalog
(G3SS)\footnote{The G3SS samples comprise GDR3 targets not included in
  the Gaia NSS catalog.} with GDR2 and the Hipparcos catalog to obtain
a sample of 77,138 single-star targets (termed ``G3SS'') for the
calibration of bright and nearby stars. This sample is further
cross-matched with GDR2, G1P2, G1P5, TYC, HIP1, and HIP2, resulting in
SS2, SS2P2, SS2P5, SS2T, SS2H2, and SS2H1 catalogs, encompassing
77,138, 6,763, 70,676, 74,216, 77,136, and 77,138 targets,
respectively. Because the Hipparcos and TYC sources were observed by
Gaia while most of the Gaia NSS sources were not observed by Hipparcos
or TYC, the SS calibration catalogs have similar sample size
while the NSS calibration catalogs have quite different sample size. 

\subsection{Calibration models}\label{sec:calmodel}
We employ the calibration sources introduced in the previous section to ascertain the relative frame rotation and offset between GDR3 and other catalogs. Given the primary objective of this series of papers—to identify companions using multiple astrometric catalogs—our focus lies solely on the differential frame rotation and parallax zero-point. Considering that our typical detection involves companions closer than 1\,kpc, the absolute parallax zero-point of GDR3, approximately 0.05\,mas \citep{lindegren21a}, does not significantly impact the detection process.

The relative difference between two reference frames is
measured by a rotation vector $\vec\omega$ and a constant offset vector
$\vec\epsilon$:
\begin{equation}
  \vec{\epsilon}(t)={\vec{\epsilon}}(t_{\rm GDR3})+{\vec{\omega}}(t-t_{\rm GDR3})~,
\end{equation}
where $t_{\rm GDR3}$ is the reference epoch of GDR3. We denote
$\vec{\epsilon}(t_{\rm GDR3})$ by
$\vec{\epsilon}_0\equiv(\epsilon_x,\epsilon_y,\epsilon_z)^T$. Additionally,
the zero point of parallax is found to be non-zero for Gaia catalogs \citep{lindegren18,arenou18,lindegren21a}. Without using
quasars to calibrate parallax zero-point like \cite{lindegren21a}, we
derive the differential zero-point parallax ($\delta\varpi$) of a catalog
relative to GDR3.

The astrometric bias caused by the differential frame roration and zero-point parallax is
\begin{eqnarray}\label{eq:rd}
  (\tilde{\alpha}-\alpha)\cos\delta&=&\epsilon_X(t)\cos\alpha\sin\delta+\epsilon_Y(t)\sin\alpha\sin\delta-\epsilon_Z(t)\cos\delta~,\nonumber\\
  \tilde{\delta}-\delta&=&-\epsilon_X(t)\sin\alpha+\epsilon_Y(t)\cos\alpha~,\nonumber\\
  \tilde{\varpi}-\varpi&=&\delta\varpi\\
  \tilde{\mu}_\alpha-\mu_\alpha&=&\omega_X\cos\alpha\sin\delta+\omega_Y\sin\alpha\sin\delta-\omega_Z\cos\delta~,\nonumber\\
  \tilde{\mu}_\delta-\mu_\delta&=&-\omega_X\sin\alpha+\omega_Y\cos\alpha~.\nonumber
\end{eqnarray}
where $\tilde{}$ denotes astrometric parameters measured in a given frame
while astrometric parameters without this subscript denotes
parameters measured in the GDR3 frame. Note that we define the
differential rotation and parallax zero-point in a convention such
that the bias is subtracted from the given catalog. 

We vectorize the above equations and get the differential astrometric
bias ($\vec{y}^d$) caused by differential frame rotation and parallax zero-point,
\begin{equation}
  \vec{y}^d=\bm{\kappa^d}\vec{\beta}^d=
  \begin{bmatrix}
    \bm{\kappa}^{\rm GDR23}\\\bm{0}\\
  \end{bmatrix}
  \cdot
  \vec{\beta}^{\rm GDR23}
\label{eq:yd}
\end{equation}
where 
\begin{equation}
  \bm{\kappa}^{\rm GDR23}=
  \begin{bmatrix}
      \cos\alpha\sin\delta&\sin\alpha\sin\delta&-\cos\delta&\Delta t\cos\alpha\sin\delta&\Delta t\sin\alpha\sin\delta&-\Delta t\cos\delta&0\\
      -\sin\alpha&\cos\alpha&0&-\Delta t\sin\alpha&\Delta t\cos\alpha&0&0\\
      0&0&0&0&0&0&1\\
0&0&0&\cos\alpha\sin\delta&\sin\alpha\sin\delta&-\cos\delta&0\\
0&0&0&-\sin\alpha&\cos\alpha&0&0\\
\end{bmatrix}
\label{eq:kappa}
\end{equation}
and
\begin{equation}
  \vec{\beta}^{\rm GDR23}=(\epsilon_x^{\rm GDR23},\epsilon_y^{\rm GDR23},\epsilon_z^{\rm GDR23},\omega_x^{\rm GDR23},\omega_y^{\rm GDR23},\omega_z^{\rm GDR23},\delta\varpi^{\rm GDR23})~.
\end{equation}
The differential frame rotation and parallax zero-point should be
subtracted from the original data $\vec{y}$ to get the corrected data
in the GDR3 reference frame. 

Because we only consider differential frame rotation and parallax
zero-point, GDR3 astrometry is treated as ``unbiased'' and thus the
last seven elements in $\vec{\beta}^d$ and the last five rows in
$\bm{\kappa^d}$ are all zeros. To model the systematics of GDR1 and
GDR2 relative to GDR3 simultaneously, we can respectively add
$\vec\beta^{\rm GDR13}$ and
$\bm{\kappa}^{\rm GDR13}$ to $\vec\beta^d$ and $\bm{\kappa}^d$. Hence
eq. \ref{eq:yd} becomes
\begin{equation}
  \vec{y}^d=\bm{\kappa}^d\vec{\beta}^d=
  \begin{bmatrix}
    \bm{\kappa}^{\rm GDR13}&\bm{0}\\
    \bm{0}&\bm{\kappa}^{\rm GDR23}\\
    \bm{0}&\bm{0}\\
  \end{bmatrix}
  \cdot
  \begin{bmatrix}
    \vec{\beta}^{\rm GDR13}\\
    \vec{\beta}^{\rm GDR23}\\
    \end{bmatrix}
\label{eq:yd1}
\end{equation}

When CAT1 is G1P5, $\bm{\kappa}^{\rm GDR13}$ is
a $5\times 7$ matrix like $\bm{\kappa}^{\rm GDR23}$, and $\vec\beta^{\rm GDR13}$ is a seven-element vector
like $\vec\beta^{\rm GDR23}$. If CAT1 is G1P2,
\begin{equation}
   \bm{\kappa}^{\rm GDR13}=
  \begin{bmatrix}
      \cos\alpha\sin\delta&\sin\alpha\sin\delta&-\cos\delta\\
      -\sin\alpha&\cos\alpha&0\\
    \end{bmatrix}
    \label{eq:gdr13}
\end{equation}
and
\begin{equation}
  \vec{\beta}^{\rm GDR13}=(\epsilon_x^{\rm GDR13},\epsilon_y^{\rm GDR13},\epsilon_z^{\rm GDR13})~.
    \label{eq:beta13}
\end{equation}

When CAT1 is TYC, $\bm{\kappa}^{\rm GDR13}$ and $\vec{\beta}^{\rm GDR13}$ are respectively replaced by
$\bm{\kappa}^{\rm TYC}$ and $\vec{\beta}^{\rm TYC}$ while the forms of
eqs. \ref{eq:yd1} and \ref{eq:gdr13} do not change. If CAT1 is HIP1 or
HIP2, the Hipparcos abscissae (HIAD) induced by frame rotation are
\begin{equation}
  \vec\xi^d={\bm \kappa}^{\rm HIAD}\vec\beta^d,
\end{equation}
and
\begin{equation}
  \bm{\kappa}^{\rm HIAD}=\bm{\kappa}^b\bm{\kappa}^{\rm HIP}~,
\end{equation}
where $\bm{\kappa}^{\rm HIP}$ has the same form as $\bm{\kappa}^{\rm GDR23}$.

Then eq. \ref{eq:beta} becomes
\begin{equation}
  \delta\vec{y}=\vec{y}-(\bm{\lambda}\vec\beta^d+\bm{\gamma}\vec\beta^r+\vec{y}^d).
  \label{eq:y2}
\end{equation}
When G3NST is used for calibration, $\beta^r$ is
given by the catalog and is subtraced from the catalog astrometry,
$\vec{y}'=\vec{y}-\bm{\gamma}\vec{\beta}^r$, and eq. \ref{eq:y2}
becomes
\begin{equation}
  \delta\vec{y}=\vec{y}'-(\bm{\lambda}\vec\beta^b+\bm{\kappa^d}\vec{\beta}^d).
\end{equation}
When G3NSB is used, the barycentric astrometry is
already given, and thus the GDR3 part of the data vector would not
change, i.e. $\vec{y}^{'\rm GDR3}=\vec{y}^{\rm GDR3}$. If G3SS is
used, the data vector would not change, i.e. $\vec{y}'=\vec{y}$.


We infer the differential calibration parameters $\beta^d$ as well as
the barycentric astrometry $\vec\beta^b$ for a number of calibration
sources simultaneously. The data vector for a sample of $n$
calibration sources is $\vec{y}'_{\rm sample}=(\vec{y}'_1,\vec{y}'_2,
\cdots,\vec{y}'_n)^T$. The residual is
\begin{equation}
  \delta\vec{y}_{\rm sample}=\vec{y}'_{\rm sample}-(\bm{\lambda}_{\rm
    sample}\vec\beta_{\rm sample}^b+\bm{\kappa}_{\rm sample}^d\vec{\beta}^d),
\end{equation}
where
\[
  \bm{\lambda}_{\rm sample}=
  \begin{bmatrix}
    \bm{\lambda_1}&&&\\
    &\bm{\lambda_2}&&\\
    &&\ddots&\\
    &&&\bm{\lambda_n}\\
  \end{bmatrix},
  \vec{\beta}_{\rm sample}^b=
  \begin{bmatrix}
    \vec{\beta}_1^b\\
    \vec{\beta}_2^b\\
    \vdots\\
    \vec{\beta}_n^b\\
  \end{bmatrix},
  \bm{\kappa}_{\rm sample}^d=
  \begin{bmatrix}
    \bm{\kappa}_1^d\\
    \bm{\kappa}_2^d\\
    \vdots\\
    \bm{\kappa}_n^d\\
  \end{bmatrix}.
\]

Similar to eq. \ref{eq:betabr}, the linear regression of the astrometric model
leads to
\begin{equation}
  \bm{\eta}_{\rm sample}\vec\beta_{\rm sample}^{bd}=\vec{b}_{\rm sample}~,
  \label{eq:brv2}
\end{equation}
where
\begin{align}
  \bm{\eta}_{\rm sample}&=
  \begin{bmatrix}
    \bm{\lambda}_{\rm sample}^T\bm{\Sigma}_{\rm
      sample}^{-1}\bm{\lambda}_{\rm sample}&\bm{\lambda}_{\rm sample}^T\bm{\Sigma}_{\rm sample}^{-1}\bm{\kappa}^d_{\rm sample}\\
    (\bm{\kappa}^d_{\rm sample})^T\bm{\Sigma}_{\rm sample}^{-1}\bm{\lambda}_{\rm sample}&(\bm{\kappa}^d_{\rm sample})^T\bm{\Sigma}_{\rm sample}^{-1}\bm{\kappa}^d_{\rm sample}
  \end{bmatrix}\\
 &=
\begin{bmatrix}
  \bm{\lambda}_1^T\bm{\Sigma}_1^{-1}\bm{\lambda}_1&&&& \bm{\lambda}_1^T\bm{\Sigma}_1^{-1}\bm{\kappa}_1^d\\
  &\bm{\lambda}_2^T\bm{\Sigma}_2^{-1}\bm{\lambda}_2&&& \bm{\lambda}_2^T\bm{\Sigma}_2^{-1}\bm{\kappa}_2^d\\
  &&\ddots&&\vdots\\
  &&&\bm{\lambda}_n^T\bm{\Sigma}_n^{-1}\bm{\lambda}_n&\bm{\lambda}_n^T\bm{\Sigma}_n^{-1}\bm{\kappa}_n^d\\
(\bm{\kappa}_1^d)^T\bm{\Sigma}_1^{-1}\bm{\lambda}_1&(\bm{\kappa}_2^d)^T\bm{\Sigma}_2^{-1}\bm{\lambda}_2&\cdots&(\bm{\kappa}_n^d)^T\bm{\Sigma}_n^{-1}\bm{\lambda}_n&\sum\limits_i (\bm{\kappa}_i^d)^T\bm{\Sigma}_i^{-1}\bm{\kappa}_i^d\\
\end{bmatrix},
\end{align}
\[
  \vec{b}_{\rm sample}=
  \begin{bmatrix}
   \bm{\lambda}_{\rm sample}^T\bm{\Sigma}_{\rm
       sample}^{-1}\vec{y}_{\rm sample}'\\
   (\bm{\kappa}^d_{\rm sample})^T\bm{\Sigma}_{\rm sample}^{-1}\vec{y}_{\rm sample}'\\
 \end{bmatrix}
=
    \begin{bmatrix}
      \bm{\lambda}_1^T\bm{\Sigma}_1^{-1}\vec{y}_1'  \\
      \bm{\lambda}_2^T\bm{\Sigma}_2^{-1}\vec{y}_2'\\
      \cdots\\
      \bm{\lambda}_n^T\bm{\Sigma}_n^{-1}\vec{y}_n'\\
      \sum\limits_i(\bm{\kappa}^d_i)^T\bm{\Sigma}_i^{-1}\vec{y}_i'\\  
  \end{bmatrix},
\]
\[
    \vec{\beta}^{bd}_{\rm sample}=
  \begin{bmatrix}
    \vec{\beta}_1^b\\
    \vec{\beta}_2^b\\
    \vdots\\
    \vec{\beta}_n^b\\
  \end{bmatrix}~,
  \bm{\Sigma}_{\rm sample}^{-1}=\begin{bmatrix}
    \bm{\Sigma_1}^{-1}&&&\\
    &\bm{\Sigma_2}^{-1}&&\\
    &&\ddots&\\
    &&&\bm{\Sigma_n}^{-1}
  \end{bmatrix},
  \bm{\Sigma_i}^{-1}=
    \begin{bmatrix}
    \bm{\Sigma_{i,\rm{CAT1}}}^{-1}&&\\
    &\bm{\Sigma_{i,\rm{GDR2}}}^{-1}&\\
    &&\bm{\Sigma_{i,\rm{GDR3}}}^{-1}
  \end{bmatrix},
  \label{eq:sigmai}
\]
 where CAT1 could be G1P2, G1P5, TYC, HIP1, or HIP2. If no CAT1 is
 provided, 
\begin{equation}
  \bm{\Sigma_i}^{-1}=
    \begin{bmatrix}
    \bm{\Sigma_{i,\rm{GDR2}}}^{-1}&\\
    &\bm{\Sigma_{i,\rm{GDR3}}}^{-1}
    \end{bmatrix}.
  \end{equation}
Considering that the covariance given by a catalog is probably
underestimated, we model the error inflation and jitter using $a$ and
$b$, respectively. The covariance of a given target becomes
$\bm{\Sigma_{jk}}=\bm{\rho_{jk}}\sqrt{a^2\sigma_j^2+b^2}\sqrt{a^2\sigma_k^2+b^2}$,
where $\bm{\rho}$ is the correlation matrix. 
  
Finally, by minimizing
\begin{equation}
  \chi_{\rm sample}^2=\delta\vec{y}^T_{\rm sample} \bm{\Sigma_{\rm sample}}^{-1} \delta\vec{y}_{\rm sample},
\end{equation}
we get the optimal calibration parameters $\vec\beta^d$. The
corresponding formulae for the above $\chi^2$ minimization is similar to that
shown in eq. \ref{eq:betabr}.

The lnBF for models with
(model 2) and without (model 1) parameters of frame rotation and
parallax zero-point (i.e. $\vec\beta^d$) is
\begin{equation}
  \ln{\rm BF}_{21}= \frac{1}{2}(\chi_{\rm 1, sample}^2-\chi_{\rm 2, sample}^2-n_d\ln{N_{\rm sample}})~,
  \label{eq:lnbf21}
\end{equation}
where $n_{b}\equiv n_{bd}-n_b$, $n_d$, and $n_{bd}$ are respectively the number of parameters
in $\vec\beta_{\rm sample}^{b}$, $\vec\beta_{\rm sample}^{d}$, and $\vec\beta_{\rm sample}^{bd}$,
$N_{\rm sample}=n\times N_{\rm single}$ is the total number of data
points for a sample of $n$ calibration sources, and $N_{\rm single}$ is
the number of data points for a single source.  If CAT1 in
eq. \ref{eq:sigmai} is not given, the differential frame rotation and
 parallax zero-point of GDR2 relative to GDR3 will be modeled by 7
 parameters (i.e., $n_d = 7$). If CAT1 is G1P5, HIP1, or
 HIP2, there would be 7 calibration parameters in addition to the GDR2
 calibration parameters. If GDR2 calibration parameters are fixed at
 their optimal values, only 7 parameters need to be optimized,
 $n_d=7$. If CAT1 is G1P2 or TYC, the proper motion of CAT1 is unknown.
Hence only the offset between frames and the differential parallax
 zero-point at the reference epoch (i.e., $\epsilon_x$, $\epsilon_y$,
 $\epsilon_z$, and $\delta\varpi$) would be considered, and $n_d=4$.

\subsection{Calibration procedure}
We derive the calibration parameters for catalogs through the
following sequential steps:
\begin{enumerate}[1)]
\item{\bf Data Collection:} Gather GOST and catalog data for various calibration sources.
\item{\bf Matrix and Vector Calculations:} Compute the matrices and vectors introduced in sections \ref{sec:method} and \ref{sec:calmodel}.
\item{\bf Random Subsampling:} Randomly select 100 targets from the
  entire sample to create a subsample. Repeat this subsampling process
  at least 1000 times to generate an ensemble of subsamples;
\item{\bf Error Inflation and Jitter Sampling:} Sample error inflation
  ($1\le a\le 2$) and jitter ($0\le b\le 2$ mas or
  mas/yr)\footnote{If the optimal solution of $b$ is found to exceed 2, we sample $b$ from 0 to 4 instead.} parameters, creating a grid with a bin size of 0.02.
\item{\bf Optimization and $\chi^2$ Calculation:} For each subsample
  and bin, infer optimal parameters and calculate the corresponding
  $\chi^2$ values for both the combined model of barycentric
  astrometry and calibration (Model 2) and the model of barycentric
  astrometry alone (Model 1). Eliminate subsamples with $\chi^2$
  values exceeding the median $\chi^2$ of the entire ensemble of
  subsamples for each bin. 
\item{\bf Maximum lnBF Calculation:} Calculate the maximum lnBF for
  each bin according to eq. \ref{eq:lnbf21}, identify the optimal $(a, b)$ at the globally maximum lnBF (lnBF${\rm max}$), and determine the uncertainty of $(a, b)$ based on the range that satisfies lnBF$_{\rm max}-$lnBF $< 0.5$ (equivalent to a 1-$\sigma$ confidence level, assuming Laplace's approximation).
\item{\bf Parameter Distribution Calculation:} With $a$ and $b$ fixed at their optimal values, repeat step 5 at least 10 times to obtain the distribution of optimal $\vec\beta_{\rm sample}^d$. Calculate the mode, 16\%, and 84\% quantiles for each parameter.
\end{enumerate}

\section{Results}\label{sec:results}
\subsection{Global calibration}
The error inflation and jitter parameters for GDR3, denoted as $(a_{\rm GDR3}$, $b_{\rm GDR3})$, are respectively determined as $(1.00_{-0.00}^{+0.04}, 0.00_{-0.00}^{+0.00})$, $(1.00_{-0.00}^{+0.00}, 0.00_{-0.00}^{+0.00})$, and $(1.00_{-0.00}^{+0.40}, 0.00_{-0.00}^{+0.00})$. These values are obtained through the simultaneous optimization of $(a_{\rm GDR2}, b_{\rm GDR2})$ and $(a_{\rm GDR3}, b_{\rm GDR3})$ using the NST, NSB, and SS calibration sources. The corresponding $(a_{\rm GDR2}, b_{\rm GDR2})$ and the differential calibration parameters of GDR2 relative to GDR3 ($\vec\beta^{\rm GDR23}$) are also determined utilizing the NST, NSB, and SS calibration sources and are detailed in Table \ref{tab:calipar}.

\begin{figure}
  \centering
  \includegraphics[scale=0.35]{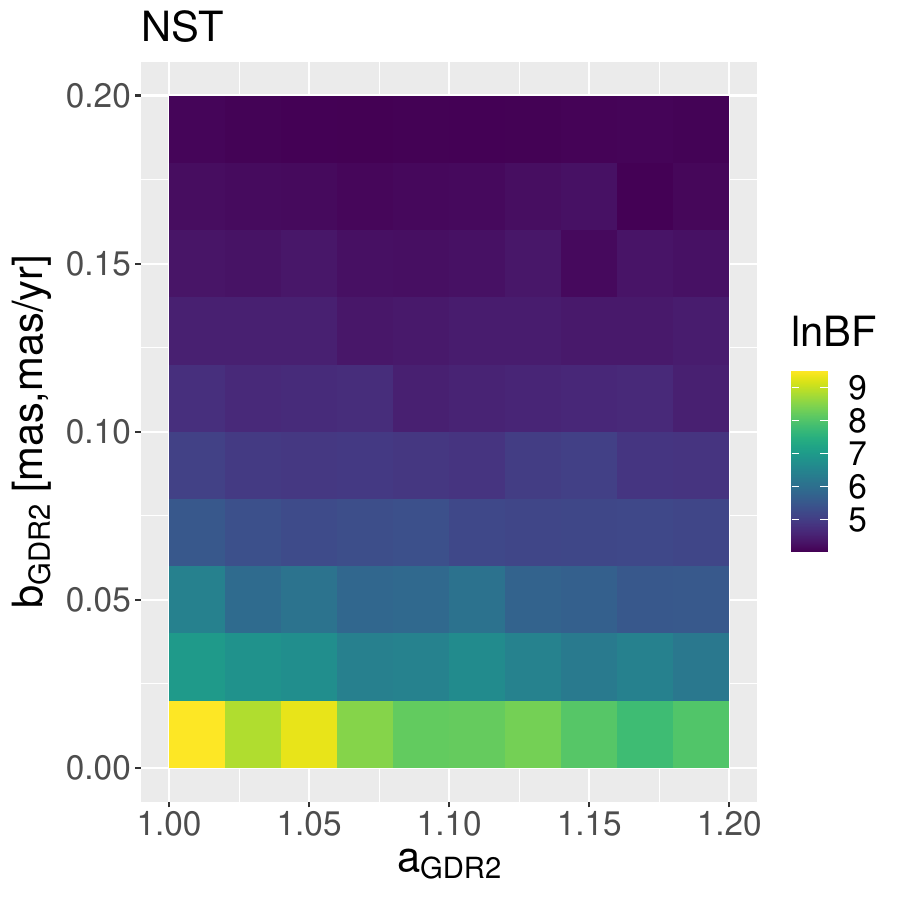}
  \includegraphics[scale=0.35]{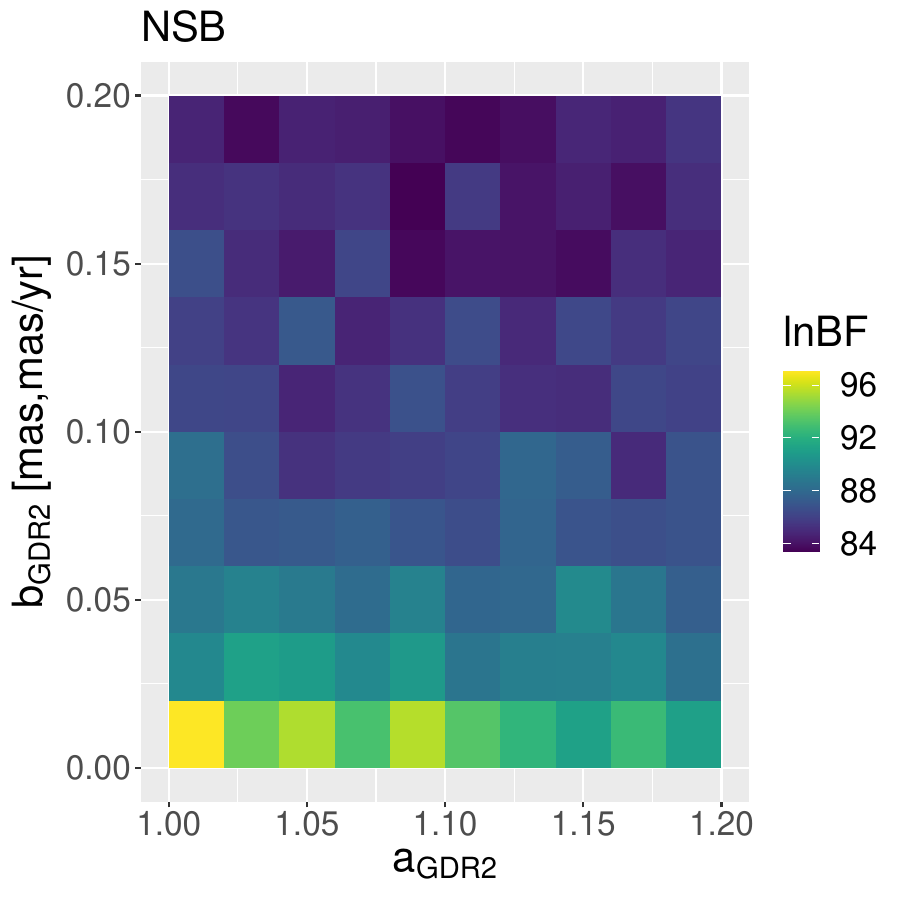}
  \includegraphics[scale=0.35]{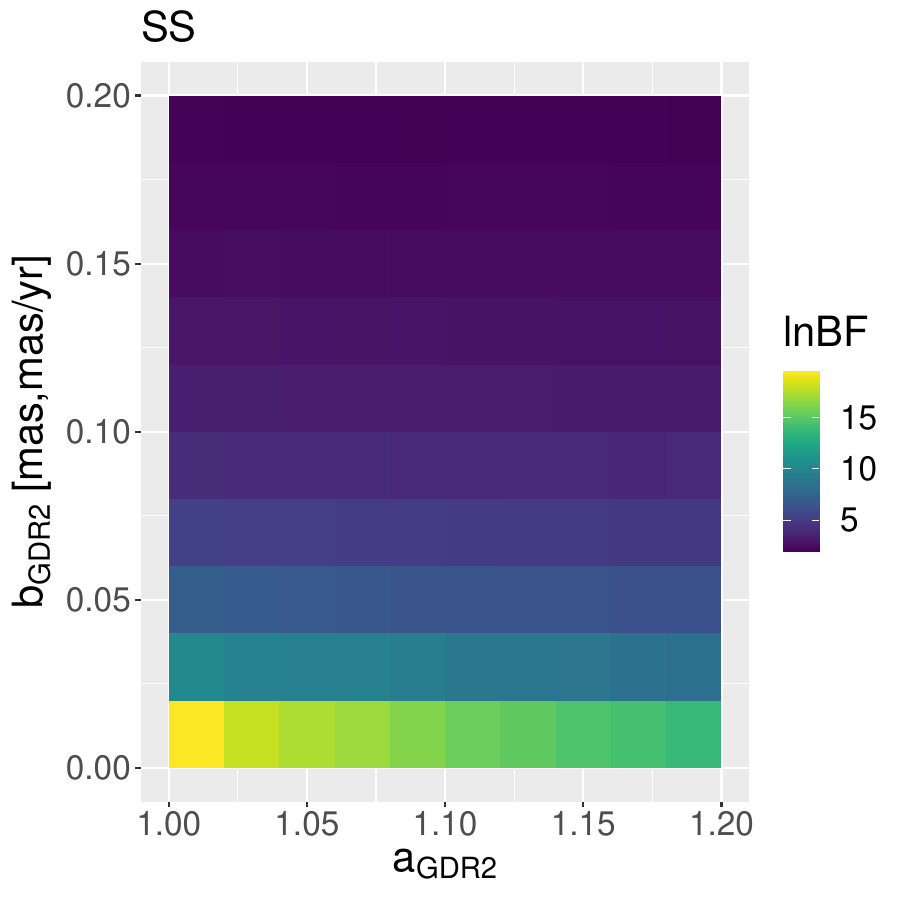}

  \includegraphics[scale=0.35]{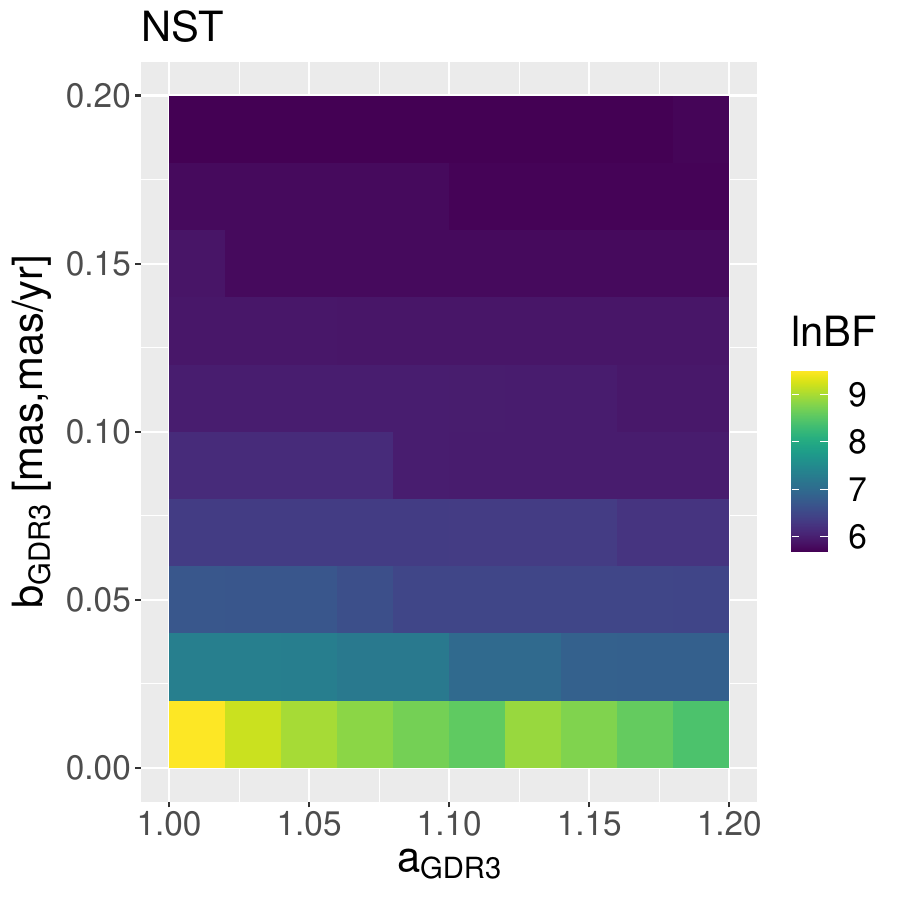}
  \includegraphics[scale=0.35]{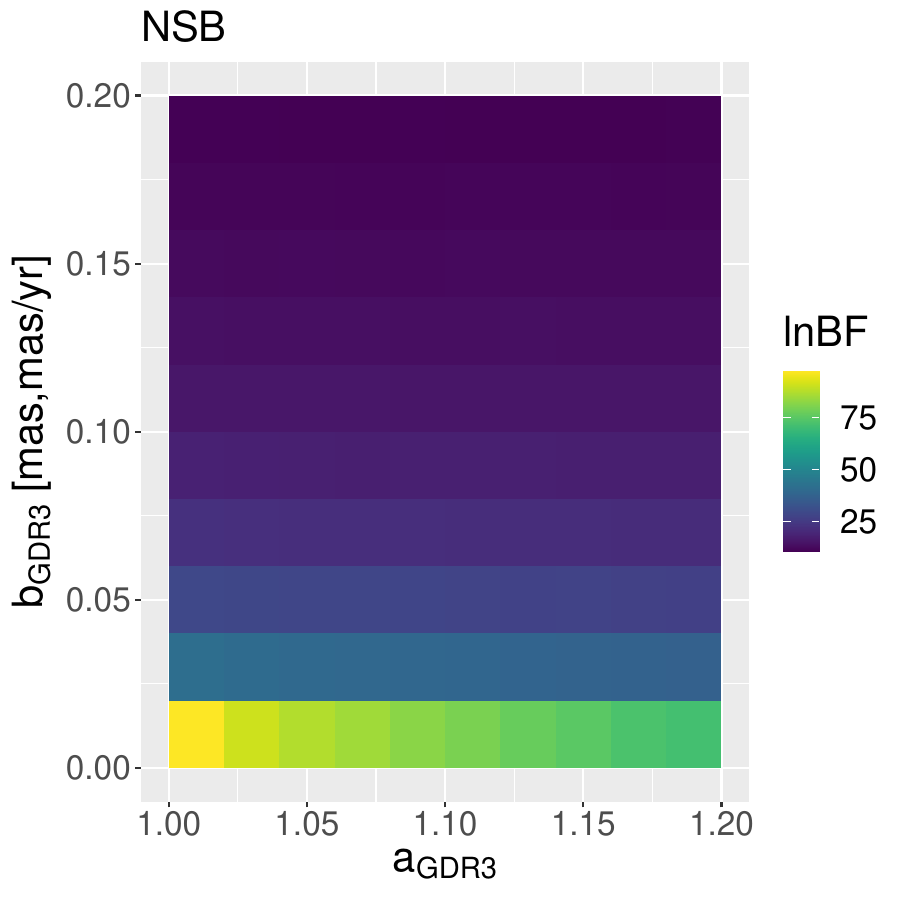}
  \includegraphics[scale=0.35]{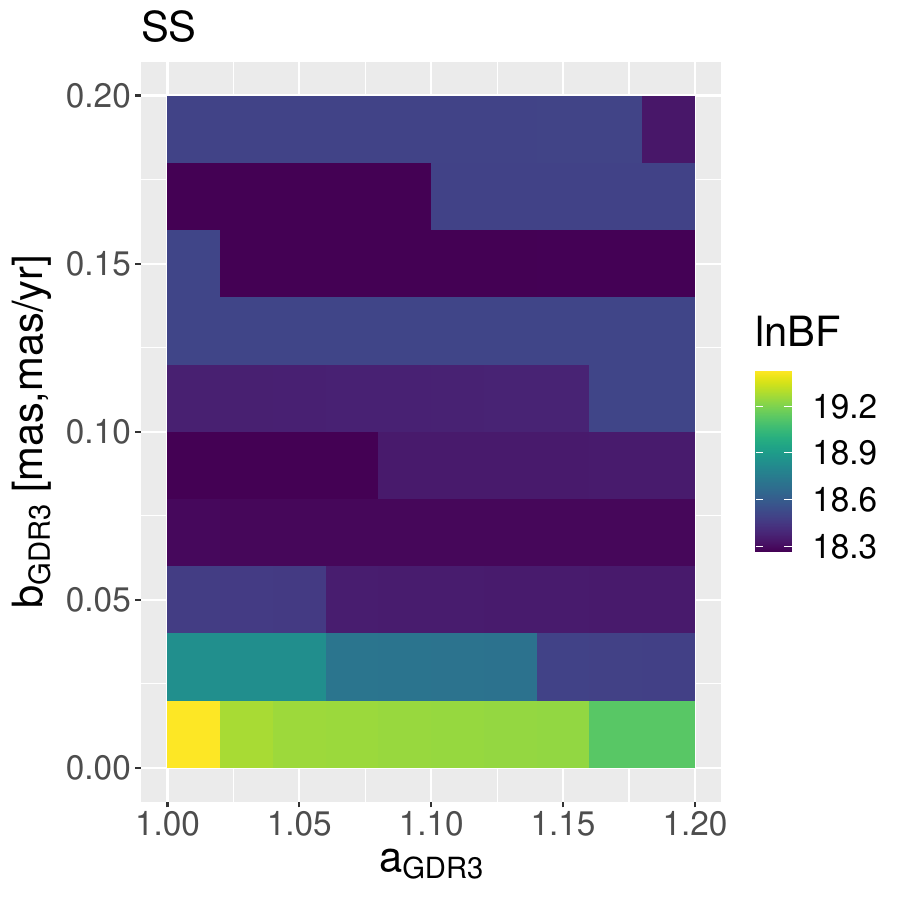}

    \includegraphics[scale=0.35]{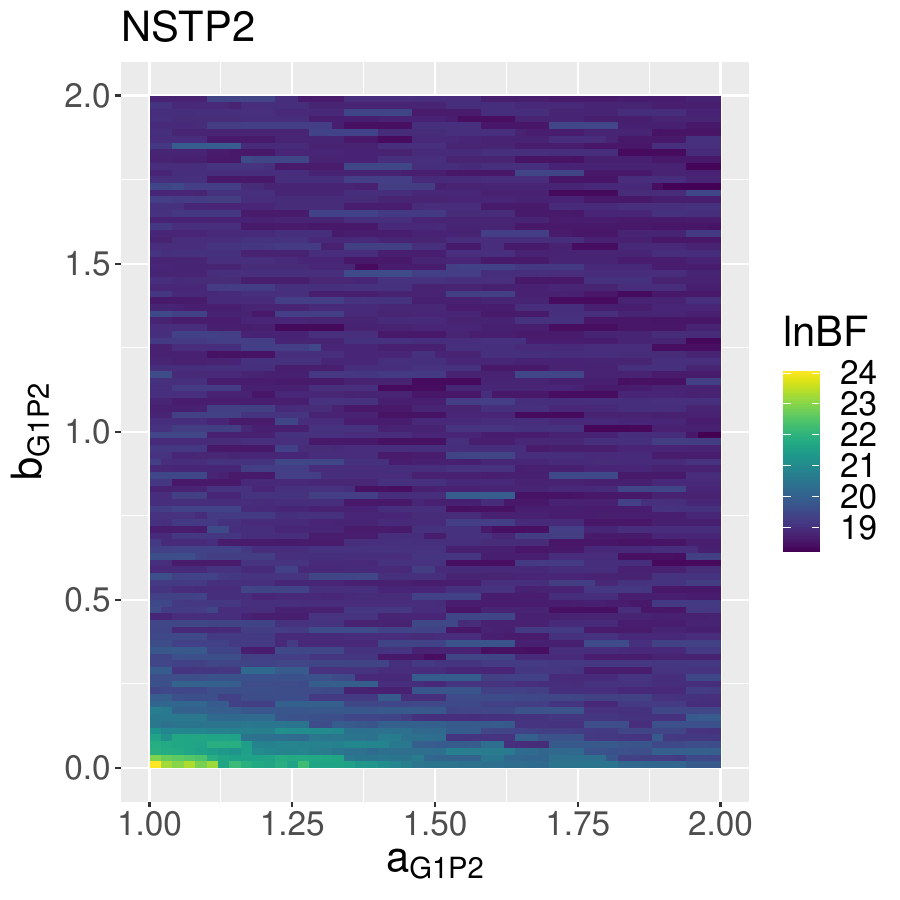}
    \includegraphics[scale=0.35]{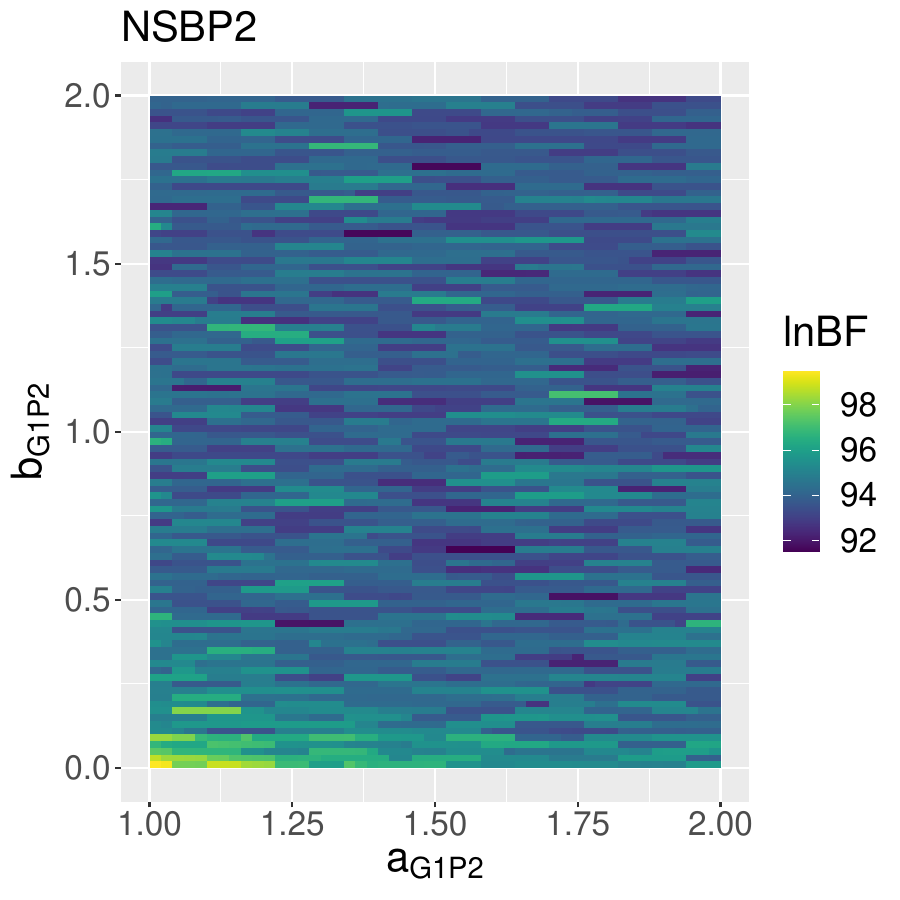}
    \includegraphics[scale=0.35]{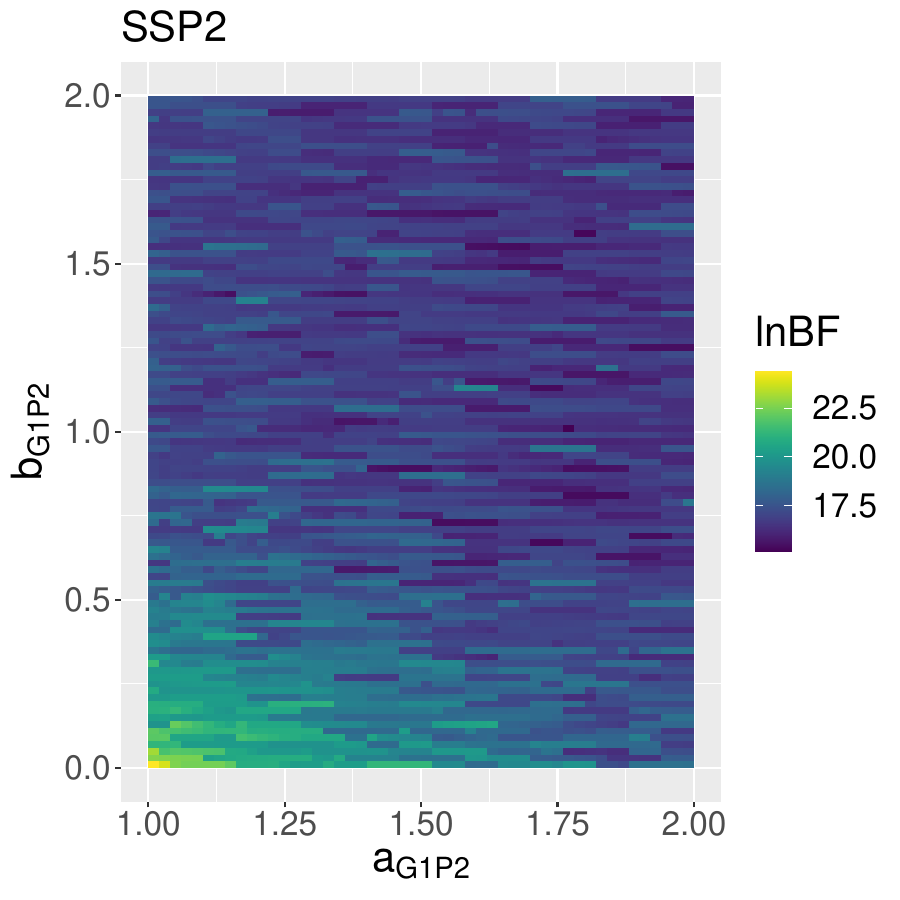}

    \includegraphics[scale=0.35]{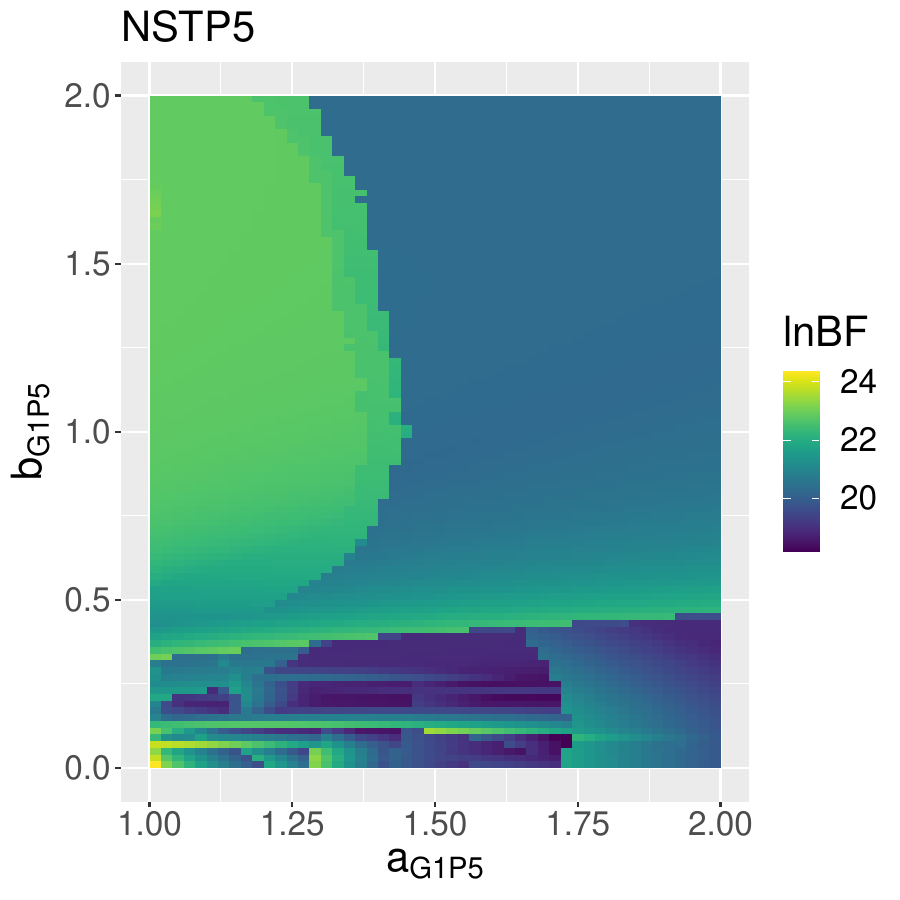}
    \includegraphics[scale=0.35]{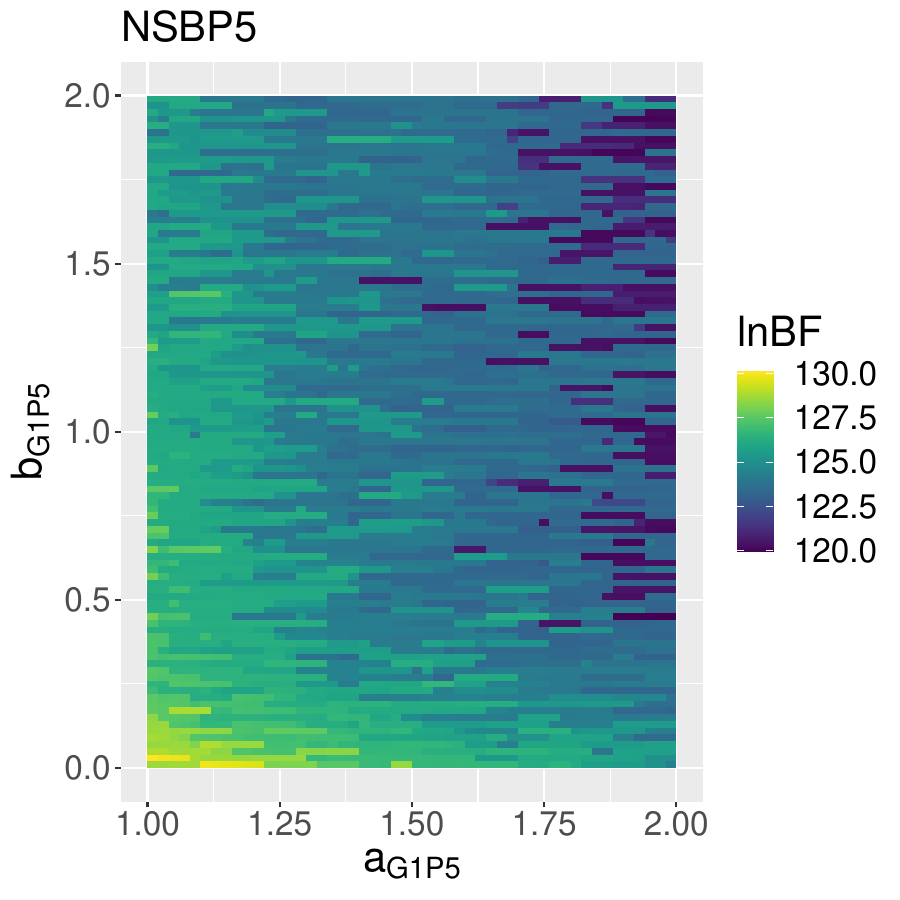}
    \includegraphics[scale=0.35]{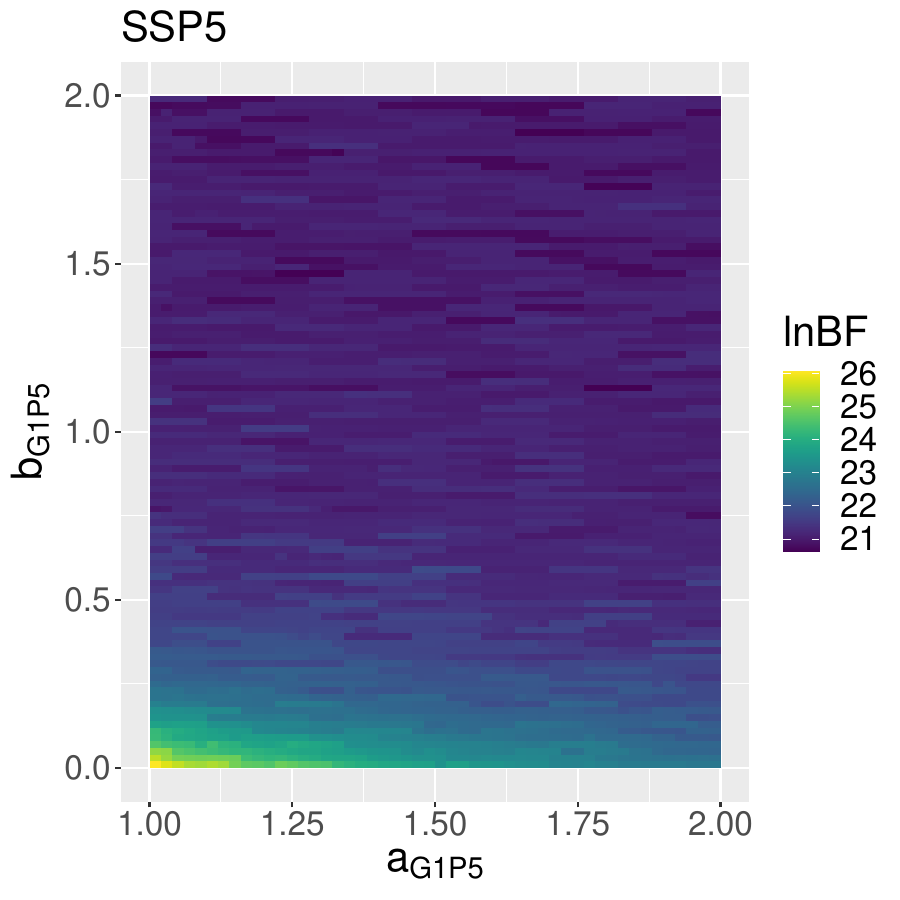}

    \caption{Distribution of lnBF over the error inflation and jitter of
    GDR2 (first row), GDR3 (second row), G1P2 (third row) and G1P5
    (fourth row) based on analyses of various calibration
    sources. The color encodes the value of lnBF of the calibration
    model relative to the model without calibration. The panels in the
    first and second rows are zoom-in versions of distributions over a
    bigger grid with $1\le a \le 2$ and $0 \le b \le
    2$, shown in the lower panels.}
  \label{fig:ab_g123}
\end{figure}

\begin{figure}
  \centering
  \includegraphics[scale=0.3]{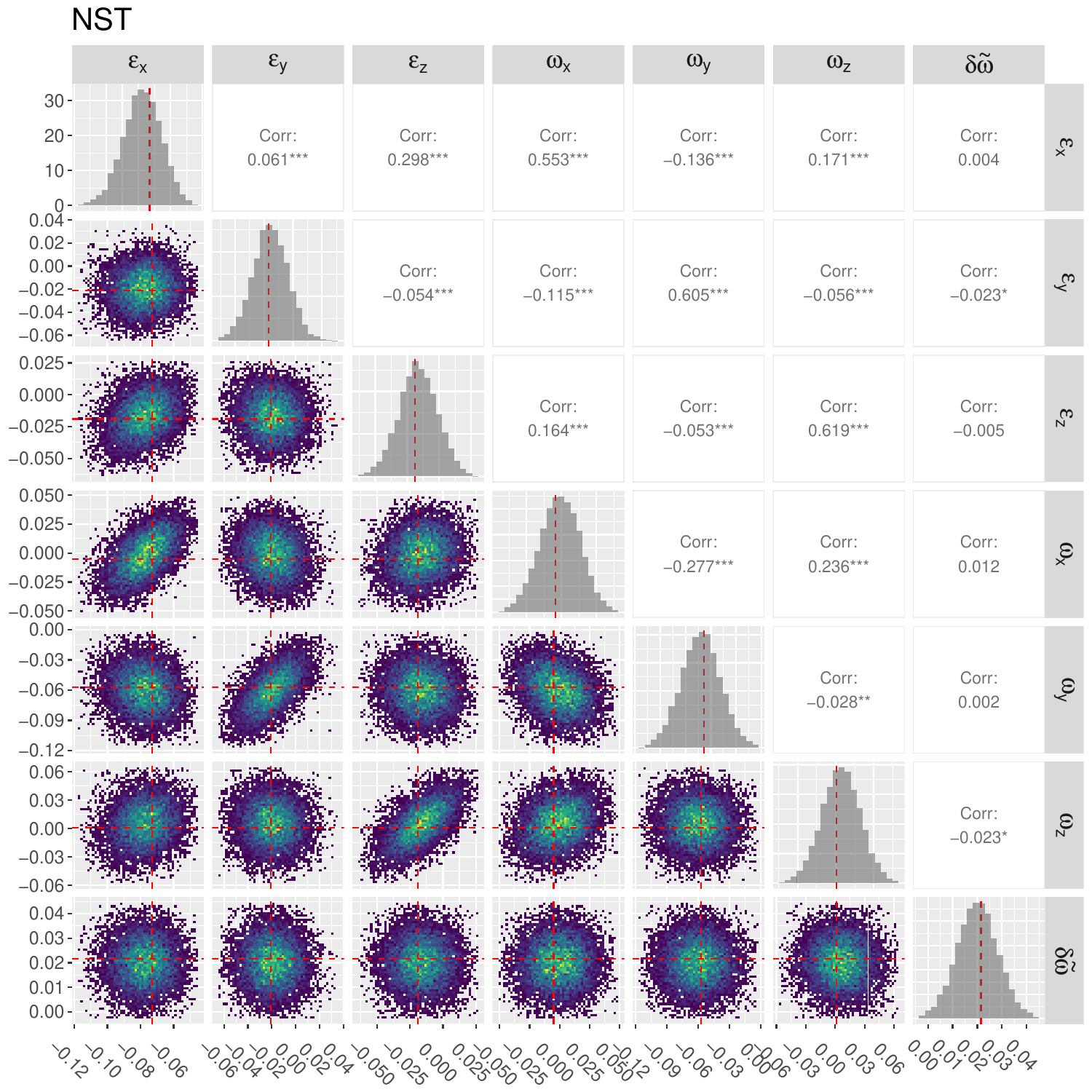}
  \includegraphics[scale=0.3]{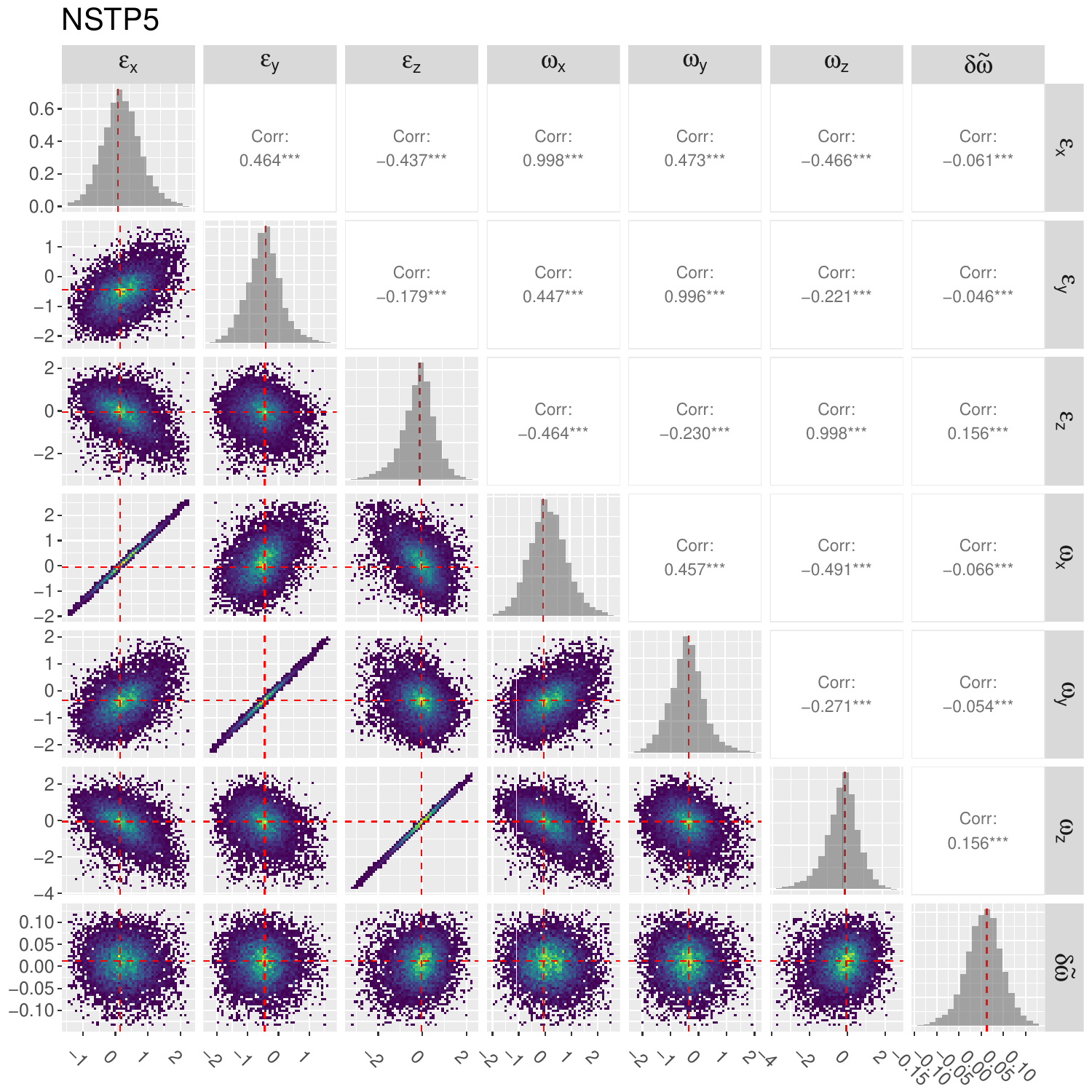}
  
  \includegraphics[scale=0.3]{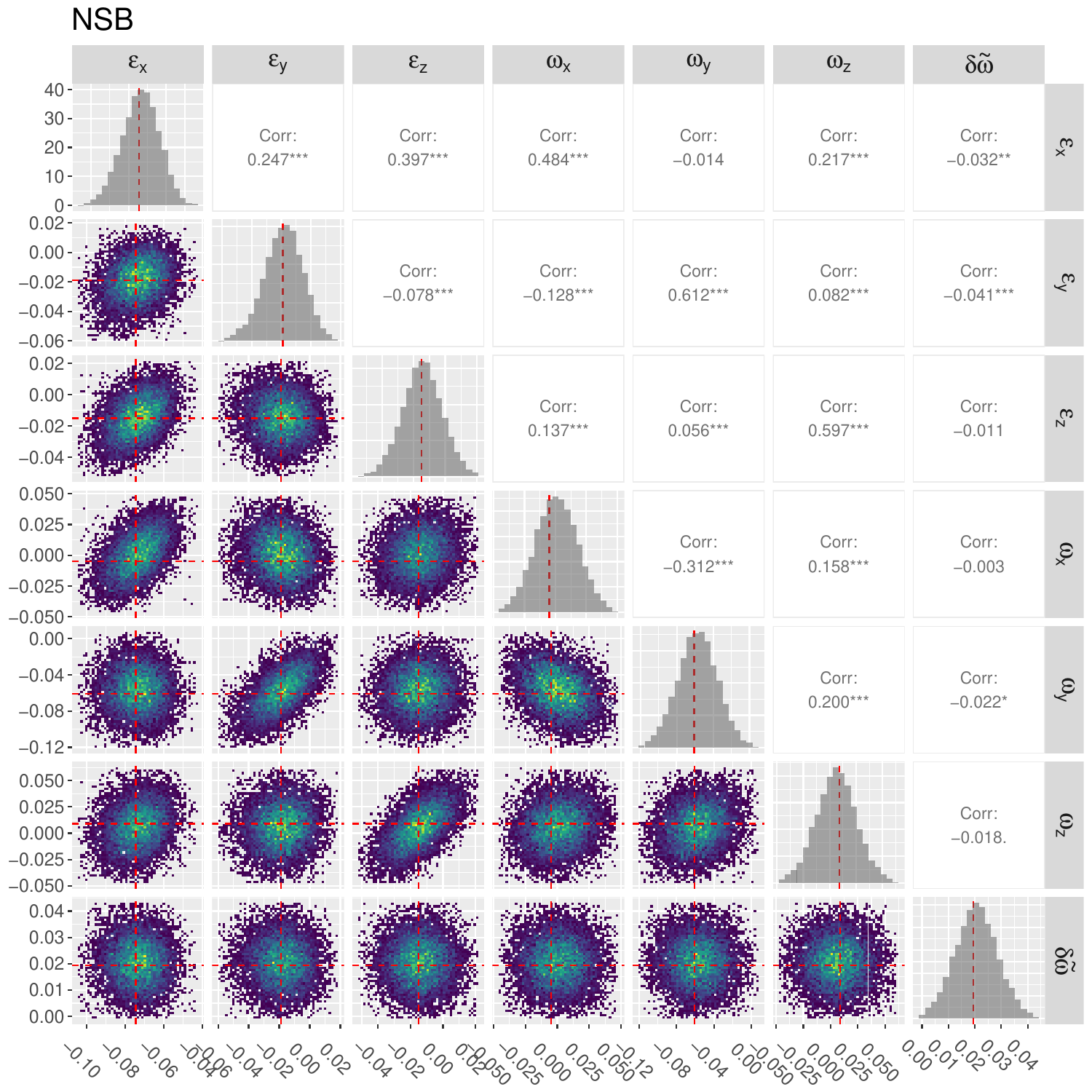}
  \includegraphics[scale=0.3]{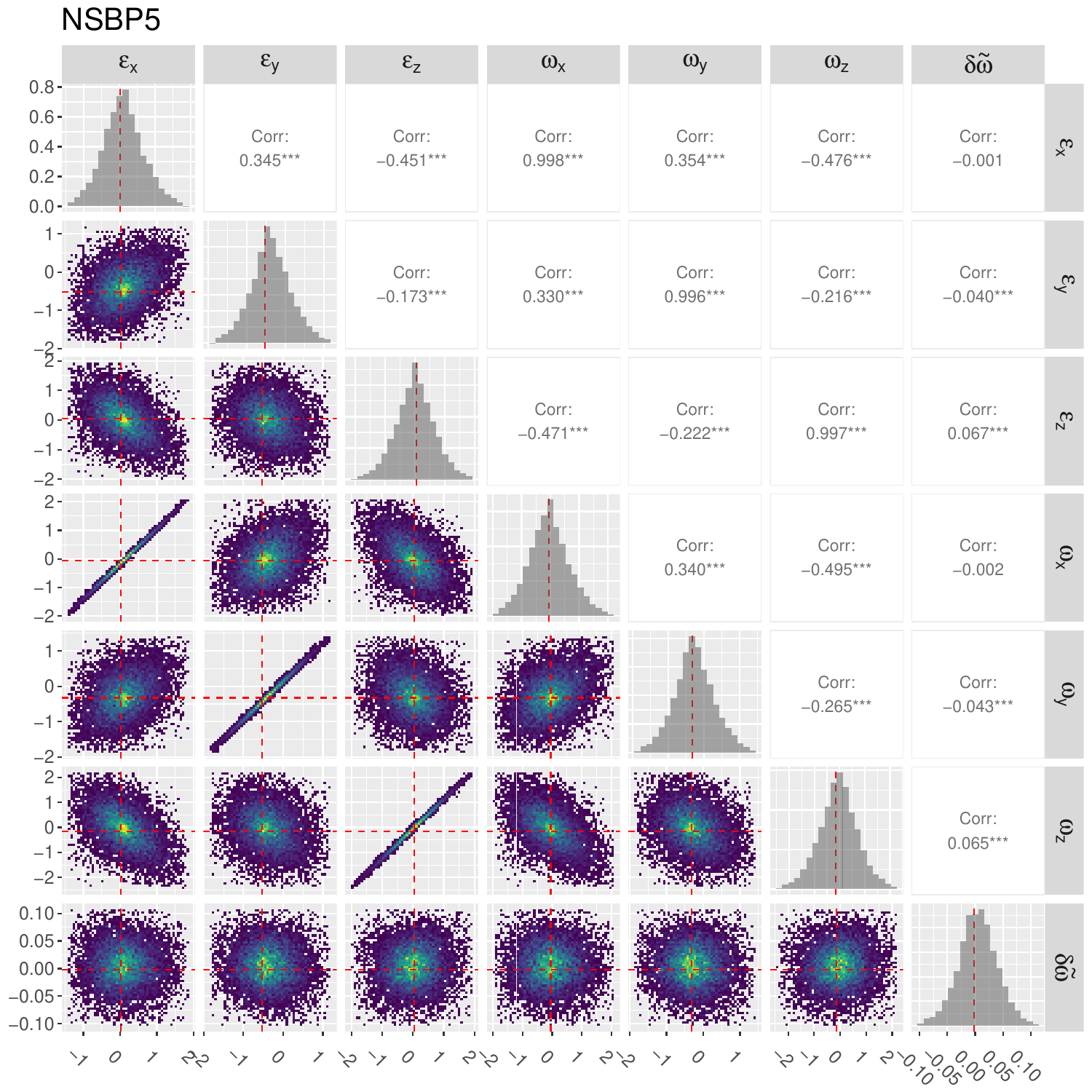}
  
  \includegraphics[scale=0.3]{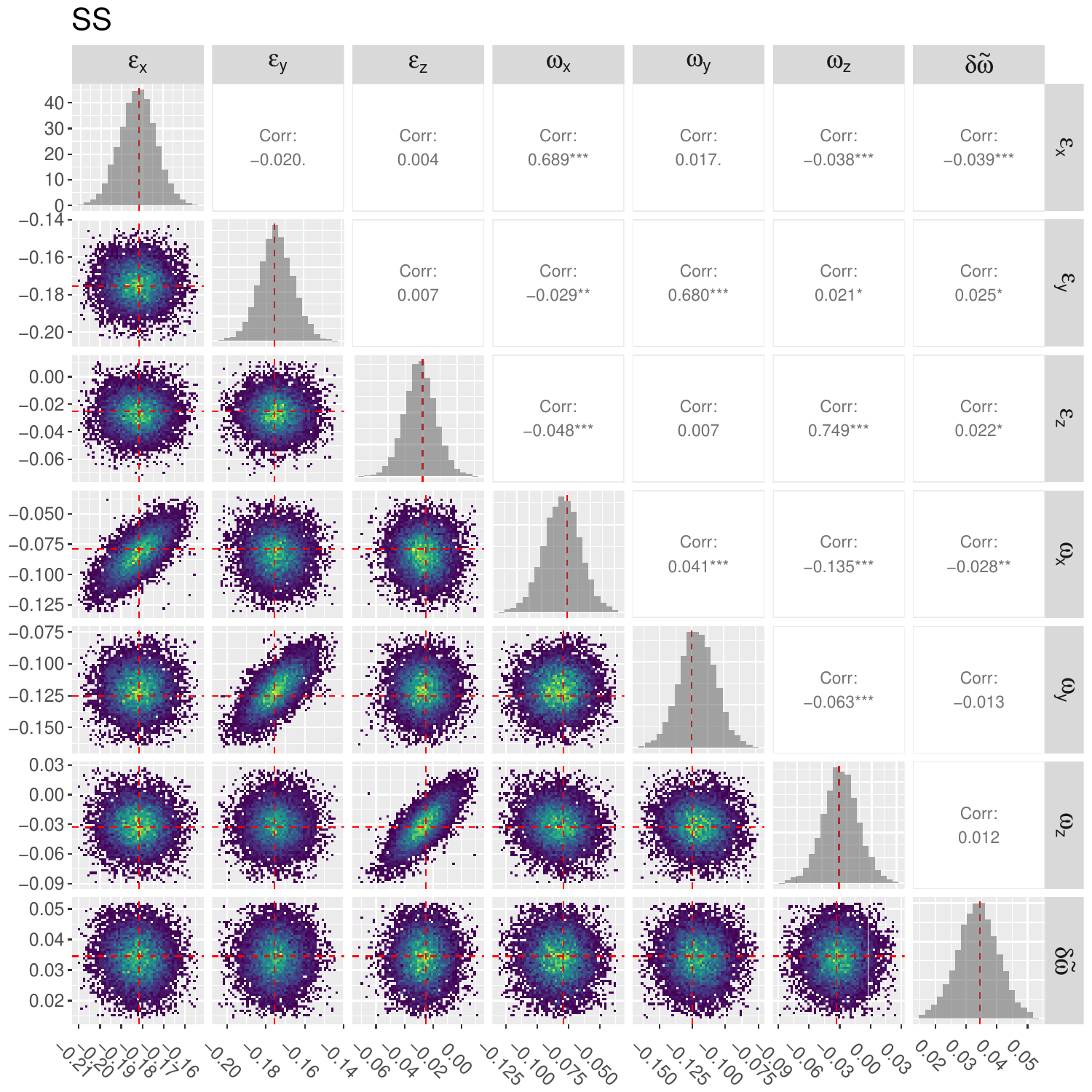}
  \includegraphics[scale=0.3]{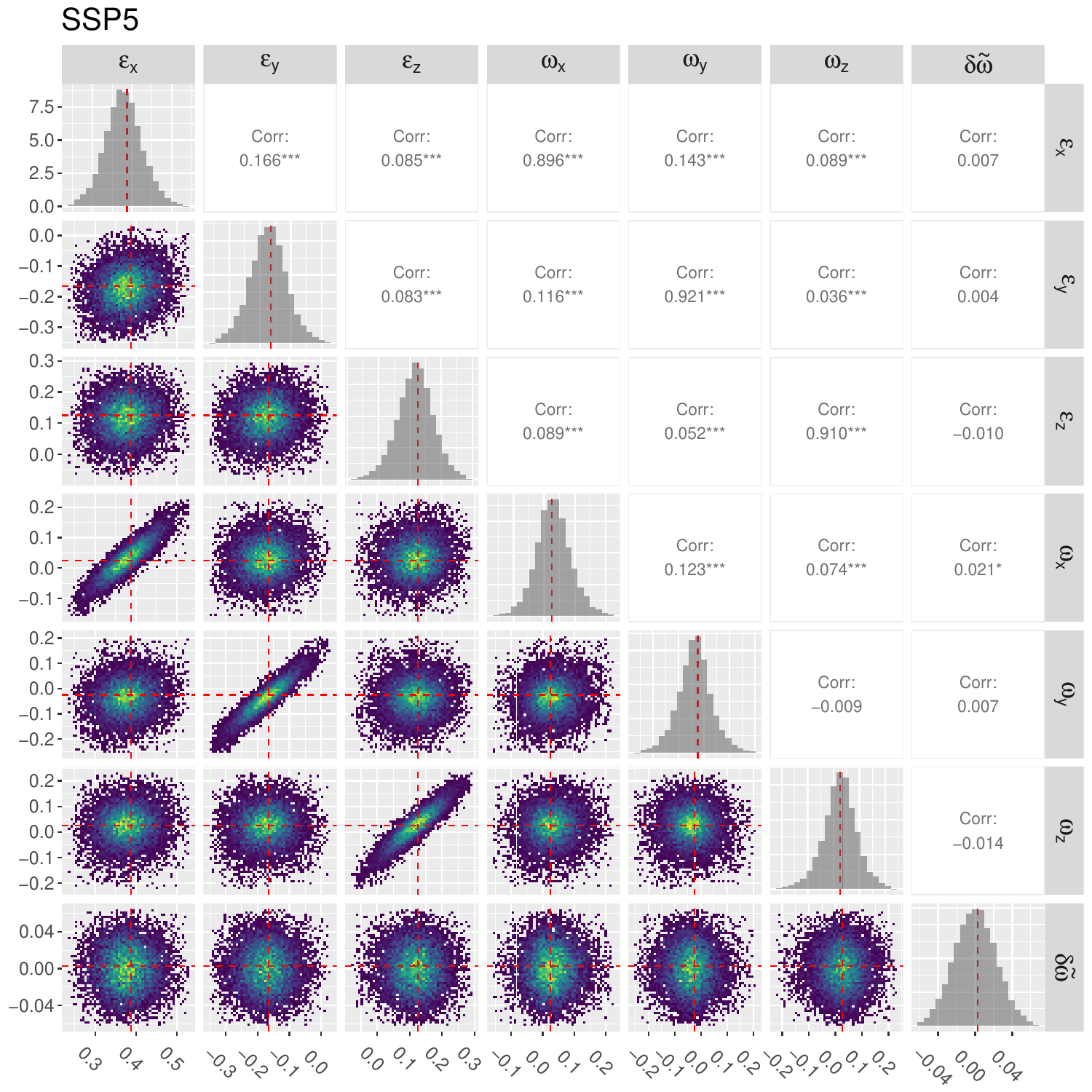}
    \caption{Distribution of calibration parameters for GDR2
      and G1P5 based on 10,000 draws of 100 samples from various
      calibration sources. The parameter
      range is divided into 50 bins. The red dashed
      lines represent the modes of the distributions. The upper panels
    show the Pearson correlation value (Corr) and its significance indicated by
    the number of star-symbols. The p-values of less than 0.1, 0.05,
    0.01, and 0.001 are respectively indicated by ``.'', ``*'',
    ``**'', and ``***''. No symbol is shown if the p-value is higher than 0.1.}
  \label{fig:pair_none_g1p5}
\end{figure}

Figures \ref{fig:ab_g123} and \ref{fig:pair_none_g1p5} illustrate the lnBF distributions over $(a,b)$ and the distribution of $\vec\beta^{\rm GDR23}$ based on analyses of the NST, NSB, and SS calibration sources, respectively. Notably, the preference for zero jitter and no error inflation holds true for both GDR2 and GDR3, and this conclusion remains robust across different choices of calibration sources, as demonstrated in Table \ref{tab:calipar}.

The analysis of NST, NSB, and SS calibration sources reveals significant frame rotation and parallax offset for GDR2 relative to GDR3. The SS values align with those reported by \cite{brandt18}, \cite{lindegren20}, and \cite{lunz23}. While NST-based and NSB-based calibrations yield similar values for $\delta\vec\beta^{\rm GDR23}$, SS-based calibration results in slightly different values, suggesting a dependency of $\vec\beta^{\rm GDR23}$ on stellar parameters, as briefly mentioned in \cite{lindegren21a}. A detailed investigation of this dependence is presented in subsection \ref{sec:dependence}.

\begin{longrotatetable}
\begin{deluxetable*}{l ccc ccc ccc cr}
   \tablecaption{Calibration parameters determined for different calibration sources.\label{tab:calipar}}
   \tabletypesize{\scriptsize}
   \tablehead{
       \colhead{Source/Catalog}&\colhead{CAT1/Ref.}& \colhead{$\epsilon_x$} &\colhead{$\epsilon_y$} &\colhead{$\epsilon_z$} &\colhead{$\omega_x$}
       &\colhead{$\omega_y$} &\colhead{$\omega_z$} &\colhead{$\delta\varpi$}&\colhead{$a$}&\colhead{$b$}&\colhead{$N$}\\
       \colhead{---}&\colhead{---}&\colhead{mas} & \colhead{mas} &
       \colhead{mas} & \colhead{\masyr} & \colhead{\masyr} &
       \colhead{\masyr} &\colhead{mas} & \colhead{---} &
       \colhead{mas,\masyr} &\colhead{---}
     }
   \startdata
      \multicolumn{12}{c}{CAT1+GDR2+G3NST}\\\hline
       NST&---&$-0.08_{-0.01}^{+0.01}$&$-0.02_{-0.02}^{+0.02}$&$-0.01_{-0.02}^{+0.01}$&$0.00_{-0.02}^{+0.02}$&$-0.07_{-0.01}^{+0.03}$&$0.01_{-0.02}^{+0.02}$&$0.02_{0.00}^{+0.01}$&$1.00_{-0.00}^{+0.06}$&$0.00_{-0.00}^{+0.00}$&130,127\\
       NSTP2&G1P2&$0.00_{-0.04}^{+0.05}$&$-0.13_{-0.03}^{+0.05}$&$-0.01_{-0.05}^{+0.05}$&---&---&---&---&$1.02_{-0.02}^{+0.00}$&$0.00_{-0.00}^{+0.02}$&116,271\\
       NSTP5&G1P5&$0.15_{-0.47}^{+0.70}$&$-0.45_{-0.61}^{+0.46}$&$-0.05_{-0.78}^{+0.63}$&$-0.05_{-0.52}^{+0.92}$&$-0.35_{-0.66}^{+0.52}$&$-0.05_{-0.99}^{+0.67}$&$0.01_{-0.04}^{+0.03}$&$1.00_{-0.00}^{+0.00}$&$0.00_{-0.00}^{+0.02}$&13,825\\
       NSTT&TYC&$1.00_{-6.00}^{+6.80}$&$3.00_{-9.11}^{+5.10}$&$-1.00_{-8.71}^{+7.85}$&---&---&---&---&$1.94_{-0.00}^{+0.00}$&$0.60_{-0.00}^{+0.00}$&14,947\\
       NSTH2&HIP2&$-8.25_{-3.57}^{+4.78}$&$12.50_{-6.39}^{+5.32}$&$23.50_{-19.58}^{+2.84}$&$-0.16_{-0.08}^{+0.09}$&$0.25_{-0.12}^{+0.10}$&$0.45_{-0.38}^{+0.05}$&$-0.25_{-0.07}^{+0.14}$&$1.00_{-0.00}^{+0.00}$&$2.16_{-0.00}^{+0.12}$&142\\
       NSTH1&HIP1&$-85.00_{-46.08}^{+59.53}$&$-5.00_{-47.92}^{+53.35}$&$75.00_{-70.87}^{+37.50}$&$-1.70_{-0.91}^{+1.23}$&$0.10_{-1.15}^{+0.89}$&$1.50_{-1.39}^{+0.79}$&$-0.15_{-1.05}^{+0.67}$&$1.00_{-0.00}^{+0.00}$&$0.08_{-0.00}^{+0.00}$&142\\\hline
       \multicolumn{12}{c}{CAT1+GDR2+G3NSB}\\\hline
       NSB&---&$-0.07_{-0.01}^{+0.01}$&$-0.02_{-0.01}^{+0.01}$&$-0.01_{-0.01}^{+0.01}$&$0.00_{-0.01}^{+0.02}$&$-0.06_{-0.02}^{+0.03}$&$0.01_{-0.02}^{+0.02}$&$0.02_{-0.01}^{+0.01}$&$1.00_{-0.00}^{+0.00}$&$0.00_{-0.00}^{+0.02}$&130,127\\
       NSBP2&G1P2&$0.02_{-0.13}^{+0.12}$&$-0.11_{-0.07}^{+0.20}$&$-0.03_{-0.02}^{+0.24}$&---&---&---&---&$1.00_{-0.00}^{+0.04}$&$0.02_{-0.02}^{+0.00}$&116,271\\
       NSBP5&G1P5&$0.02_{-0.47}^{+0.65}$&$-0.52_{-0.34}^{+0.65}$&$0.05_{-0.66}^{+0.58}$&$-0.05_{-0.67}^{+0.71}$&$-0.33_{-0.49}^{+0.61}$&$-0.15_{-0.65}^{+0.84}$&$0.00_{-0.03}^{+0.04}$&$1.00_{-0.00}^{+0.18}$&$0.04_{-0.04}^{+0.00}$&13,825\\
       NSBT&TYC&$0.50_{-4.64}^{+4.10}$&$0.50_{-3.61}^{+4.69}$&$0.50_{-5.86}^{+4.10}$&---&---&---&---&$1.30_{-0.00}^{+0.10}$&$1.70_{-0.00}^{+0.00}$&14,947\\
       NSBH2&HIP2&$-13.50_{-2.61}^{+11.06}$&$-7.50_{-4.05}^{+20.53}$&$33.00_{-23.54}^{+4.31}$&$-0.25_{-0.05}^{+0.20}$&$-0.09_{-0.09}^{+0.34}$&$0.58_{-0.40}^{+0.11}$&$-0.17_{-0.09}^{+0.12}$&$1.00_{-0.00}^{+0.00}$&$0.48_{-0.00}^{+0.00}$&142\\
       NSBH1&HIP1&$13.00_{-27.72}^{+9.64}$&$2.50_{-16.80}^{+20.33}$&$-82.50_{-11.82}^{+48.90}$&$0.28_{-0.56}^{+0.20}$&$-0.02_{-0.24}^{+0.49}$&$-1.65_{-0.25}^{+0.97}$&$0.27_{-0.20}^{+0.19}$&$1.00_{-0.00}^{+0.10}$&$0.20_{-0.10}^{+0.90}$&142\\\hline
       \multicolumn{12}{c}{CAT1+GDR2+G3SS}\\\hline
       SS&---&$-0.18_{-0.01}^{+0.01}$&$-0.18_{-0.01}^{+0.01}$&$-0.03_{-0.01}^{+0.01}$&$-0.08_{-0.02}^{+0.01}$&$-0.12_{-0.01}^{+0.02}$&$-0.03_{-0.02}^{+0.02}$&$0.03_{-0.01}^{+0.01}$&$1.00_{-0.00}^{+0.00}$&$0.00_{-0.00}^{+0.00}$&77,138\\
       SSP2&G1P2&$2.50_{-27.09}^{+36.30}$&$2.50_{-26.74}^{+27.42}$&$-2.50_{-19.31}^{+33.98}$&---&---&---&---&$1.00_{-0.00}^{+0.02}$&$0.02_{-0.00}^{+0.00}$&6,763\\
       SSP5&G1P5&$0.39_{-0.05}^{+0.04}$&$-0.17_{-0.07}^{+0.05}$&$0.12_{-0.06}^{+0.05}$&$0.02_{-0.05}^{+0.06}$&$-0.03_{-0.07}^{+0.05}$&$0.02_{-0.06}^{+0.06}$&$0.00_{-0.03}^{+0.02}$&$1.00_{-0.00}^{+0.06}$&$0.02_{-0.02}^{+0.00}$&70,676\\
       SST&TYC&$0.25_{-2.37}^{+1.75}$&$-0.25_{-1.75}^{+2.30}$&$-0.25_{-2.03}^{+2.21}$&---&---&---&---&$1.00_{-0.00}^{+0.14}$&$0.44_{-0.44}^{+1.42}$&74,216\\
       SSH2&HIP2&$5.00_{-11.24}^{+10.94}$&$-9.00_{-9.73}^{+11.82}$&$-2.50_{-15.96}^{+9.62}$&$0.13_{-0.26}^{+0.20}$&$-0.13_{-0.25}^{+0.21}$&$-0.17_{-0.22}^{+0.32}$&$0.13_{-0.22}^{+0.15}$&$1.00_{-0.00}^{+0.00}$&$0.06_{-0.00}^{+0.00}$&70,358\\
       SSH1&HIP1&$-10.00_{-85.39}^{+107.76}$&$-10.00_{-95.88}^{+97.87}$&$-10.00_{-101.15}^{+101.38}$&$-0.25_{-1.68}^{+2.21}$&$0.25_{-2.37}^{+1.54}$&$-0.25_{-2.00}^{+2.08}$&$0.30_{-1.75}^{+1.22}$&$1.00_{-0.00}^{+0.02}$&$0.02_{-0.00}^{+0.00}$&77,136\\\hline\hline
       \multicolumn{12}{c}{Values in Literature\tablenotemark{$\dag\dag$}}\\\hline
       ICRF3-GDR3&L23\tablenotemark{a}&$0.226\pm0.165$&$0.327\pm0.213$&$0.168\pm0.128$&$0.022\pm0.024$&$0.065\pm0.024$&$-0.016\pm0.024$&---&---&---&55\\
       HG-GDR3&L21\tablenotemark{b}&---&---&---&$0.017\pm0.024$&$0.095\pm0.024$&$-0.028\pm0.024$&---&---&---&$\sim$90,000\\
       ICRF3-GDR2&L23\tablenotemark{a}&$0.093\pm0.180$&$0.463\pm0.243$&$0.028\pm0.141$&$-0.056\pm0.046$&$-0.113\pm0.058$&$0.033\pm0.053$&---&---&---&55\\
       VLBI-GDR2&L20\tablenotemark{c} &$-0.347\pm 0.137$&$0.358\pm 0.245$&$0.050\pm 0.045$&$-0.077\pm 0.051$&$-0.096\pm0.042$&$-0.002\pm0.036$&---&---&---&26\\
       HG2-GDR2&B18\tablenotemark{d}&---&---&---&$-0.081$&$-0.113$&$-0.038$&---&1.743&---&83,034\\
       HG2-HIP12&B18\tablenotemark{d}&---&---&---&$-0.098$&$0.170$&$0.089$&---&---&0.226&83,034\\
       GDR3-HIP1&F21\tablenotemark{e}&---&---&---&---&---&---&$-0.089\pm0.003$&---&---&62,484\\
       ICRF2-HIP1&L16\tablenotemark{f}&$-2.99\pm 0.04$&$4.39\pm0.04$&$1.81\pm0.04$&$-0.126\pm 0.03$&$0.185\pm 0.03$&$0.076\pm0.03$&---&---&---&262\\
       HG3-HIP2&B23\tablenotemark{g}&---&---&---&---&---&---&---&---&$2.25\pm0.04$&62,484\\
       \enddata
       \tablenotetext{$\dag$}{The asymmetry in the uncertainty of parameters $a$ and $b$ is attributable to the restricted bin size of 0.02 employed during the sampling process. }
       \tablenotetext{$\dag\dag$}{The first column displays the
         differential rotation and parallax zero-point between the RF0
         and RF1 frames. For example, RF0-RF1 means that the
         astrometric offset induced by $\vec\beta^d$ should be added
         to the astrometry in the RF1 frame to derive the astrometry
         in the RF0 frame. In this work,
         the data ($\vec{y}$) in the original reference frame of CAT1 is subtracted by $\bm{\kappa^d}\vec{\beta}^d$ to derive the astrometry in the GDR3 frame, which can be represented by CAT1-GDR3.}
       \tablenotetext{a}{GDR3 Using optically bright radio stars observed by the very long
         baseline interferometry (VLBI), \cite{lunz23} measure the rotation of
         GDR3 frame (equivalent to GEDR3 frame) relative to the Third International Celestial Reference Frame
         (ICRF3; \citealt{charlot20}) at reference epoch J2016.0.}
         \tablenotetext{b}{\cite{lindegren21b} conduct an ad hoc calibration using the
       difference between Hipparcos and GDR3 proper motions.}
     \tablenotetext{c}{This calibration is
       conducted by \cite{lindegren20} using the VLBI observations of 26 radio stars. Hence
       the parameters shown in this row measure the frame rotation of GDR2 bright
       sources relative to the frame determined by the 26 radio stars
       at epoch J2015.5. }
     \tablenotetext{d}{The parameters are
       determined by \cite{brandt18} by mixing HIP2 and HIP1 with a
       ratio of $60/40$ (HIP12) and calibrating the proper motions of HIP12 and
       GDR2 using their positional differences (HG2). They claim negligible
       uncertainties on the parameters. }
     \tablenotetext{e}{The value is obtained from table 1 of \cite{fabricius21}. }
     \tablenotetext{f}{The
       parameters measures the frame rotation of the original
       Hipparcos catalog relative to the ICRF2 \citep{fey15} realized by radio
       observation of 262 sources \cite{lindegren16} at epoch
       J1991.25. However, $\vec{\epsilon}$ and $\vec{\omega}$ cannot
       be used together to correct for bias in the Hipparcos data
       because they are calculated under different assumptions. }
     \tablenotetext{g}{\cite{brandt23} calibrate the HIP2 IAD using
       Hipparcos-GEDR3 (HG3) long-term proper motions and parallax
       differences.}
     \tablecomments{Note that $(a,b)$ for NST, NSB, and SS represent
       the error inflation and jitter for GDR2 while $(a,b)$ for
       other calibration sources is for the corresponding CAT1. }
       \end{deluxetable*}
\end{longrotatetable}

Having set $(a_{\rm GDR2}, b_{\rm GDR2})$, $(a_{\rm GDR3}, b_{\rm
  GDR3})$, and $\vec\beta^{\rm GDR23}$ to their optimal values, we
proceed to optimize the calibration parameters for CAT1 using various
combinations of CAT1, GDR2, and GDR3-based catalogs. The analysis,
detailed in Table \ref{tab:calipar} and depicted in
Fig. \ref{fig:pair_g1p2_tyc}, reveals negligible error inflation and
jitter for G1P2 and G1P5 across different calibration sources. The
G1P2 frame exhibits an offset of -0.13$_{-0.03}^{+0.05}$\,mas relative
to GDR3 along the $y$-axis, as determined through NSTP2 analysis. This
result is consistent with values obtained from NSTP2 and notably more
precise than those derived from SSP2. This precision improvement is
likely attributed to SSP2 having fewer calibration sources compared to
NSTP2 and NSBP2.

The G1P5 frame exhibits a consistent offset of $(\epsilon_x, \epsilon_y, \epsilon_z) = (0.39_{-0.05}^{+0.04}, -0.17_{-0.07}^{+0.05}, 0.12_{-0.06}^{+0.05})$\,mas relative to GDR3, as determined through analysis of SSP5, as depicted in Fig. \ref{fig:pair_none_g1p5}. Notably, the frame-rotation speed $\omega$ of G1P5 relative to GDR3 aligns with zero.
The NSTP5 and NSBP5 calibration sources, however, do not ascertain
these parameters with the same high precision, likely due to their
limited sample size and the inappropriate subtraction of a
photocentric motion from TGAS proper motions derived partially from
TYC \citep{michalik15}. Given the potential "contamination" from TYC,
we recommend utilizing only the reference position in G1P5 for
detecting companions.

As indicated in Table \ref{tab:calipar}, the error inflation and jitter for TYC are not in alignment with zero at the 1$\sigma$ level based on analyses of NSTT and NSBT. However, upon inspecting the lnBF distributions for NSTT and NSBT depicted in Fig. \ref{fig:ab_th12}, it becomes evident that these distributions exhibit significant non-Gaussian characteristics, with a seemingly random occurrence of high lnBF regions. In contrast, the lnBF distribution for SST appears smoother, with lnBF variations consistently below 3, indicative of negligible error inflation and jitter. The frame rotation of TYC relative to GDR3 is also found to be consistent with zero, as detailed in Table \ref{tab:calipar} and illustrated in Fig. \ref{fig:pair_g1p2_tyc}.

HIP2 exhibits negligible error inflation, aligning with the less than
20\% error inflation previously determined by \cite{perryman97} and
\cite{brandt23}. An analysis of NSTH2 reveals a jitter of
$2.16_{-0.00}^{+0.12}$\,mas (or \masyr) for HIP2, a result consistent
with the value of 2.25$\pm$0.04\,mas (or \masyr) reported by
\cite{brandt23}. Fig. \ref{fig:ab_th12} illustrates that NSBH2 and
SSH2 also exhibit elevated lnBF around $b=2$\,mas (or \masyr), although
not as prominently as NSTH2. As presented in Table \ref{tab:calipar} and Fig. \ref{fig:pair_hip2_hip1}, the differential calibration parameters of HIP2 deviate from zero at the 2$\sigma$ level and align with the values provided by \cite{lindegren16} at the 1.5$\sigma$ level. Additionally, our analysis indicates a negative differential zero-point parallax based on NSTH2 and NSBH2 calibrations, consistent with the value determined by \cite{fabricius21}.

As outlined in Table \ref{tab:calipar}, the error inflation of HIP1 is
negligible, while the jitter of HIP1 is determined to be 0.08, 0.20,
and 0.02 mas (or mas/yr) based on analyses conducted with NSTH1,
NSBH1, and SSH1, respectively. Examining Fig. \ref{fig:ab_th12}, it is
evident that the $(a, b)$ values for HIP1 are largely consistent with
$(1, 0)$. However, it's noteworthy that the differential calibration
parameters for HIP1 are not as precisely determined as those for
HIP2. Conversely, Figs. \ref{fig:pair_none_g1p5} and \ref{fig:pair_hip2_hip1} reveal a pronounced correlation between $\vec\omega$ and $\vec\epsilon$ for G1P5, HIP1, and HIP2. This correlation is likely influenced by the 24-year positional difference between TYC, HIP2, HIP1, and GDR3, which imposes a more stringent constraint on $\vec\beta^d$ than the difference in proper motion does. This leads to a degeneracy between $\vec\omega$ and $\vec\epsilon$ (see eq. \ref{eq:rd}).

\begin{figure}
  \centering
  \includegraphics[scale=0.6]{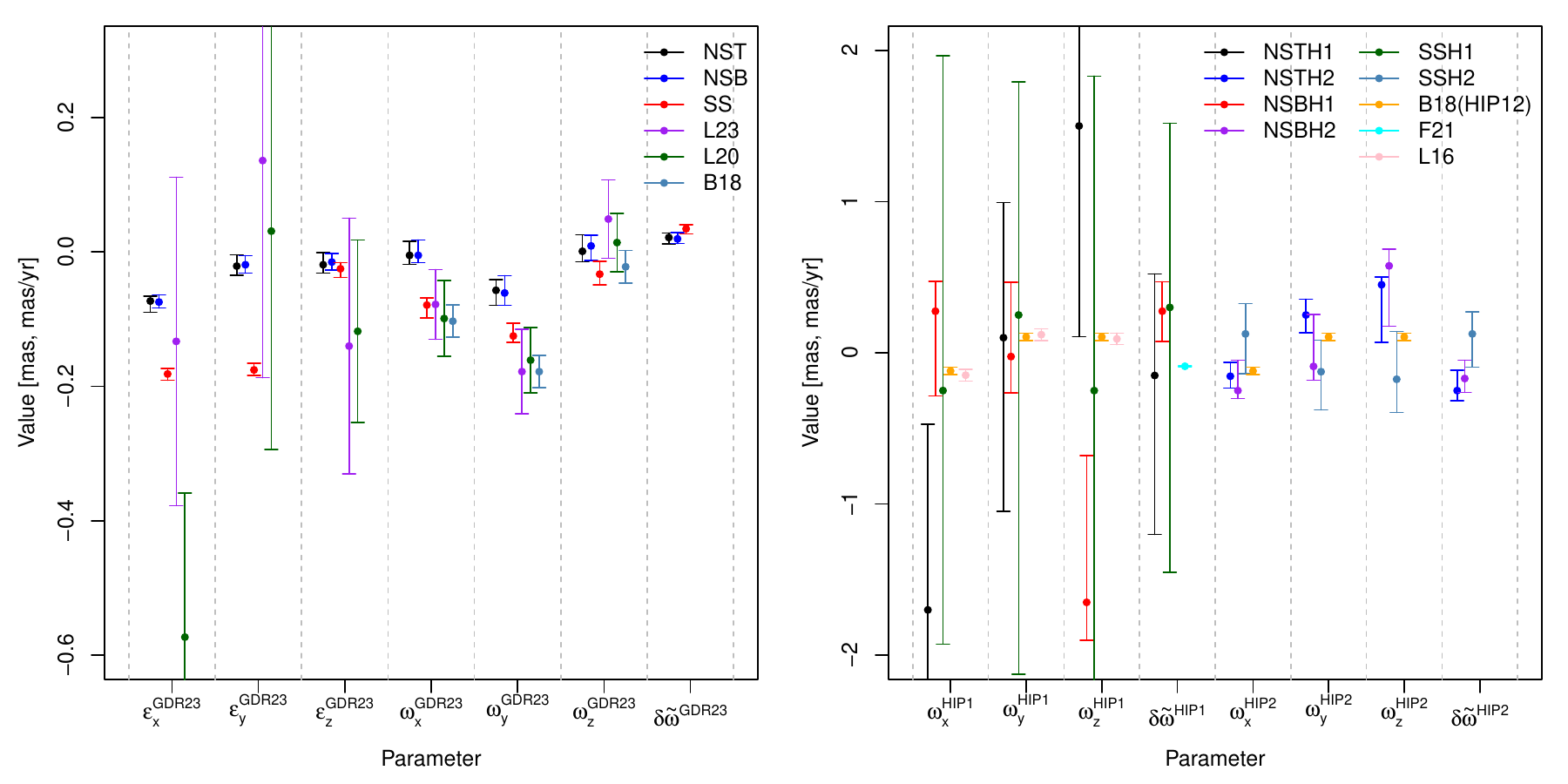}
    \caption{Comparison of differential calibration parameters determined in this
      work and in literature. The left panel shows the values of
      $\beta^{\rm GDR23}$ while the right panel shows $\vec\beta^d$ of
      HIP1 and HIP2 relative to GDR3. Because \cite{brandt18} determines the
      calibration parameters for a hybrid of HIP1 and HIP2, the same B18
      values are presented for both HIP1 and HIP2. }
  \label{fig:compare}
\end{figure}

In Fig. \ref{fig:compare}, we juxtapose the calibration parameters derived in this study with values reported in the literature. Given that literature values for frame rotation and zero-point parallax are referenced to specific realizations of the International Celestial Reference Frame (ICRF), we adjust the literature values by subtracting the GDR3 values determined by \cite{lunz23} for a meaningful comparison. Notably, due to the absence of literature calibration for the frame rotation of TYC and GDR1, our comparison in Fig. \ref{fig:compare} focuses solely on calibration parameters for GDR2 and Hipparcos.

In the left panel, it is evident that the values of $\vec\beta^{\rm GDR23}$ obtained through analyses of NST, NSB, and SS calibration sources align well with those from previous studies within a 2$\sigma$ range. Importantly, our parameters exhibit higher precision compared to previous studies, underscoring the robustness of our calibration methodology. Noteworthy consistency is observed between NST and NSB, as expected since both correct GDR3 astrometry to the binary barycenter albeit with different methods. While NST (or NSB) and SS yield similar values for $\epsilon_z$, $\omega_x$, $\omega_y$, $\omega_z$, and $\delta\varpi$, they diverge in $\epsilon_x$ and $\epsilon_y$, indicating a potential dependence of frame rotation on calibration sources.

The right panel of Fig. \ref{fig:compare} reveals significant scatter in $\vec\omega$ and $\delta\varpi$ for NSTH1, NSTH2, NSBH1, NSBH2, SSH1, and SSH2 in our study. In contrast, \cite{lindegren16} provide more precise values for these parameters by comparing Hipparcos with VLBI observations of 262 radio stars. Additionally, \cite{brandt18} presents similar values for $\vec\omega$ by comparing the Hipparcos catalog with long-term proper motion derived from the positional difference between Hipparcos and Gaia, although the fitting does not include $\vec\epsilon$.

\subsection{Dependence of calibration parameters on magnitude and color}\label{sec:dependence}
\begin{figure}
  \centering
  \includegraphics[scale=0.55]{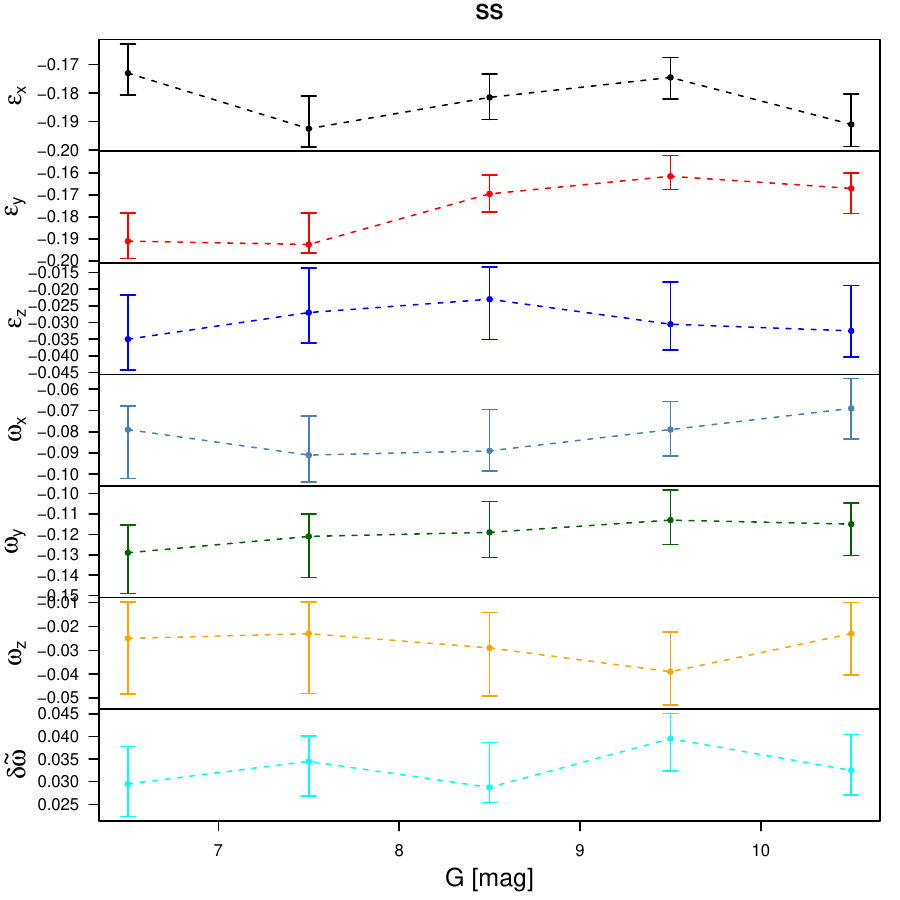}
  \includegraphics[scale=0.55]{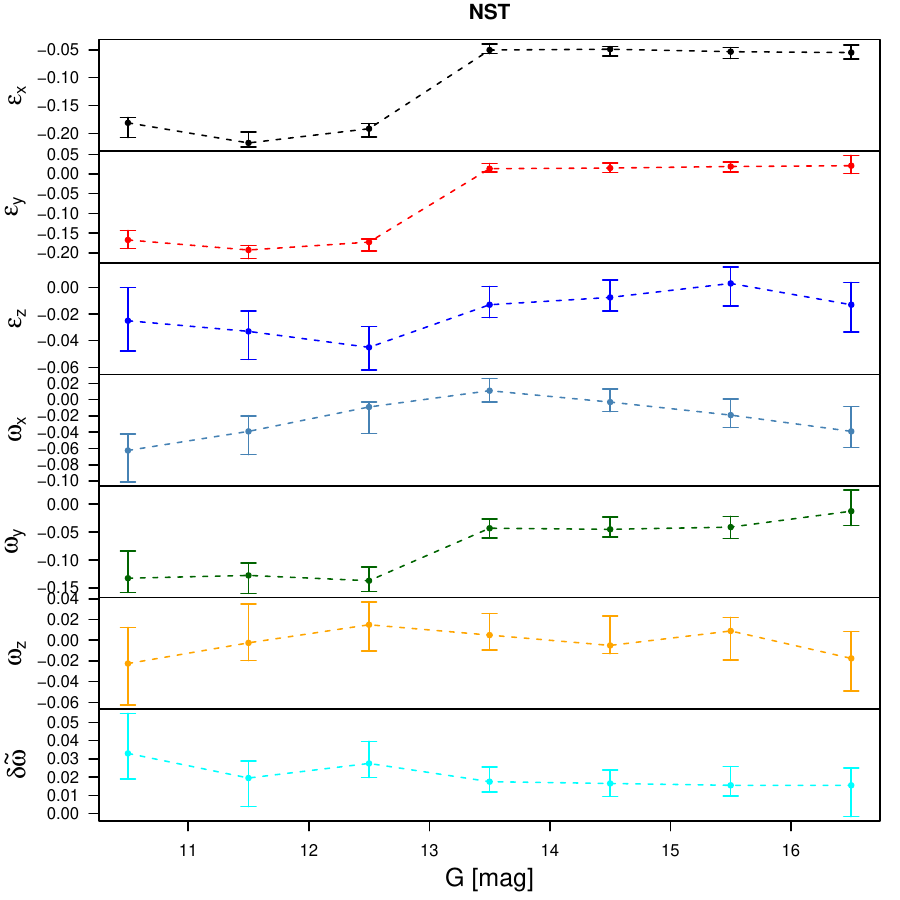}
  \includegraphics[scale=0.55]{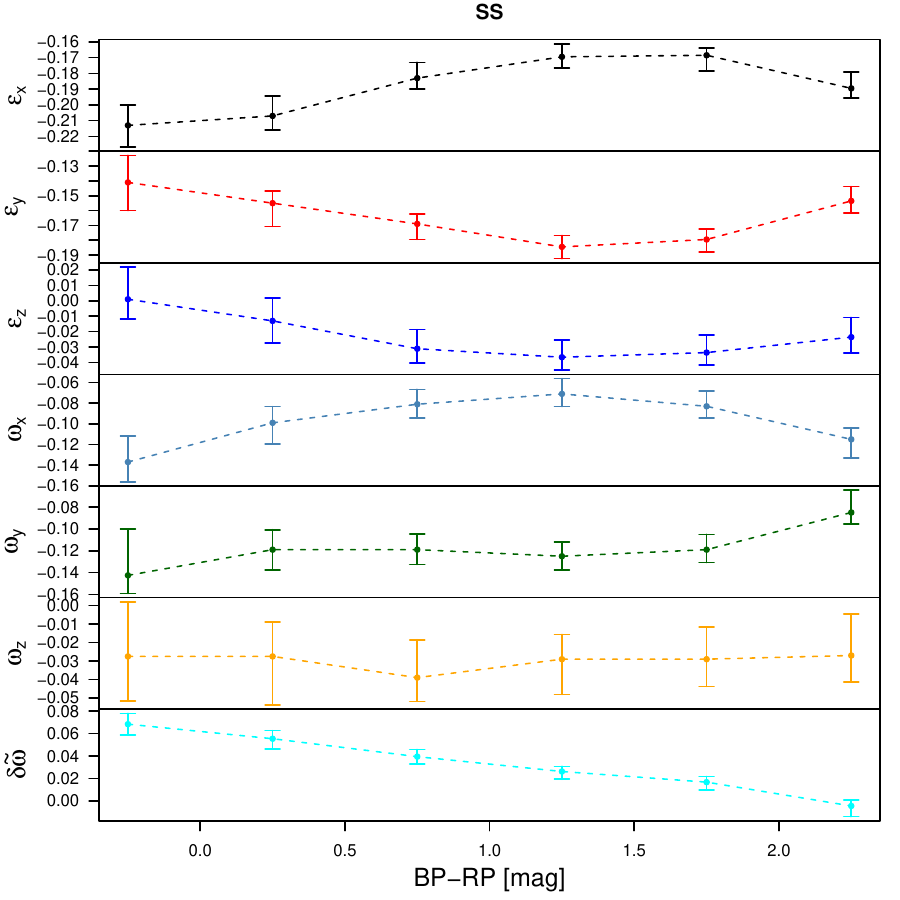}
  \includegraphics[scale=0.55]{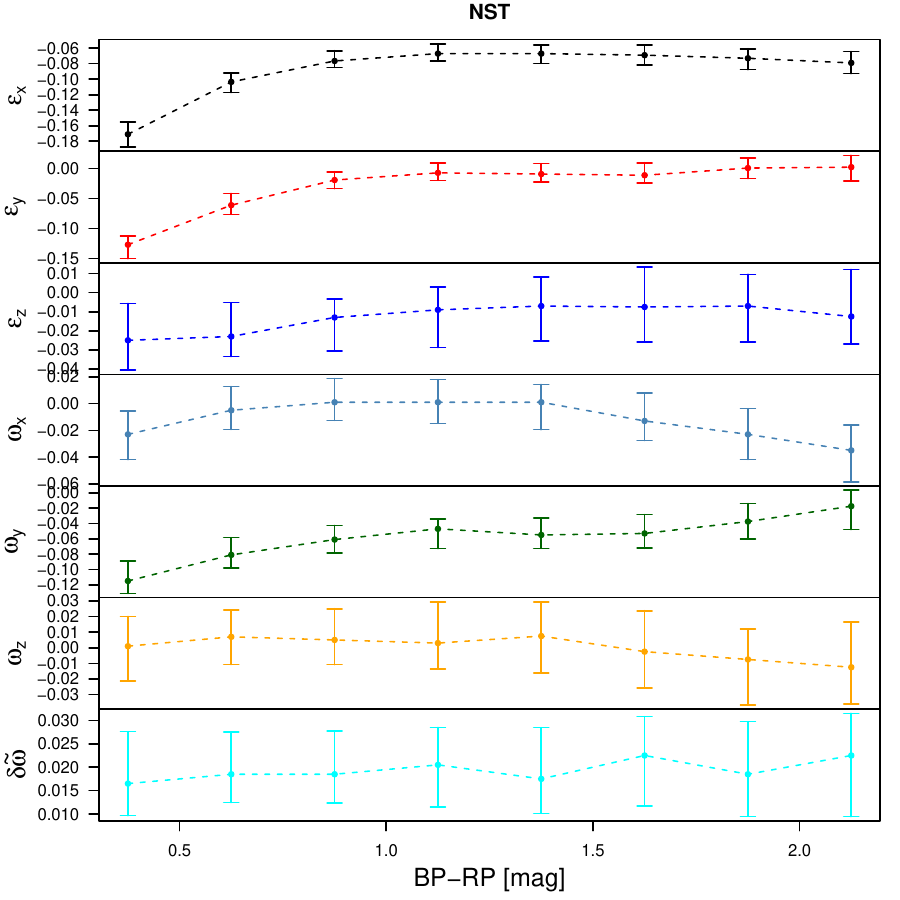}
    \caption{Dependence of differential calibration parameters
      ($\vec\beta^{\rm GDR23}$) on G magnitude (top) and BP-RP
      color (bottom) for the SS (left) and NST (right) calibration
      sources. Different colors represent different parameters.}
  \label{fig:mag_color}
\end{figure}

The relationship between calibration parameters in Gaia Data Releases
(DRs) and stellar magnitude and color has been explored in previous
studies (e.g., \citealt{lindegren21a} and \citealt{gaudin21}). Given
the notably precise determination of calibration parameters for GDR2
relative to GDR3 ($\vec\beta^{\rm GDR23}$) compared to other catalogs,
our investigation focuses exclusively on exploring the dependence of
$\vec\beta^{\rm GDR23}$ on the G magnitude and BP-RP color of Gaia
sources. This dependence is illustrated for NST and SS calibration
sources in Fig. \ref{fig:mag_color}.

The top two panels exhibit the correlation between calibration
parameters and G magnitude. Notably, the faint NST calibration sources
display more pronounced parameter variations than their bright SS
counterparts. A distinct shift in GDR2-GDR3 frame offsets is observed
around $G=13$ mag, a phenomenon also noted by \cite{brandt18},
\cite{lindegren18}, and \cite{gaudin21}. This abrupt change is likely
attributed to Hipparcos being unable to serve as a reference catalog
for calibrating GDR3 for targets with $11<G<13$ mag
\citep{gaudin21}. Specifically, we identify a frame offset of
$\vec\epsilon^{\rm GDR23}=(0.14,0.19,0.03)\pm(0.01,0.02,0.02)$ mas
between the faint (G$>13$ mag) and bright (G$<13$ mag) frames of GDR2
relative to GDR3. However, the bright frame does not exhibit a sudden
change in the rate of frame rotation (i.e., $\vec\omega$) relative to
the faint frame of GDR2, despite continuous variation across the G
magnitude range. In comparison to NST calibration sources, the SS sources do not demonstrate a significant dependence of differential calibration on G magnitude.

The lower panels of Fig. \ref{fig:mag_color} illustrate the color-dependent behavior of the differential calibration parameters for both SS and NST sources. Specifically, the color-dependent variations for SS and NST are respectively $(0.02, 0.02, 0.02, 0.03, 0.02, 0.02, 0.03)$\,mas (or \masyr) and $(0.04, 0.05, 0.02, 0.02, 0.03, 0.02, 0.01)$\,mas (or \masyr), exhibiting a significance level of approximately $3\sigma$. Notably, we observe a consistent monotonic decrease in the zero-point parallax ($\delta\varpi$) with BP-RP for the SS sources, aligning with the color-dependent trend of $\delta\varpi$ depicted in the top-left panel of \cite{lindegren21a}. 

Following a thorough comparison of various calibration results, we
offer the following differential calibration recommendations relative
to GDR3 for different catalogs:
\begin{itemize}
\item For straightforward calibration of GDR2 data, employ SS-based values for sources with $G<$10.5\,mag and NST-based values for sources with $G>$10.5\,mag.

\item For precise calibration of GDR2 data, utilize the {\small Python} scripts provided in the appendix to address bias on a case-by-case basis.

\item Calibrate the 2-parameter GDR1 solutions using parameters determined with NSTP2.

\item For GDR1 targets with 5-parameter solutions, use the SSP5 values
  to calibrate $G<10.5$\,mag stars and use the NSTP2 values to
  calibrate $G>10.5$\,mag stars. 
  
\item Use TYC positions without correction, in conjunction with GDR2 and GDR3, to constrain orbits.

\item Calibrate HIP1 and HIP2 data using either non-zero
  $\vec{\epsilon}$ or non-zero $\vec{\omega}$ given by
  \cite{lindegren16} but not both, the ICRF2 could be transformed to
  GDR3 frame using the calibration parameters given by \cite{charlot20,lunz23}.

\item Quadratically add an astrometric jitter of 2.16 mas (or mas/yr) to HIP2 astrometry.
\end{itemize}
These recommendations are automatically implemented in the Python script provided in the appendix.

\subsection{Dependence of calibration parameters on RUWE}\label{sec:ruwe}
Investigating potential biases in the SS astrometric solutions
  arising from undetected binaries, we examine the impact of the RUWE
  parameter on calibration parameters. RUWE serves as an indicator of binarity or significant excess noise. Similar to the approach outlined in section \ref{sec:dependence}, we categorize the SS calibration sources into subsamples based on different RUWE values.

Specifically, we partition the RUWE range of SS sources into five bins: [0, 0.8], [0.8, 1], [1, 1.2], [1.2, 1.4], and [1.4, 60]. The corresponding subsample sizes are 987, 34,743, 31,762, 5,073, and 5,338, respectively. For each subsample, we derive the calibration parameters using linear regression.

\begin{figure}
  \centering
  \includegraphics[scale=0.8]{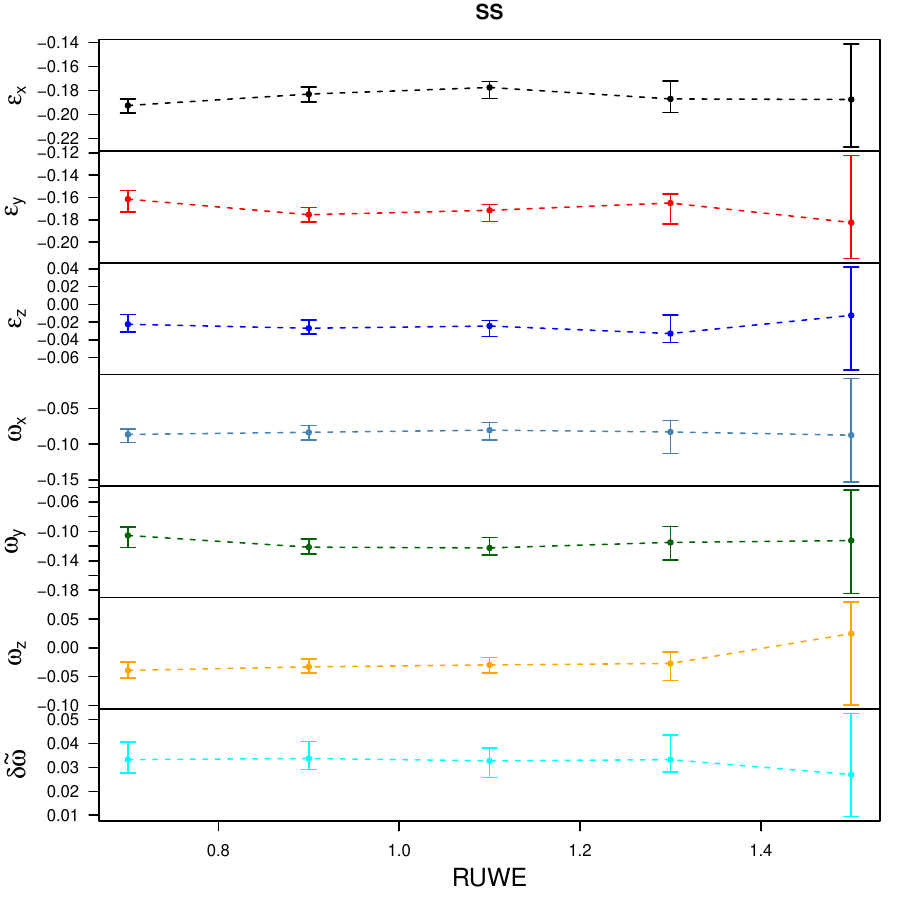}
    \caption{\bf Dependence of differential calibration parameters
      ($\vec\beta^{\rm GDR23}$) on GDR3 RUWE for the SS calibration
      sources. Different colors represent different parameters.}
  \label{fig:ruwe}
\end{figure}

The dependence of calibration parameters on RUWE is visually
represented in Fig. \ref{fig:ruwe}. Remarkably, the parameters derived from the SS subsamples demonstrate minimal sensitivity to RUWE, indicating that our SS-based calibration remains largely unbiased even in the presence of undetected binaries. However, the uncertainty associated with calibration parameters increases, particularly for RUWE values exceeding 1.2. This outcome aligns with expectations, as sources with higher RUWEs or elevated astrometric jitters contribute substantial uncertainty to the calibration parameters. This reinforces the robustness of our approach as an effective calibration method.

\section{Conclusion}\label{sec:conclusion}
In order to identify dark companions and binaries across multiple Gaia
DRs, we have developed an astrometric modeling framework, designed for the combined analysis of Gaia, Hipparcos, and
Tycho-2 catalogs. To address biases inherent in these catalogs, we
ascertain error inflation, astrometric jitter, differential frame
rotation, and parallax zero-point relative to GDR3. This involves
utilizing calibration sources selected from GDR3, both with
and without orbital solutions. Through the simultaneous fitting of
calibration parameters and barycentric astrometry to the calibration
sources, our analysis reveals negligible error inflation across all
catalogs. Notably, a significant jitter of 2.16\,mas is identified for
HIP2 IAD, a frame offset ranging from 0.12 to 0.26\,mas, and a frame
rotation of 0.07-0.15\masyr for GDR2 relative to GDR3. Additionally, a
substantial frame offset of approximately -0.12\,mas along the y-axis
is observed between G1P2 and GDR3. While our estimation of calibration parameters for HIP1 and HIP2 aligns with previous studies, it's worth noting that the precision of values determined in this work falls short compared to those derived from VLBI observations of radio stars, as demonstrated by \cite{lindegren16}.

Given the higher precision with which we determine the differential
calibration parameters for GDR2 in this study compared to prior
research, we delve deeper into investigating the dependency of these
parameters on G-magnitude and BP-RP color. Our findings reveal a
magnitude-dependent frame offset of $0.08\pm0.02$ mas and frame
rotation of $0.26\pm 0.02$\masyr for faint stars (G$>11$ mag) in
GDR2. Additionally, we identify a noteworthy 0.24\,mas offset between
bright (G$<13$\,mag) and faint (G$>13$\,mag) frames, aligning closely
with findings from earlier studies such as \cite{brandt18},
\cite{lindegren18}, and \cite{gaudin21}. Additionally, a significant dependence of zero-point parallax on color is found in the SS calibration sources,
consistent with previous studies by \cite{lindegren21a}. A pipeline to
calculate the magnitude-and-color dependence of calibration parameters
is provided in the appendix. 

After comparing the calibration conducted in this study with previous
research, we propose the utilization of NST and SS values for the
global calibration of GDR2. For the calibration of G1P2 and G1P5, we
suggest employing NSTP2 values. Additionally, we recommend adopting
the calibration values provided by \cite{lindegren16} and
\cite{lunz23} for the calibration of HIP1 and HIP2. Between the two versions of Hipparcos data, we advocate for the use of HIP2, incorporating an astrometric jitter of 2.16 mas. To mitigate unknown systematics in constraining stellar orbits, our recommendation is to utilize TYC data without correction. Furthermore, for GDR1 sources, we propose using the position instead of five-parameter astrometry.

While we conduct global calibrations for the most widely used
astrometric catalogs, our current focus is solely on the examination
of the dependence of differential calibration parameters on magnitude
and color specifically for GDR2. To enhance the a priori calibration
of astrometric bias, future efforts should delve into a detailed
investigation of how systematics vary with additional parameters such
as position and distance across various catalogs. However, it is
important to note that our ability to develop a comprehensive
calibration function for different catalogs is constrained. Therefore,
adopting an approach of analyzing astrometric data on a case-by-case
basis is likely the optimal solution for bias mitigation, as
demonstrated by studies such as \citealt{snellen18} and
\citealt{feng19b}. Nonetheless, our work ensures a robust calibration
of the most prevalent all-sky astrometric catalogs, facilitating the
efficient detection of dark companions and binaries in extensive
datasets using our algorithm. 

\section*{Acknowledgements}
This work is supported by Shanghai Jiao Tong University 2030
Initiative. We extend our gratitude to the anonymous referee and the statistical editor for their valuable comments, which have significantly contributed to the enhancement of the manuscript. This work has made use of data from
the European Space Agency (ESA) mission Gaia
(https://www.cosmos.esa.int/gaia), processed by the Gaia Data
Processing and Analysis Consortium (DPAC, https://www.cosmos.esa.int/web/gaia/dpac/consortium). Funding for the DPAC has been provided by national institutions, in particular the
institutions participating in the Gaia Multilateral Agreement. This
research has also made use of the SIMBAD database, operated at CDS, Strasbourg, France. 

\software{R Statistical Software \citep{r23}, GGally \citep{ggally}, viridis \citep{viridis}, ggplot2 \citep{ggplot2}}
\bibliographystyle{aasjournal}
\bibliography{nm}

\begin{thebibliography}{}
\expandafter\ifx\csname natexlab\endcsname\relax\def\natexlab#1{#1}\fi
\providecommand{\url}[1]{\href{#1}{#1}}
\providecommand{\dodoi}[1]{doi:~\href{http://doi.org/#1}{\nolinkurl{#1}}}
\providecommand{\doeprint}[1]{\href{http://ascl.net/#1}{\nolinkurl{http://ascl.net/#1}}}
\providecommand{\doarXiv}[1]{\href{https://arxiv.org/abs/#1}{\nolinkurl{https://arxiv.org/abs/#1}}}

\bibitem[{{Arenou} {et~al.}(2018){Arenou}, {Luri}, {Babusiaux}, {Fabricius},
  {Helmi}, {Muraveva}, {Robin}, {Spoto}, {Vallenari}, {Antoja},
  {Cantat-Gaudin}, {Jordi}, {Leclerc}, {Reyl{\'e}}, {Romero-G{\'o}mez}, {Shih},
  {Soria}, {Barache}, {Bossini}, {Bragaglia}, {Breddels}, {Fabrizio},
  {Lambert}, {Marrese}, {Massari}, {Moitinho}, {Robichon}, {Ruiz-Dern},
  {Sordo}, {Veljanoski}, {Eyer}, {Jasniewicz}, {Pancino}, {Soubiran}, {Spagna},
  {Tanga}, {Turon}, \& {Zurbach}}]{arenou18}
{Arenou}, F., {Luri}, X., {Babusiaux}, C., {et~al.} 2018, \aap, 616, A17,
  \dodoi{10.1051/0004-6361/201833234}

\bibitem[{{Brandt} {et~al.}(2023){Brandt}, {Michalik}, \& {Brandt}}]{brandt23}
{Brandt}, G.~M., {Michalik}, D., \& {Brandt}, T.~D. 2023, RAS Techniques and
  Instruments, 2, 218, \dodoi{10.1093/rasti/rzad011}

\bibitem[{{Brandt} {et~al.}(2021){Brandt}, {Michalik}, {Brandt}, {Li}, {Dupuy},
  \& {Zeng}}]{brandt21}
{Brandt}, G.~M., {Michalik}, D., {Brandt}, T.~D., {et~al.} 2021, \aj, 162, 230,
  \dodoi{10.3847/1538-3881/ac12d0}

\bibitem[{{Brandt}(2018)}]{brandt18}
{Brandt}, T.~D. 2018, \apjs, 239, 31, \dodoi{10.3847/1538-4365/aaec06}

\bibitem[{{Brandt} {et~al.}(2019){Brandt}, {Dupuy}, \& {Bowler}}]{brandt19}
{Brandt}, T.~D., {Dupuy}, T.~J., \& {Bowler}, B.~P. 2019, \aj, 158, 140,
  \dodoi{10.3847/1538-3881/ab04a8}

\bibitem[{{Cantat-Gaudin} \& {Brandt}(2021)}]{gaudin21}
{Cantat-Gaudin}, T., \& {Brandt}, T.~D. 2021, \aap, 649, A124,
  \dodoi{10.1051/0004-6361/202140807}

\bibitem[{{Charlot} {et~al.}(2020){Charlot}, {Jacobs}, {Gordon}, {Lambert}, {de
  Witt}, {B{\"o}hm}, {Fey}, {Heinkelmann}, {Skurikhina}, {Titov}, {Arias},
  {Bolotin}, {Bourda}, {Ma}, {Malkin}, {Nothnagel}, {Mayer}, {MacMillan},
  {Nilsson}, \& {Gaume}}]{charlot20}
{Charlot}, P., {Jacobs}, C.~S., {Gordon}, D., {et~al.} 2020, \aap, 644, A159,
  \dodoi{10.1051/0004-6361/202038368}

\bibitem[{{El-Badry} {et~al.}(2023){El-Badry}, {Rix}, {Cendes}, {Rodriguez},
  {Conroy}, {Quataert}, {Hawkins}, {Zari}, {Hobson}, {Breivik}, {Rau},
  {Berger}, {Shahaf}, {Seeburger}, {Burdge}, {Latham}, {Buchhave}, {Bieryla},
  {Bashi}, {Mazeh}, \& {Faigler}}]{elbadry23}
{El-Badry}, K., {Rix}, H.-W., {Cendes}, Y., {et~al.} 2023, \mnras, 521, 4323,
  \dodoi{10.1093/mnras/stad799}

\bibitem[{{Fabricius} {et~al.}(2021){Fabricius}, {Luri}, {Arenou}, {Babusiaux},
  {Helmi}, {Muraveva}, {Reyl{\'e}}, {Spoto}, {Vallenari}, {Antoja}, {Balbinot},
  {Barache}, {Bauchet}, {Bragaglia}, {Busonero}, {Cantat-Gaudin}, {Carrasco},
  {Diakit{\'e}}, {Fabrizio}, {Figueras}, {Garcia-Gutierrez}, {Garofalo},
  {Jordi}, {Kervella}, {Khanna}, {Leclerc}, {Licata}, {Lambert}, {Marrese},
  {Masip}, {Ramos}, {Robichon}, {Robin}, {Romero-G{\'o}mez}, {Rubele}, \&
  {Weiler}}]{fabricius21}
{Fabricius}, C., {Luri}, X., {Arenou}, F., {et~al.} 2021, \aap, 649, A5,
  \dodoi{10.1051/0004-6361/202039834}

\bibitem[{{Feng} {et~al.}(2019){Feng}, {Anglada-Escud{\'e}}, {Tuomi}, {Jones},
  {Chanam{\'e}}, {Butler}, \& {Janson}}]{feng19b}
{Feng}, F., {Anglada-Escud{\'e}}, G., {Tuomi}, M., {et~al.} 2019, \mnras, 490,
  5002, \dodoi{10.1093/mnras/stz2912}

\bibitem[{{Feng} {et~al.}(2023){Feng}, {Butler}, {Vogt}, {Holden}, \&
  {Rui}}]{feng23}
{Feng}, F., {Butler}, R.~P., {Vogt}, S.~S., {Holden}, B., \& {Rui}, Y. 2023,
  \mnras, 525, 607, \dodoi{10.1093/mnras/stad2297}

\bibitem[{{Feng} {et~al.}(2016){Feng}, {Tuomi}, {Jones}, {Butler}, \&
  {Vogt}}]{feng16}
{Feng}, F., {Tuomi}, M., {Jones}, H.~R.~A., {Butler}, R.~P., \& {Vogt}, S.
  2016, \mnras, 461, 2440, \dodoi{10.1093/mnras/stw1478}

\bibitem[{{Feng} {et~al.}(2021){Feng}, {Butler}, {Jones}, {Phillips}, {Vogt},
  {Oppenheimer}, {Holden}, {Burt}, \& {Boss}}]{feng21}
{Feng}, F., {Butler}, R.~P., {Jones}, H. R.~A., {et~al.} 2021, \mnras, 507,
  2856, \dodoi{10.1093/mnras/stab2225}

\bibitem[{{Feng} {et~al.}(2022){Feng}, {Butler}, {Vogt}, {Clement}, {Tinney},
  {Cui}, {Aizawa}, {Jones}, {Bailey}, {Burt}, {Carter}, {Crane}, {Dotti},
  {Holden}, {Ma}, {Ogihara}, {Oppenheimer}, {O'Toole}, {Shectman},
  {Wittenmyer}, {Wang}, {Wright}, \& {Xuan}}]{feng22}
{Feng}, F., {Butler}, R.~P., {Vogt}, S.~S., {et~al.} 2022, \apjs, 262, 21,
  \dodoi{10.3847/1538-4365/ac7e57}

\bibitem[{{Fey} {et~al.}(2015){Fey}, {Gordon}, {Jacobs}, {Ma}, {Gaume},
  {Arias}, {Bianco}, {Boboltz}, {B{\"o}ckmann}, {Bolotin}, {Charlot},
  {Collioud}, {Engelhardt}, {Gipson}, {Gontier}, {Heinkelmann}, {Kurdubov},
  {Lambert}, {Lytvyn}, {MacMillan}, {Malkin}, {Nothnagel}, {Ojha},
  {Skurikhina}, {Sokolova}, {Souchay}, {Sovers}, {Tesmer}, {Titov}, {Wang}, \&
  {Zharov}}]{fey15}
{Fey}, A.~L., {Gordon}, D., {Jacobs}, C.~S., {et~al.} 2015, \aj, 150, 58,
  \dodoi{10.1088/0004-6256/150/2/58}

\bibitem[{{Gaia Collaboration} {et~al.}(2016{\natexlab{a}}){Gaia
  Collaboration}, {Prusti}, {de Bruijne}, {Brown}, {Vallenari}, {Babusiaux},
  {Bailer-Jones}, {Bastian}, {Biermann}, {Evans}, {Eyer}, {Jansen}, {Jordi},
  {Klioner}, {Lammers}, {Lindegren}, {Luri}, {Mignard}, {Milligan}, {Panem},
  {Poinsignon}, {Pourbaix}, {Randich}, {Sarri}, {Sartoretti}, {Siddiqui},
  {Soubiran}, {Valette}, {van Leeuwen}, {Walton}, {Aerts}, {Arenou}, {Cropper},
  {Drimmel}, {H{\o}g}, {Katz}, {Lattanzi}, {O'Mullane}, {Grebel}, {Holland},
  {Huc}, {Passot}, {Bramante}, {Cacciari}, {Casta{\~n}eda}, {Chaoul}, {Cheek},
  {De Angeli}, {Fabricius}, {Guerra}, {Hern{\'a}ndez}, {Jean-Antoine-Piccolo},
  {Masana}, {Messineo}, {Mowlavi}, {Nienartowicz}, {Ord{\'o}{\~n}ez-Blanco},
  {Panuzzo}, {Portell}, {Richards}, {Riello}, {Seabroke}, {Tanga},
  {Th{\'e}venin}, {Torra}, {Els}, {Gracia-Abril}, {Comoretto},
  {Garcia-Reinaldos}, {Lock}, {Mercier}, {Altmann}, {Andrae}, {Astraatmadja},
  {Bellas-Velidis}, {Benson}, {Berthier}, {Blomme}, {Busso}, {Carry},
  {Cellino}, {Clementini}, {Cowell}, {Creevey}, {Cuypers}, {Davidson}, {De
  Ridder}, {de Torres}, {Delchambre}, {Dell'Oro}, {Ducourant}, {Fr{\'e}mat},
  {Garc{\'\i}a-Torres}, {Gosset}, {Halbwachs}, {Hambly}, {Harrison}, {Hauser},
  {Hestroffer}, {Hodgkin}, {Huckle}, {Hutton}, {Jasniewicz}, {Jordan},
  {Kontizas}, {Korn}, {Lanzafame}, {Manteiga}, {Moitinho}, {Muinonen},
  {Osinde}, {Pancino}, {Pauwels}, {Petit}, {Recio-Blanco}, {Robin}, {Sarro},
  {Siopis}, {Smith}, {Smith}, {Sozzetti}, {Thuillot}, {van Reeven}, {Viala},
  {Abbas}, {Abreu Aramburu}, {Accart}, {Aguado}, {Allan}, {Allasia},
  {Altavilla}, {{\'A}lvarez}, {Alves}, {Anderson}, {Andrei}, {Anglada Varela},
  {Antiche}, {Antoja}, {Ant{\'o}n}, {Arcay}, {Atzei}, {Ayache}, {Bach},
  {Baker}, {Balaguer-N{\'u}{\~n}ez}, {Barache}, {Barata}, {Barbier}, {Barblan},
  {Baroni}, {Barrado y Navascu{\'e}s}, {Barros}, {Barstow}, {Becciani},
  {Bellazzini}, {Bellei}, {Bello Garc{\'\i}a}, {Belokurov}, {Bendjoya},
  {Berihuete}, {Bianchi}, {Bienaym{\'e}}, {Billebaud}, {Blagorodnova},
  {Blanco-Cuaresma}, {Boch}, {Bombrun}, {Borrachero}, {Bouquillon}, {Bourda},
  {Bouy}, {Bragaglia}, {Breddels}, {Brouillet}, {Br{\"u}semeister},
  {Bucciarelli}, {Budnik}, {Burgess}, {Burgon}, {Burlacu}, {Busonero}, {Buzzi},
  {Caffau}, {Cambras}, {Campbell}, {Cancelliere}, {Cantat-Gaudin}, {Carlucci},
  {Carrasco}, {Castellani}, {Charlot}, {Charnas}, {Charvet}, {Chassat},
  {Chiavassa}, {Clotet}, {Cocozza}, {Collins}, {Collins}, {Costigan}, {Crifo},
  {Cross}, {Crosta}, {Crowley}, {Dafonte}, {Damerdji}, {Dapergolas}, {David},
  {David}, {De Cat}, {de Felice}, {de Laverny}, {De Luise}, {De March}, {de
  Martino}, {de Souza}, {Debosscher}, {del Pozo}, {Delbo}, {Delgado},
  {Delgado}, {di Marco}, {Di Matteo}, {Diakite}, {Distefano}, {Dolding}, {Dos
  Anjos}, {Drazinos}, {Dur{\'a}n}, {Dzigan}, {Ecale}, {Edvardsson}, {Enke},
  {Erdmann}, {Escolar}, {Espina}, {Evans}, {Eynard Bontemps}, {Fabre},
  {Fabrizio}, {Faigler}, {Falc{\~a}o}, {Farr{\`a}s Casas}, {Faye}, {Federici},
  {Fedorets}, {Fern{\'a}ndez-Hern{\'a}ndez}, {Fernique}, {Fienga}, {Figueras},
  {Filippi}, {Findeisen}, {Fonti}, {Fouesneau}, {Fraile}, {Fraser}, {Fuchs},
  {Furnell}, {Gai}, {Galleti}, {Galluccio}, {Garabato}, {Garc{\'\i}a-Sedano},
  {Gar{\'e}}, {Garofalo}, {Garralda}, {Gavras}, {Gerssen}, {Geyer}, {Gilmore},
  {Girona}, {Giuffrida}, {Gomes}, {Gonz{\'a}lez-Marcos},
  {Gonz{\'a}lez-N{\'u}{\~n}ez}, {Gonz{\'a}lez-Vidal}, {Granvik}, {Guerrier},
  {Guillout}, {Guiraud}, {G{\'u}rpide}, {Guti{\'e}rrez-S{\'a}nchez}, {Guy},
  {Haigron}, {Hatzidimitriou}, {Haywood}, {Heiter}, {Helmi}, {Hobbs},
  {Hofmann}, {Holl}, {Holland}, {Hunt}, {Hypki}, {Icardi}, {Irwin}, {Jevardat
  de Fombelle}, {Jofr{\'e}}, {Jonker}, {Jorissen}, {Julbe}, {Karampelas},
  {Kochoska}, {Kohley}, {Kolenberg}, {Kontizas}, {Koposov}, {Kordopatis},
  {Koubsky}, {Kowalczyk}, {Krone-Martins}, {Kudryashova}, {Kull}, {Bachchan},
  {Lacoste-Seris}, {Lanza}, {Lavigne}, {Le Poncin-Lafitte}, {Lebreton},
  {Lebzelter}, {Leccia}, {Leclerc}, {Lecoeur-Taibi}, {Lemaitre}, {Lenhardt},
  {Leroux}, {Liao}, {Licata}, {Lindstr{\o}m}, {Lister}, {Livanou}, {Lobel},
  {L{\"o}ffler}, {L{\'o}pez}, {Lopez-Lozano}, {Lorenz}, {Loureiro},
  {MacDonald}, {Magalh{\~a}es Fernandes}, {Managau}, {Mann}, {Mantelet},
  {Marchal}, {Marchant}, {Marconi}, {Marie}, {Marinoni}, {Marrese},
  {Marschalk{\'o}}, {Marshall}, {Mart{\'\i}n-Fleitas}, {Martino}, {Mary},
  {Matijevi{\v{c}}}, {Mazeh}, {McMillan}, {Messina}, {Mestre}, {Michalik},
  {Millar}, {Miranda}, {Molina}, {Molinaro}, {Molinaro}, {Moln{\'a}r},
  {Moniez}, {Montegriffo}, {Monteiro}, {Mor}, {Mora}, {Morbidelli}, {Morel},
  {Morgenthaler}, {Morley}, {Morris}, {Mulone}, {Muraveva}, {Musella},
  {Narbonne}, {Nelemans}, {Nicastro}, {Noval}, {Ord{\'e}novic},
  {Ordieres-Mer{\'e}}, {Osborne}, {Pagani}, {Pagano}, {Pailler}, {Palacin},
  {Palaversa}, {Parsons}, {Paulsen}, {Pecoraro}, {Pedrosa}, {Pentik{\"a}inen},
  {Pereira}, {Pichon}, {Piersimoni}, {Pineau}, {Plachy}, {Plum}, {Poujoulet},
  {Pr{\v{s}}a}, {Pulone}, {Ragaini}, {Rago}, {Rambaux}, {Ramos-Lerate},
  {Ranalli}, {Rauw}, {Read}, {Regibo}, {Renk}, {Reyl{\'e}}, {Ribeiro},
  {Rimoldini}, {Ripepi}, {Riva}, {Rixon}, {Roelens}, {Romero-G{\'o}mez},
  {Rowell}, {Royer}, {Rudolph}, {Ruiz-Dern}, {Sadowski}, {Sagrist{\`a}
  Sell{\'e}s}, {Sahlmann}, {Salgado}, {Salguero}, {Sarasso}, {Savietto},
  {Schnorhk}, {Schultheis}, {Sciacca}, {Segol}, {Segovia}, {Segransan},
  {Serpell}, {Shih}, {Smareglia}, {Smart}, {Smith}, {Solano}, {Solitro},
  {Sordo}, {Soria Nieto}, {Souchay}, {Spagna}, {Spoto}, {Stampa}, {Steele},
  {Steidelm{\"u}ller}, {Stephenson}, {Stoev}, {Suess}, {S{\"u}veges}, {Surdej},
  {Szabados}, {Szegedi-Elek}, {Tapiador}, {Taris}, {Tauran}, {Taylor},
  {Teixeira}, {Terrett}, {Tingley}, {Trager}, {Turon}, {Ulla}, {Utrilla},
  {Valentini}, {van Elteren}, {Van Hemelryck}, {van Leeuwen}, {Varadi},
  {Vecchiato}, {Veljanoski}, {Via}, {Vicente}, {Vogt}, {Voss}, {Votruba},
  {Voutsinas}, {Walmsley}, {Weiler}, {Weingrill}, {Werner}, {Wevers},
  {Whitehead}, {Wyrzykowski}, {Yoldas}, {{\v{Z}}erjal}, {Zucker}, {Zurbach},
  {Zwitter}, {Alecu}, {Allen}, {Allende Prieto}, {Amorim},
  {Anglada-Escud{\'e}}, {Arsenijevic}, {Azaz}, {Balm}, {Beck}, {Bernstein},
  {Bigot}, {Bijaoui}, {Blasco}, {Bonfigli}, {Bono}, {Boudreault}, {Bressan},
  {Brown}, {Brunet}, {Bunclark}, {Buonanno}, {Butkevich}, {Carret}, {Carrion},
  {Chemin}, {Ch{\'e}reau}, {Corcione}, {Darmigny}, {de Boer}, {de Teodoro}, {de
  Zeeuw}, {Delle Luche}, {Domingues}, {Dubath}, {Fodor}, {Fr{\'e}zouls},
  {Fries}, {Fustes}, {Fyfe}, {Gallardo}, {Gallegos}, {Gardiol}, {Gebran},
  {Gomboc}, {G{\'o}mez}, {Grux}, {Gueguen}, {Heyrovsky}, {Hoar}, {Iannicola},
  {Isasi Parache}, {Janotto}, {Joliet}, {Jonckheere}, {Keil}, {Kim},
  {Klagyivik}, {Klar}, {Knude}, {Kochukhov}, {Kolka}, {Kos}, {Kutka}, {Lainey},
  {LeBouquin}, {Liu}, {Loreggia}, {Makarov}, {Marseille}, {Martayan},
  {Martinez-Rubi}, {Massart}, {Meynadier}, {Mignot}, {Munari}, {Nguyen},
  {Nordlander}, {Ocvirk}, {O'Flaherty}, {Olias Sanz}, {Ortiz}, {Osorio},
  {Oszkiewicz}, {Ouzounis}, {Palmer}, {Park}, {Pasquato}, {Peltzer}, {Peralta},
  {P{\'e}turaud}, {Pieniluoma}, {Pigozzi}, {Poels}, {Prat}, {Prod'homme},
  {Raison}, {Rebordao}, {Risquez}, {Rocca-Volmerange}, {Rosen}, {Ruiz-Fuertes},
  {Russo}, {Sembay}, {Serraller Vizcaino}, {Short}, {Siebert}, {Silva},
  {Sinachopoulos}, {Slezak}, {Soffel}, {Sosnowska}, {Strai{\v{z}}ys}, {ter
  Linden}, {Terrell}, {Theil}, {Tiede}, {Troisi}, {Tsalmantza}, {Tur},
  {Vaccari}, {Vachier}, {Valles}, {Van Hamme}, {Veltz}, {Virtanen}, {Wallut},
  {Wichmann}, {Wilkinson}, {Ziaeepour}, \& {Zschocke}}]{gaia16a}
{Gaia Collaboration}, {Prusti}, T., {de Bruijne}, J.~H.~J., {et~al.}
  2016{\natexlab{a}}, \aap, 595, A1, \dodoi{10.1051/0004-6361/201629272}

\bibitem[{{Gaia Collaboration} {et~al.}(2016{\natexlab{b}}){Gaia
  Collaboration}, {Brown}, {Vallenari}, {Prusti}, {de Bruijne}, {Mignard},
  {Drimmel}, {Babusiaux}, {Bailer-Jones}, {Bastian}, {Biermann}, {Evans},
  {Eyer}, {Jansen}, {Jordi}, {Katz}, {Klioner}, {Lammers}, {Lindegren}, {Luri},
  {O'Mullane}, {Panem}, {Pourbaix}, {Randich}, {Sartoretti}, {Siddiqui},
  {Soubiran}, {Valette}, {van Leeuwen}, {Walton}, {Aerts}, {Arenou}, {Cropper},
  {H{\o}g}, {Lattanzi}, {Grebel}, {Holland}, {Huc}, {Passot}, {Perryman},
  {Bramante}, {Cacciari}, {Casta{\~n}eda}, {Chaoul}, {Cheek}, {De Angeli},
  {Fabricius}, {Guerra}, {Hern{\'a}ndez}, {Jean-Antoine-Piccolo}, {Masana},
  {Messineo}, {Mowlavi}, {Nienartowicz}, {Ord{\'o}{\~n}ez-Blanco}, {Panuzzo},
  {Portell}, {Richards}, {Riello}, {Seabroke}, {Tanga}, {Th{\'e}venin},
  {Torra}, {Els}, {Gracia-Abril}, {Comoretto}, {Garcia-Reinaldos}, {Lock},
  {Mercier}, {Altmann}, {Andrae}, {Astraatmadja}, {Bellas-Velidis}, {Benson},
  {Berthier}, {Blomme}, {Busso}, {Carry}, {Cellino}, {Clementini}, {Cowell},
  {Creevey}, {Cuypers}, {Davidson}, {De Ridder}, {de Torres}, {Delchambre},
  {Dell'Oro}, {Ducourant}, {Fr{\'e}mat}, {Garc{\'\i}a-Torres}, {Gosset},
  {Halbwachs}, {Hambly}, {Harrison}, {Hauser}, {Hestroffer}, {Hodgkin},
  {Huckle}, {Hutton}, {Jasniewicz}, {Jordan}, {Kontizas}, {Korn}, {Lanzafame},
  {Manteiga}, {Moitinho}, {Muinonen}, {Osinde}, {Pancino}, {Pauwels}, {Petit},
  {Recio-Blanco}, {Robin}, {Sarro}, {Siopis}, {Smith}, {Smith}, {Sozzetti},
  {Thuillot}, {van Reeven}, {Viala}, {Abbas}, {Abreu Aramburu}, {Accart},
  {Aguado}, {Allan}, {Allasia}, {Altavilla}, {{\'A}lvarez}, {Alves},
  {Anderson}, {Andrei}, {Anglada Varela}, {Antiche}, {Antoja}, {Ant{\'o}n},
  {Arcay}, {Bach}, {Baker}, {Balaguer-N{\'u}{\~n}ez}, {Barache}, {Barata},
  {Barbier}, {Barblan}, {Barrado y Navascu{\'e}s}, {Barros}, {Barstow},
  {Becciani}, {Bellazzini}, {Bello Garc{\'\i}a}, {Belokurov}, {Bendjoya},
  {Berihuete}, {Bianchi}, {Bienaym{\'e}}, {Billebaud}, {Blagorodnova},
  {Blanco-Cuaresma}, {Boch}, {Bombrun}, {Borrachero}, {Bouquillon}, {Bourda},
  {Bouy}, {Bragaglia}, {Breddels}, {Brouillet}, {Br{\"u}semeister},
  {Bucciarelli}, {Burgess}, {Burgon}, {Burlacu}, {Busonero}, {Buzzi}, {Caffau},
  {Cambras}, {Campbell}, {Cancelliere}, {Cantat-Gaudin}, {Carlucci},
  {Carrasco}, {Castellani}, {Charlot}, {Charnas}, {Chiavassa}, {Clotet},
  {Cocozza}, {Collins}, {Costigan}, {Crifo}, {Cross}, {Crosta}, {Crowley},
  {Dafonte}, {Damerdji}, {Dapergolas}, {David}, {David}, {De Cat}, {de Felice},
  {de Laverny}, {De Luise}, {De March}, {de Martino}, {de Souza}, {Debosscher},
  {del Pozo}, {Delbo}, {Delgado}, {Delgado}, {Di Matteo}, {Diakite},
  {Distefano}, {Dolding}, {Dos Anjos}, {Drazinos}, {Duran}, {Dzigan},
  {Edvardsson}, {Enke}, {Evans}, {Eynard Bontemps}, {Fabre}, {Fabrizio},
  {Faigler}, {Falc{\~a}o}, {Farr{\`a}s Casas}, {Federici}, {Fedorets},
  {Fern{\'a}ndez-Hern{\'a}ndez}, {Fernique}, {Fienga}, {Figueras}, {Filippi},
  {Findeisen}, {Fonti}, {Fouesneau}, {Fraile}, {Fraser}, {Fuchs}, {Gai},
  {Galleti}, {Galluccio}, {Garabato}, {Garc{\'\i}a-Sedano}, {Garofalo},
  {Garralda}, {Gavras}, {Gerssen}, {Geyer}, {Gilmore}, {Girona}, {Giuffrida},
  {Gomes}, {Gonz{\'a}lez-Marcos}, {Gonz{\'a}lez-N{\'u}{\~n}ez},
  {Gonz{\'a}lez-Vidal}, {Granvik}, {Guerrier}, {Guillout}, {Guiraud},
  {G{\'u}rpide}, {Guti{\'e}rrez-S{\'a}nchez}, {Guy}, {Haigron},
  {Hatzidimitriou}, {Haywood}, {Heiter}, {Helmi}, {Hobbs}, {Hofmann}, {Holl},
  {Holland}, {Hunt}, {Hypki}, {Icardi}, {Irwin}, {Jevardat de Fombelle},
  {Jofr{\'e}}, {Jonker}, {Jorissen}, {Julbe}, {Karampelas}, {Kochoska},
  {Kohley}, {Kolenberg}, {Kontizas}, {Koposov}, {Kordopatis}, {Koubsky},
  {Krone-Martins}, {Kudryashova}, {Kull}, {Bachchan}, {Lacoste-Seris}, {Lanza},
  {Lavigne}, {Le Poncin-Lafitte}, {Lebreton}, {Lebzelter}, {Leccia}, {Leclerc},
  {Lecoeur-Taibi}, {Lemaitre}, {Lenhardt}, {Leroux}, {Liao}, {Licata},
  {Lindstr{\o}m}, {Lister}, {Livanou}, {Lobel}, {L{\"o}ffler}, {L{\'o}pez},
  {Lorenz}, {MacDonald}, {Magalh{\~a}es Fernandes}, {Managau}, {Mann},
  {Mantelet}, {Marchal}, {Marchant}, {Marconi}, {Marinoni}, {Marrese},
  {Marschalk{\'o}}, {Marshall}, {Mart{\'\i}n-Fleitas}, {Martino}, {Mary},
  {Matijevi{\v{c}}}, {Mazeh}, {McMillan}, {Messina}, {Michalik}, {Millar},
  {Miranda}, {Molina}, {Molinaro}, {Molinaro}, {Moln{\'a}r}, {Moniez},
  {Montegriffo}, {Mor}, {Mora}, {Morbidelli}, {Morel}, {Morgenthaler},
  {Morris}, {Mulone}, {Muraveva}, {Musella}, {Narbonne}, {Nelemans},
  {Nicastro}, {Noval}, {Ord{\'e}novic}, {Ordieres-Mer{\'e}}, {Osborne},
  {Pagani}, {Pagano}, {Pailler}, {Palacin}, {Palaversa}, {Parsons}, {Pecoraro},
  {Pedrosa}, {Pentik{\"a}inen}, {Pichon}, {Piersimoni}, {Pineau}, {Plachy},
  {Plum}, {Poujoulet}, {Pr{\v{s}}a}, {Pulone}, {Ragaini}, {Rago}, {Rambaux},
  {Ramos-Lerate}, {Ranalli}, {Rauw}, {Read}, {Regibo}, {Reyl{\'e}}, {Ribeiro},
  {Rimoldini}, {Ripepi}, {Riva}, {Rixon}, {Roelens}, {Romero-G{\'o}mez},
  {Rowell}, {Royer}, {Ruiz-Dern}, {Sadowski}, {Sagrist{\`a} Sell{\'e}s},
  {Sahlmann}, {Salgado}, {Salguero}, {Sarasso}, {Savietto}, {Schultheis},
  {Sciacca}, {Segol}, {Segovia}, {Segransan}, {Shih}, {Smareglia}, {Smart},
  {Solano}, {Solitro}, {Sordo}, {Soria Nieto}, {Souchay}, {Spagna}, {Spoto},
  {Stampa}, {Steele}, {Steidelm{\"u}ller}, {Stephenson}, {Stoev}, {Suess},
  {S{\"u}veges}, {Surdej}, {Szabados}, {Szegedi-Elek}, {Tapiador}, {Taris},
  {Tauran}, {Taylor}, {Teixeira}, {Terrett}, {Tingley}, {Trager}, {Turon},
  {Ulla}, {Utrilla}, {Valentini}, {van Elteren}, {Van Hemelryck}, {van
  Leeuwen}, {Varadi}, {Vecchiato}, {Veljanoski}, {Via}, {Vicente}, {Vogt},
  {Voss}, {Votruba}, {Voutsinas}, {Walmsley}, {Weiler}, {Weingrill}, {Wevers},
  {Wyrzykowski}, {Yoldas}, {{\v{Z}}erjal}, {Zucker}, {Zurbach}, {Zwitter},
  {Alecu}, {Allen}, {Allende Prieto}, {Amorim}, {Anglada-Escud{\'e}},
  {Arsenijevic}, {Azaz}, {Balm}, {Beck}, {Bernstein}, {Bigot}, {Bijaoui},
  {Blasco}, {Bonfigli}, {Bono}, {Boudreault}, {Bressan}, {Brown}, {Brunet},
  {Bunclark}, {Buonanno}, {Butkevich}, {Carret}, {Carrion}, {Chemin},
  {Ch{\'e}reau}, {Corcione}, {Darmigny}, {de Boer}, {de Teodoro}, {de Zeeuw},
  {Delle Luche}, {Domingues}, {Dubath}, {Fodor}, {Fr{\'e}zouls}, {Fries},
  {Fustes}, {Fyfe}, {Gallardo}, {Gallegos}, {Gardiol}, {Gebran}, {Gomboc},
  {G{\'o}mez}, {Grux}, {Gueguen}, {Heyrovsky}, {Hoar}, {Iannicola}, {Isasi
  Parache}, {Janotto}, {Joliet}, {Jonckheere}, {Keil}, {Kim}, {Klagyivik},
  {Klar}, {Knude}, {Kochukhov}, {Kolka}, {Kos}, {Kutka}, {Lainey}, {LeBouquin},
  {Liu}, {Loreggia}, {Makarov}, {Marseille}, {Martayan}, {Martinez-Rubi},
  {Massart}, {Meynadier}, {Mignot}, {Munari}, {Nguyen}, {Nordlander}, {Ocvirk},
  {O'Flaherty}, {Olias Sanz}, {Ortiz}, {Osorio}, {Oszkiewicz}, {Ouzounis},
  {Palmer}, {Park}, {Pasquato}, {Peltzer}, {Peralta}, {P{\'e}turaud},
  {Pieniluoma}, {Pigozzi}, {Poels}, {Prat}, {Prod'homme}, {Raison}, {Rebordao},
  {Risquez}, {Rocca-Volmerange}, {Rosen}, {Ruiz-Fuertes}, {Russo}, {Sembay},
  {Serraller Vizcaino}, {Short}, {Siebert}, {Silva}, {Sinachopoulos}, {Slezak},
  {Soffel}, {Sosnowska}, {Strai{\v{z}}ys}, {ter Linden}, {Terrell}, {Theil},
  {Tiede}, {Troisi}, {Tsalmantza}, {Tur}, {Vaccari}, {Vachier}, {Valles}, {Van
  Hamme}, {Veltz}, {Virtanen}, {Wallut}, {Wichmann}, {Wilkinson}, {Ziaeepour},
  \& {Zschocke}}]{gaia16}
{Gaia Collaboration}, {Brown}, A.~G.~A., {Vallenari}, A., {et~al.}
  2016{\natexlab{b}}, \aap, 595, A2, \dodoi{10.1051/0004-6361/201629512}

\bibitem[{{Gaia Collaboration} {et~al.}(2018){Gaia Collaboration}, {Brown},
  {Vallenari}, {Prusti}, {de Bruijne}, {Babusiaux}, {Bailer-Jones}, {Biermann},
  {Evans}, {Eyer}, {Jansen}, {Jordi}, {Klioner}, {Lammers}, {Lindegren},
  {Luri}, {Mignard}, {Panem}, {Pourbaix}, {Randich}, {Sartoretti}, {Siddiqui},
  {Soubiran}, {van Leeuwen}, {Walton}, {Arenou}, {Bastian}, {Cropper},
  {Drimmel}, {Katz}, {Lattanzi}, {Bakker}, {Cacciari}, {Casta{\~n}eda},
  {Chaoul}, {Cheek}, {De Angeli}, {Fabricius}, {Guerra}, {Holl}, {Masana},
  {Messineo}, {Mowlavi}, {Nienartowicz}, {Panuzzo}, {Portell}, {Riello},
  {Seabroke}, {Tanga}, {Th{\'e}venin}, {Gracia-Abril}, {Comoretto},
  {Garcia-Reinaldos}, {Teyssier}, {Altmann}, {Andrae}, {Audard},
  {Bellas-Velidis}, {Benson}, {Berthier}, {Blomme}, {Burgess}, {Busso},
  {Carry}, {Cellino}, {Clementini}, {Clotet}, {Creevey}, {Davidson}, {De
  Ridder}, {Delchambre}, {Dell'Oro}, {Ducourant},
  {Fern{\'a}ndez-Hern{\'a}ndez}, {Fouesneau}, {Fr{\'e}mat}, {Galluccio},
  {Garc{\'\i}a-Torres}, {Gonz{\'a}lez-N{\'u}{\~n}ez}, {Gonz{\'a}lez-Vidal},
  {Gosset}, {Guy}, {Halbwachs}, {Hambly}, {Harrison}, {Hern{\'a}ndez},
  {Hestroffer}, {Hodgkin}, {Hutton}, {Jasniewicz}, {Jean-Antoine-Piccolo},
  {Jordan}, {Korn}, {Krone-Martins}, {Lanzafame}, {Lebzelter}, {L{\"o}ffler},
  {Manteiga}, {Marrese}, {Mart{\'\i}n-Fleitas}, {Moitinho}, {Mora}, {Muinonen},
  {Osinde}, {Pancino}, {Pauwels}, {Petit}, {Recio-Blanco}, {Richards},
  {Rimoldini}, {Robin}, {Sarro}, {Siopis}, {Smith}, {Sozzetti}, {S{\"u}veges},
  {Torra}, {van Reeven}, {Abbas}, {Abreu Aramburu}, {Accart}, {Aerts},
  {Altavilla}, {{\'A}lvarez}, {Alvarez}, {Alves}, {Anderson}, {Andrei},
  {Anglada Varela}, {Antiche}, {Antoja}, {Arcay}, {Astraatmadja}, {Bach},
  {Baker}, {Balaguer-N{\'u}{\~n}ez}, {Balm}, {Barache}, {Barata}, {Barbato},
  {Barblan}, {Barklem}, {Barrado}, {Barros}, {Barstow}, {Bartholom{\'e}
  Mu{\~n}oz}, {Bassilana}, {Becciani}, {Bellazzini}, {Berihuete}, {Bertone},
  {Bianchi}, {Bienaym{\'e}}, {Blanco-Cuaresma}, {Boch}, {Boeche}, {Bombrun},
  {Borrachero}, {Bossini}, {Bouquillon}, {Bourda}, {Bragaglia}, {Bramante},
  {Breddels}, {Bressan}, {Brouillet}, {Br{\"u}semeister}, {Brugaletta},
  {Bucciarelli}, {Burlacu}, {Busonero}, {Butkevich}, {Buzzi}, {Caffau},
  {Cancelliere}, {Cannizzaro}, {Cantat-Gaudin}, {Carballo}, {Carlucci},
  {Carrasco}, {Casamiquela}, {Castellani}, {Castro-Ginard}, {Charlot},
  {Chemin}, {Chiavassa}, {Cocozza}, {Costigan}, {Cowell}, {Crifo}, {Crosta},
  {Crowley}, {Cuypers}, {Dafonte}, {Damerdji}, {Dapergolas}, {David}, {David},
  {de Laverny}, {De Luise}, {De March}, {de Martino}, {de Souza}, {de Torres},
  {Debosscher}, {del Pozo}, {Delbo}, {Delgado}, {Delgado}, {Di Matteo},
  {Diakite}, {Diener}, {Distefano}, {Dolding}, {Drazinos}, {Dur{\'a}n},
  {Edvardsson}, {Enke}, {Eriksson}, {Esquej}, {Eynard Bontemps}, {Fabre},
  {Fabrizio}, {Faigler}, {Falc{\~a}o}, {Farr{\`a}s Casas}, {Federici},
  {Fedorets}, {Fernique}, {Figueras}, {Filippi}, {Findeisen}, {Fonti},
  {Fraile}, {Fraser}, {Fr{\'e}zouls}, {Gai}, {Galleti}, {Garabato},
  {Garc{\'\i}a-Sedano}, {Garofalo}, {Garralda}, {Gavel}, {Gavras}, {Gerssen},
  {Geyer}, {Giacobbe}, {Gilmore}, {Girona}, {Giuffrida}, {Glass}, {Gomes},
  {Granvik}, {Gueguen}, {Guerrier}, {Guiraud}, {Guti{\'e}rrez-S{\'a}nchez},
  {Haigron}, {Hatzidimitriou}, {Hauser}, {Haywood}, {Heiter}, {Helmi}, {Heu},
  {Hilger}, {Hobbs}, {Hofmann}, {Holland}, {Huckle}, {Hypki}, {Icardi},
  {Jan{\ss}en}, {Jevardat de Fombelle}, {Jonker}, {Juh{\'a}sz}, {Julbe},
  {Karampelas}, {Kewley}, {Klar}, {Kochoska}, {Kohley}, {Kolenberg},
  {Kontizas}, {Kontizas}, {Koposov}, {Kordopatis}, {Kostrzewa-Rutkowska},
  {Koubsky}, {Lambert}, {Lanza}, {Lasne}, {Lavigne}, {Le Fustec}, {Le
  Poncin-Lafitte}, {Lebreton}, {Leccia}, {Leclerc}, {Lecoeur-Taibi},
  {Lenhardt}, {Leroux}, {Liao}, {Licata}, {Lindstr{\o}m}, {Lister}, {Livanou},
  {Lobel}, {L{\'o}pez}, {Managau}, {Mann}, {Mantelet}, {Marchal}, {Marchant},
  {Marconi}, {Marinoni}, {Marschalk{\'o}}, {Marshall}, {Martino}, {Marton},
  {Mary}, {Massari}, {Matijevi{\v{c}}}, {Mazeh}, {McMillan}, {Messina},
  {Michalik}, {Millar}, {Molina}, {Molinaro}, {Moln{\'a}r}, {Montegriffo},
  {Mor}, {Morbidelli}, {Morel}, {Morris}, {Mulone}, {Muraveva}, {Musella},
  {Nelemans}, {Nicastro}, {Noval}, {O'Mullane}, {Ord{\'e}novic},
  {Ord{\'o}{\~n}ez-Blanco}, {Osborne}, {Pagani}, {Pagano}, {Pailler},
  {Palacin}, {Palaversa}, {Panahi}, {Pawlak}, {Piersimoni}, {Pineau}, {Plachy},
  {Plum}, {Poggio}, {Poujoulet}, {Pr{\v{s}}a}, {Pulone}, {Racero}, {Ragaini},
  {Rambaux}, {Ramos-Lerate}, {Regibo}, {Reyl{\'e}}, {Riclet}, {Ripepi}, {Riva},
  {Rivard}, {Rixon}, {Roegiers}, {Roelens}, {Romero-G{\'o}mez}, {Rowell},
  {Royer}, {Ruiz-Dern}, {Sadowski}, {Sagrist{\`a} Sell{\'e}s}, {Sahlmann},
  {Salgado}, {Salguero}, {Sanna}, {Santana-Ros}, {Sarasso}, {Savietto},
  {Schultheis}, {Sciacca}, {Segol}, {Segovia}, {S{\'e}gransan}, {Shih},
  {Siltala}, {Silva}, {Smart}, {Smith}, {Solano}, {Solitro}, {Sordo}, {Soria
  Nieto}, {Souchay}, {Spagna}, {Spoto}, {Stampa}, {Steele},
  {Steidelm{\"u}ller}, {Stephenson}, {Stoev}, {Suess}, {Surdej}, {Szabados},
  {Szegedi-Elek}, {Tapiador}, {Taris}, {Tauran}, {Taylor}, {Teixeira},
  {Terrett}, {Teyssandier}, {Thuillot}, {Titarenko}, {Torra Clotet}, {Turon},
  {Ulla}, {Utrilla}, {Uzzi}, {Vaillant}, {Valentini}, {Valette}, {van Elteren},
  {Van Hemelryck}, {van Leeuwen}, {Vaschetto}, {Vecchiato}, {Veljanoski},
  {Viala}, {Vicente}, {Vogt}, {von Essen}, {Voss}, {Votruba}, {Voutsinas},
  {Walmsley}, {Weiler}, {Wertz}, {Wevers}, {Wyrzykowski}, {Yoldas},
  {{\v{Z}}erjal}, {Ziaeepour}, {Zorec}, {Zschocke}, {Zucker}, {Zurbach}, \&
  {Zwitter}}]{gaia18}
---. 2018, \aap, 616, A1, \dodoi{10.1051/0004-6361/201833051}

\bibitem[{{Gaia Collaboration} {et~al.}(2021){Gaia Collaboration}, {Brown},
  {Vallenari}, {Prusti}, {de Bruijne}, {Babusiaux}, {Biermann}, {Creevey},
  {Evans}, {Eyer}, {Hutton}, {Jansen}, {Jordi}, {Klioner}, {Lammers},
  {Lindegren}, {Luri}, {Mignard}, {Panem}, {Pourbaix}, {Randich}, {Sartoretti},
  {Soubiran}, {Walton}, {Arenou}, {Bailer-Jones}, {Bastian}, {Cropper},
  {Drimmel}, {Katz}, {Lattanzi}, {van Leeuwen}, {Bakker}, {Cacciari},
  {Casta{\~n}eda}, {De Angeli}, {Ducourant}, {Fabricius}, {Fouesneau},
  {Fr{\'e}mat}, {Guerra}, {Guerrier}, {Guiraud}, {Jean-Antoine Piccolo},
  {Masana}, {Messineo}, {Mowlavi}, {Nicolas}, {Nienartowicz}, {Pailler},
  {Panuzzo}, {Riclet}, {Roux}, {Seabroke}, {Sordo}, {Tanga}, {Th{\'e}venin},
  {Gracia-Abril}, {Portell}, {Teyssier}, {Altmann}, {Andrae}, {Bellas-Velidis},
  {Benson}, {Berthier}, {Blomme}, {Brugaletta}, {Burgess}, {Busso}, {Carry},
  {Cellino}, {Cheek}, {Clementini}, {Damerdji}, {Davidson}, {Delchambre},
  {Dell'Oro}, {Fern{\'a}ndez-Hern{\'a}ndez}, {Galluccio}, {Garc{\'\i}a-Lario},
  {Garcia-Reinaldos}, {Gonz{\'a}lez-N{\'u}{\~n}ez}, {Gosset}, {Haigron},
  {Halbwachs}, {Hambly}, {Harrison}, {Hatzidimitriou}, {Heiter},
  {Hern{\'a}ndez}, {Hestroffer}, {Hodgkin}, {Holl}, {Jan{\ss}en}, {Jevardat de
  Fombelle}, {Jordan}, {Krone-Martins}, {Lanzafame}, {L{\"o}ffler}, {Lorca},
  {Manteiga}, {Marchal}, {Marrese}, {Moitinho}, {Mora}, {Muinonen}, {Osborne},
  {Pancino}, {Pauwels}, {Petit}, {Recio-Blanco}, {Richards}, {Riello},
  {Rimoldini}, {Robin}, {Roegiers}, {Rybizki}, {Sarro}, {Siopis}, {Smith},
  {Sozzetti}, {Ulla}, {Utrilla}, {van Leeuwen}, {van Reeven}, {Abbas}, {Abreu
  Aramburu}, {Accart}, {Aerts}, {Aguado}, {Ajaj}, {Altavilla}, {{\'A}lvarez},
  {{\'A}lvarez Cid-Fuentes}, {Alves}, {Anderson}, {Anglada Varela}, {Antoja},
  {Audard}, {Baines}, {Baker}, {Balaguer-N{\'u}{\~n}ez}, {Balbinot}, {Balog},
  {Barache}, {Barbato}, {Barros}, {Barstow}, {Bartolom{\'e}}, {Bassilana},
  {Bauchet}, {Baudesson-Stella}, {Becciani}, {Bellazzini}, {Bernet}, {Bertone},
  {Bianchi}, {Blanco-Cuaresma}, {Boch}, {Bombrun}, {Bossini}, {Bouquillon},
  {Bragaglia}, {Bramante}, {Breedt}, {Bressan}, {Brouillet}, {Bucciarelli},
  {Burlacu}, {Busonero}, {Butkevich}, {Buzzi}, {Caffau}, {Cancelliere},
  {C{\'a}novas}, {Cantat-Gaudin}, {Carballo}, {Carlucci}, {Carnerero},
  {Carrasco}, {Casamiquela}, {Castellani}, {Castro-Ginard}, {Castro Sampol},
  {Chaoul}, {Charlot}, {Chemin}, {Chiavassa}, {Cioni}, {Comoretto}, {Cooper},
  {Cornez}, {Cowell}, {Crifo}, {Crosta}, {Crowley}, {Dafonte}, {Dapergolas},
  {David}, {David}, {de Laverny}, {De Luise}, {De March}, {De Ridder}, {de
  Souza}, {de Teodoro}, {de Torres}, {del Peloso}, {del Pozo}, {Delbo},
  {Delgado}, {Delgado}, {Delisle}, {Di Matteo}, {Diakite}, {Diener},
  {Distefano}, {Dolding}, {Eappachen}, {Edvardsson}, {Enke}, {Esquej}, {Fabre},
  {Fabrizio}, {Faigler}, {Fedorets}, {Fernique}, {Fienga}, {Figueras},
  {Fouron}, {Fragkoudi}, {Fraile}, {Franke}, {Gai}, {Garabato},
  {Garcia-Gutierrez}, {Garc{\'\i}a-Torres}, {Garofalo}, {Gavras}, {Gerlach},
  {Geyer}, {Giacobbe}, {Gilmore}, {Girona}, {Giuffrida}, {Gomel}, {Gomez},
  {Gonzalez-Santamaria}, {Gonz{\'a}lez-Vidal}, {Granvik},
  {Guti{\'e}rrez-S{\'a}nchez}, {Guy}, {Hauser}, {Haywood}, {Helmi}, {Hidalgo},
  {Hilger}, {H{\l}adczuk}, {Hobbs}, {Holland}, {Huckle}, {Jasniewicz},
  {Jonker}, {Juaristi Campillo}, {Julbe}, {Karbevska}, {Kervella}, {Khanna},
  {Kochoska}, {Kontizas}, {Kordopatis}, {Korn}, {Kostrzewa-Rutkowska},
  {Kruszy{\'n}ska}, {Lambert}, {Lanza}, {Lasne}, {Le Campion}, {Le Fustec},
  {Lebreton}, {Lebzelter}, {Leccia}, {Leclerc}, {Lecoeur-Taibi}, {Liao},
  {Licata}, {Lindstr{\o}m}, {Lister}, {Livanou}, {Lobel}, {Madrero Pardo},
  {Managau}, {Mann}, {Marchant}, {Marconi}, {Marcos Santos}, {Marinoni},
  {Marocco}, {Marshall}, {Martin Polo}, {Mart{\'\i}n-Fleitas}, {Masip},
  {Massari}, {Mastrobuono-Battisti}, {Mazeh}, {McMillan}, {Messina},
  {Michalik}, {Millar}, {Mints}, {Molina}, {Molinaro}, {Moln{\'a}r},
  {Montegriffo}, {Mor}, {Morbidelli}, {Morel}, {Morris}, {Mulone}, {Munoz},
  {Muraveva}, {Murphy}, {Musella}, {Noval}, {Ord{\'e}novic}, {Orr{\`u}},
  {Osinde}, {Pagani}, {Pagano}, {Palaversa}, {Palicio}, {Panahi}, {Pawlak},
  {Pe{\~n}alosa Esteller}, {Penttil{\"a}}, {Piersimoni}, {Pineau}, {Plachy},
  {Plum}, {Poggio}, {Poretti}, {Poujoulet}, {Pr{\v{s}}a}, {Pulone}, {Racero},
  {Ragaini}, {Rainer}, {Raiteri}, {Rambaux}, {Ramos}, {Ramos-Lerate}, {Re
  Fiorentin}, {Regibo}, {Reyl{\'e}}, {Ripepi}, {Riva}, {Rixon}, {Robichon},
  {Robin}, {Roelens}, {Rohrbasser}, {Romero-G{\'o}mez}, {Rowell}, {Royer},
  {Rybicki}, {Sadowski}, {Sagrist{\`a} Sell{\'e}s}, {Sahlmann}, {Salgado},
  {Salguero}, {Samaras}, {Sanchez Gimenez}, {Sanna}, {Santove{\~n}a},
  {Sarasso}, {Schultheis}, {Sciacca}, {Segol}, {Segovia}, {S{\'e}gransan},
  {Semeux}, {Shahaf}, {Siddiqui}, {Siebert}, {Siltala}, {Slezak}, {Smart},
  {Solano}, {Solitro}, {Souami}, {Souchay}, {Spagna}, {Spoto}, {Steele},
  {Steidelm{\"u}ller}, {Stephenson}, {S{\"u}veges}, {Szabados}, {Szegedi-Elek},
  {Taris}, {Tauran}, {Taylor}, {Teixeira}, {Thuillot}, {Tonello}, {Torra},
  {Torra}, {Turon}, {Unger}, {Vaillant}, {van Dillen}, {Vanel}, {Vecchiato},
  {Viala}, {Vicente}, {Voutsinas}, {Weiler}, {Wevers}, {Wyrzykowski}, {Yoldas},
  {Yvard}, {Zhao}, {Zorec}, {Zucker}, {Zurbach}, \& {Zwitter}}]{gaia21}
---. 2021, \aap, 649, A1, \dodoi{10.1051/0004-6361/202039657}

\bibitem[{{Gaia Collaboration} {et~al.}(2023{\natexlab{a}}){Gaia
  Collaboration}, {Vallenari}, {Brown}, {Prusti}, {de Bruijne}, {Arenou},
  {Babusiaux}, {Biermann}, {Creevey}, {Ducourant}, {Evans}, {Eyer}, {Guerra},
  {Hutton}, {Jordi}, {Klioner}, {Lammers}, {Lindegren}, {Luri}, {Mignard},
  {Panem}, {Pourbaix}, {Randich}, {Sartoretti}, {Soubiran}, {Tanga}, {Walton},
  {Bailer-Jones}, {Bastian}, {Drimmel}, {Jansen}, {Katz}, {Lattanzi}, {van
  Leeuwen}, {Bakker}, {Cacciari}, {Casta{\~n}eda}, {De Angeli}, {Fabricius},
  {Fouesneau}, {Fr{\'e}mat}, {Galluccio}, {Guerrier}, {Heiter}, {Masana},
  {Messineo}, {Mowlavi}, {Nicolas}, {Nienartowicz}, {Pailler}, {Panuzzo},
  {Riclet}, {Roux}, {Seabroke}, {Sordo}, {Th{\'e}venin}, {Gracia-Abril},
  {Portell}, {Teyssier}, {Altmann}, {Andrae}, {Audard}, {Bellas-Velidis},
  {Benson}, {Berthier}, {Blomme}, {Burgess}, {Busonero}, {Busso},
  {C{\'a}novas}, {Carry}, {Cellino}, {Cheek}, {Clementini}, {Damerdji},
  {Davidson}, {de Teodoro}, {Nu{\~n}ez Campos}, {Delchambre}, {Dell'Oro},
  {Esquej}, {Fern{\'a}ndez-Hern{\'a}ndez}, {Fraile}, {Garabato},
  {Garc{\'\i}a-Lario}, {Gosset}, {Haigron}, {Halbwachs}, {Hambly}, {Harrison},
  {Hern{\'a}ndez}, {Hestroffer}, {Hodgkin}, {Holl}, {Jan{\ss}en}, {Jevardat de
  Fombelle}, {Jordan}, {Krone-Martins}, {Lanzafame}, {L{\"o}ffler}, {Marchal},
  {Marrese}, {Moitinho}, {Muinonen}, {Osborne}, {Pancino}, {Pauwels},
  {Recio-Blanco}, {Reyl{\'e}}, {Riello}, {Rimoldini}, {Roegiers}, {Rybizki},
  {Sarro}, {Siopis}, {Smith}, {Sozzetti}, {Utrilla}, {van Leeuwen}, {Abbas},
  {{\'A}brah{\'a}m}, {Abreu Aramburu}, {Aerts}, {Aguado}, {Ajaj},
  {Aldea-Montero}, {Altavilla}, {{\'A}lvarez}, {Alves}, {Anders}, {Anderson},
  {Anglada Varela}, {Antoja}, {Baines}, {Baker}, {Balaguer-N{\'u}{\~n}ez},
  {Balbinot}, {Balog}, {Barache}, {Barbato}, {Barros}, {Barstow},
  {Bartolom{\'e}}, {Bassilana}, {Bauchet}, {Becciani}, {Bellazzini},
  {Berihuete}, {Bernet}, {Bertone}, {Bianchi}, {Binnenfeld}, {Blanco-Cuaresma},
  {Blazere}, {Boch}, {Bombrun}, {Bossini}, {Bouquillon}, {Bragaglia},
  {Bramante}, {Breedt}, {Bressan}, {Brouillet}, {Brugaletta}, {Bucciarelli},
  {Burlacu}, {Butkevich}, {Buzzi}, {Caffau}, {Cancelliere}, {Cantat-Gaudin},
  {Carballo}, {Carlucci}, {Carnerero}, {Carrasco}, {Casamiquela}, {Castellani},
  {Castro-Ginard}, {Chaoul}, {Charlot}, {Chemin}, {Chiaramida}, {Chiavassa},
  {Chornay}, {Comoretto}, {Contursi}, {Cooper}, {Cornez}, {Cowell}, {Crifo},
  {Cropper}, {Crosta}, {Crowley}, {Dafonte}, {Dapergolas}, {David}, {David},
  {de Laverny}, {De Luise}, {De March}, {De Ridder}, {de Souza}, {de Torres},
  {del Peloso}, {del Pozo}, {Delbo}, {Delgado}, {Delisle}, {Demouchy},
  {Dharmawardena}, {Di Matteo}, {Diakite}, {Diener}, {Distefano}, {Dolding},
  {Edvardsson}, {Enke}, {Fabre}, {Fabrizio}, {Faigler}, {Fedorets}, {Fernique},
  {Fienga}, {Figueras}, {Fournier}, {Fouron}, {Fragkoudi}, {Gai},
  {Garcia-Gutierrez}, {Garcia-Reinaldos}, {Garc{\'\i}a-Torres}, {Garofalo},
  {Gavel}, {Gavras}, {Gerlach}, {Geyer}, {Giacobbe}, {Gilmore}, {Girona},
  {Giuffrida}, {Gomel}, {Gomez}, {Gonz{\'a}lez-N{\'u}{\~n}ez},
  {Gonz{\'a}lez-Santamar{\'\i}a}, {Gonz{\'a}lez-Vidal}, {Granvik}, {Guillout},
  {Guiraud}, {Guti{\'e}rrez-S{\'a}nchez}, {Guy}, {Hatzidimitriou}, {Hauser},
  {Haywood}, {Helmer}, {Helmi}, {Sarmiento}, {Hidalgo}, {Hilger},
  {H{\l}adczuk}, {Hobbs}, {Holland}, {Huckle}, {Jardine}, {Jasniewicz},
  {Jean-Antoine Piccolo}, {Jim{\'e}nez-Arranz}, {Jorissen}, {Juaristi
  Campillo}, {Julbe}, {Karbevska}, {Kervella}, {Khanna}, {Kontizas},
  {Kordopatis}, {Korn}, {K{\'o}sp{\'a}l}, {Kostrzewa-Rutkowska},
  {Kruszy{\'n}ska}, {Kun}, {Laizeau}, {Lambert}, {Lanza}, {Lasne}, {Le
  Campion}, {Lebreton}, {Lebzelter}, {Leccia}, {Leclerc}, {Lecoeur-Taibi},
  {Liao}, {Licata}, {Lindstr{\o}m}, {Lister}, {Livanou}, {Lobel}, {Lorca},
  {Loup}, {Madrero Pardo}, {Magdaleno Romeo}, {Managau}, {Mann}, {Manteiga},
  {Marchant}, {Marconi}, {Marcos}, {Marcos Santos}, {Mar{\'\i}n Pina},
  {Marinoni}, {Marocco}, {Marshall}, {Martin Polo}, {Mart{\'\i}n-Fleitas},
  {Marton}, {Mary}, {Masip}, {Massari}, {Mastrobuono-Battisti}, {Mazeh},
  {McMillan}, {Messina}, {Michalik}, {Millar}, {Mints}, {Molina}, {Molinaro},
  {Moln{\'a}r}, {Monari}, {Mongui{\'o}}, {Montegriffo}, {Montero}, {Mor},
  {Mora}, {Morbidelli}, {Morel}, {Morris}, {Muraveva}, {Murphy}, {Musella},
  {Nagy}, {Noval}, {Oca{\~n}a}, {Ogden}, {Ordenovic}, {Osinde}, {Pagani},
  {Pagano}, {Palaversa}, {Palicio}, {Pallas-Quintela}, {Panahi},
  {Payne-Wardenaar}, {Pe{\~n}alosa Esteller}, {Penttil{\"a}}, {Pichon},
  {Piersimoni}, {Pineau}, {Plachy}, {Plum}, {Poggio}, {Pr{\v{s}}a}, {Pulone},
  {Racero}, {Ragaini}, {Rainer}, {Raiteri}, {Rambaux}, {Ramos}, {Ramos-Lerate},
  {Re Fiorentin}, {Regibo}, {Richards}, {Rios Diaz}, {Ripepi}, {Riva}, {Rix},
  {Rixon}, {Robichon}, {Robin}, {Robin}, {Roelens}, {Rogues}, {Rohrbasser},
  {Romero-G{\'o}mez}, {Rowell}, {Royer}, {Ruz Mieres}, {Rybicki}, {Sadowski},
  {S{\'a}ez N{\'u}{\~n}ez}, {Sagrist{\`a} Sell{\'e}s}, {Sahlmann}, {Salguero},
  {Samaras}, {Sanchez Gimenez}, {Sanna}, {Santove{\~n}a}, {Sarasso},
  {Schultheis}, {Sciacca}, {Segol}, {Segovia}, {S{\'e}gransan}, {Semeux},
  {Shahaf}, {Siddiqui}, {Siebert}, {Siltala}, {Silvelo}, {Slezak}, {Slezak},
  {Smart}, {Snaith}, {Solano}, {Solitro}, {Souami}, {Souchay}, {Spagna},
  {Spina}, {Spoto}, {Steele}, {Steidelm{\"u}ller}, {Stephenson}, {S{\"u}veges},
  {Surdej}, {Szabados}, {Szegedi-Elek}, {Taris}, {Taylor}, {Teixeira},
  {Tolomei}, {Tonello}, {Torra}, {Torra}, {Torralba Elipe}, {Trabucchi},
  {Tsounis}, {Turon}, {Ulla}, {Unger}, {Vaillant}, {van Dillen}, {van Reeven},
  {Vanel}, {Vecchiato}, {Viala}, {Vicente}, {Voutsinas}, {Weiler}, {Wevers},
  {Wyrzykowski}, {Yoldas}, {Yvard}, {Zhao}, {Zorec}, {Zucker}, \&
  {Zwitter}}]{gaia23}
{Gaia Collaboration}, {Vallenari}, A., {Brown}, A.~G.~A., {et~al.}
  2023{\natexlab{a}}, \aap, 674, A1, \dodoi{10.1051/0004-6361/202243940}

\bibitem[{{Gaia Collaboration} {et~al.}(2023{\natexlab{b}}){Gaia
  Collaboration}, {Arenou}, {Babusiaux}, {Barstow}, {Faigler}, {Jorissen},
  {Kervella}, {Mazeh}, {Mowlavi}, {Panuzzo}, {Sahlmann}, {Shahaf}, {Sozzetti},
  {Bauchet}, {Damerdji}, {Gavras}, {Giacobbe}, {Gosset}, {Halbwachs}, {Holl},
  {Lattanzi}, {Leclerc}, {Morel}, {Pourbaix}, {Re Fiorentin}, {Sadowski},
  {S{\'e}gransan}, {Siopis}, {Teyssier}, {Zwitter}, {Planquart}, {Brown},
  {Vallenari}, {Prusti}, {de Bruijne}, {Biermann}, {Creevey}, {Ducourant},
  {Evans}, {Eyer}, {Guerra}, {Hutton}, {Jordi}, {Klioner}, {Lammers},
  {Lindegren}, {Luri}, {Mignard}, {Panem}, {Randich}, {Sartoretti}, {Soubiran},
  {Tanga}, {Walton}, {Bailer-Jones}, {Bastian}, {Drimmel}, {Jansen}, {Katz},
  {van Leeuwen}, {Bakker}, {Cacciari}, {Casta{\~n}eda}, {De Angeli},
  {Fabricius}, {Fouesneau}, {Fr{\'e}mat}, {Galluccio}, {Guerrier}, {Heiter},
  {Masana}, {Messineo}, {Nicolas}, {Nienartowicz}, {Pailler}, {Riclet}, {Roux},
  {Seabroke}, {Sordo}, {Th{\'e}venin}, {Gracia-Abril}, {Portell}, {Altmann},
  {Andrae}, {Audard}, {Bellas-Velidis}, {Benson}, {Berthier}, {Blomme},
  {Burgess}, {Busonero}, {Busso}, {C{\'a}novas}, {Carry}, {Cellino}, {Cheek},
  {Clementini}, {Davidson}, {de Teodoro}, {Nu{\~n}ez Campos}, {Delchambre},
  {Dell'Oro}, {Esquej}, {Fern{\'a}ndez-Hern{\'a}ndez}, {Fraile}, {Garabato},
  {Garc{\'\i}a-Lario}, {Haigron}, {Hambly}, {Harrison}, {Hern{\'a}ndez},
  {Hestroffer}, {Hodgkin}, {Jan{\ss}en}, {Jevardat de Fombelle}, {Jordan},
  {Krone-Martins}, {Lanzafame}, {L{\"o}ffler}, {Marchal}, {Marrese},
  {Moitinho}, {Muinonen}, {Osborne}, {Pancino}, {Pauwels}, {Recio-Blanco},
  {Reyl{\'e}}, {Riello}, {Rimoldini}, {Roegiers}, {Rybizki}, {Sarro}, {Smith},
  {Utrilla}, {van Leeuwen}, {Abbas}, {{\'A}brah{\'a}m}, {Abreu Aramburu},
  {Aerts}, {Aguado}, {Ajaj}, {Aldea-Montero}, {Altavilla}, {{\'A}lvarez},
  {Alves}, {Anders}, {Anderson}, {Anglada Varela}, {Antoja}, {Baines}, {Baker},
  {Balaguer-N{\'u}{\~n}ez}, {Balbinot}, {Balog}, {Barache}, {Barbato},
  {Barros}, {Bartolom{\'e}}, {Bassilana}, {Becciani}, {Bellazzini},
  {Berihuete}, {Bernet}, {Bertone}, {Bianchi}, {Binnenfeld}, {Blanco-Cuaresma},
  {Blazere}, {Boch}, {Bombrun}, {Bossini}, {Bouquillon}, {Bragaglia},
  {Bramante}, {Breedt}, {Bressan}, {Brouillet}, {Brugaletta}, {Bucciarelli},
  {Burlacu}, {Butkevich}, {Buzzi}, {Caffau}, {Cancelliere}, {Cantat-Gaudin},
  {Carballo}, {Carlucci}, {Carnerero}, {Carrasco}, {Casamiquela}, {Castellani},
  {Castro-Ginard}, {Chaoul}, {Charlot}, {Chemin}, {Chiaramida}, {Chiavassa},
  {Chornay}, {Comoretto}, {Contursi}, {Cooper}, {Cornez}, {Cowell}, {Crifo},
  {Cropper}, {Crosta}, {Crowley}, {Dafonte}, {Dapergolas}, {David}, {de
  Laverny}, {De Luise}, {De March}, {De Ridder}, {de Souza}, {de Torres}, {del
  Peloso}, {del Pozo}, {Delbo}, {Delgado}, {Delisle}, {Demouchy},
  {Dharmawardena}, {Diakite}, {Diener}, {Distefano}, {Dolding}, {Enke},
  {Fabre}, {Fabrizio}, {Fedorets}, {Fernique}, {Figueras}, {Fournier},
  {Fouron}, {Fragkoudi}, {Gai}, {Garcia-Gutierrez}, {Garcia-Reinaldos},
  {Garc{\'\i}a-Torres}, {Garofalo}, {Gavel}, {Gerlach}, {Geyer}, {Gilmore},
  {Girona}, {Giuffrida}, {Gomel}, {Gomez}, {Gonz{\'a}lez-N{\'u}{\~n}ez},
  {Gonz{\'a}lez-Santamar{\'\i}a}, {Gonz{\'a}lez-Vidal}, {Granvik}, {Guillout},
  {Guiraud}, {Guti{\'e}rrez-S{\'a}nchez}, {Guy}, {Hatzidimitriou}, {Hauser},
  {Haywood}, {Helmer}, {Helmi}, {Sarmiento}, {Hidalgo}, {Hilger},
  {H{\l}adczuk}, {Hobbs}, {Holland}, {Huckle}, {Jardine}, {Jasniewicz},
  {Jean-Antoine Piccolo}, {Jim{\'e}nez-Arranz}, {Juaristi Campillo}, {Julbe},
  {Karbevska}, {Khanna}, {Kordopatis}, {Korn}, {K{\'o}sp{\'a}l},
  {Kostrzewa-Rutkowska}, {Kruszy{\'n}ska}, {Kun}, {Laizeau}, {Lambert},
  {Lanza}, {Lasne}, {Le Campion}, {Lebreton}, {Lebzelter}, {Leccia},
  {Lecoeur-Taibi}, {Liao}, {Licata}, {Lindstr{\o}m}, {Lister}, {Livanou},
  {Lobel}, {Lorca}, {Loup}, {Madrero Pardo}, {Magdaleno Romeo}, {Managau},
  {Mann}, {Manteiga}, {Marchant}, {Marconi}, {Marcos}, {Marcos Santos},
  {Mar{\'\i}n Pina}, {Marinoni}, {Marocco}, {Marshall}, {Martin Polo},
  {Mart{\'\i}n-Fleitas}, {Marton}, {Mary}, {Masip}, {Massari},
  {Mastrobuono-Battisti}, {McMillan}, {Messina}, {Michalik}, {Millar}, {Mints},
  {Molina}, {Molinaro}, {Moln{\'a}r}, {Monari}, {Mongui{\'o}}, {Montegriffo},
  {Montero}, {Mor}, {Mora}, {Morbidelli}, {Morris}, {Muraveva}, {Murphy},
  {Musella}, {Nagy}, {Noval}, {Oca{\~n}a}, {Ogden}, {Ordenovic}, {Osinde},
  {Pagani}, {Pagano}, {Palaversa}, {Palicio}, {Pallas-Quintela}, {Panahi},
  {Payne-Wardenaar}, {Pe{\~n}alosa Esteller}, {Penttil{\"a}}, {Pichon},
  {Piersimoni}, {Pineau}, {Plachy}, {Plum}, {Poggio}, {Pr{\v{s}}a}, {Pulone},
  {Racero}, {Ragaini}, {Rainer}, {Raiteri}, {Ramos}, {Ramos-Lerate}, {Regibo},
  {Richards}, {Rios Diaz}, {Ripepi}, {Riva}, {Rix}, {Rixon}, {Robichon},
  {Robin}, {Robin}, {Roelens}, {Rogues}, {Rohrbasser}, {Romero-G{\'o}mez},
  {Rowell}, {Royer}, {Ruz Mieres}, {Rybicki}, {S{\'a}ez N{\'u}{\~n}ez},
  {Sagrist{\`a} Sell{\'e}s}, {Salguero}, {Samaras}, {Sanchez Gimenez}, {Sanna},
  {Santove{\~n}a}, {Sarasso}, {Schultheis}, {Sciacca}, {Segol}, {Segovia},
  {Semeux}, {Siddiqui}, {Siebert}, {Siltala}, {Silvelo}, {Slezak}, {Slezak},
  {Smart}, {Snaith}, {Solano}, {Solitro}, {Souami}, {Souchay}, {Spagna},
  {Spina}, {Spoto}, {Steele}, {Steidelm{\"u}ller}, {Stephenson}, {S{\"u}veges},
  {Surdej}, {Szabados}, {Szegedi-Elek}, {Taris}, {Taylor}, {Teixeira},
  {Tolomei}, {Tonello}, {Torra}, {Torra}, {Torralba Elipe}, {Trabucchi},
  {Tsounis}, {Turon}, {Ulla}, {Unger}, {Vaillant}, {van Dillen}, {van Reeven},
  {Vanel}, {Vecchiato}, {Viala}, {Vicente}, {Voutsinas}, {Weiler}, {Wevers},
  {Wyrzykowski}, {Yoldas}, {Yvard}, {Zhao}, {Zorec}, \& {Zucker}}]{arenou23}
{Gaia Collaboration}, {Arenou}, F., {Babusiaux}, C., {et~al.}
  2023{\natexlab{b}}, \aap, 674, A34, \dodoi{10.1051/0004-6361/202243782}

\bibitem[{{Ganguly} {et~al.}(2023){Ganguly}, {Nayak}, \&
  {Chatterjee}}]{ganguly23}
{Ganguly}, A., {Nayak}, P.~K., \& {Chatterjee}, S. 2023, \apj, 954, 4,
  \dodoi{10.3847/1538-4357/ace42f}

\bibitem[{{Garnier} {et~al.}(2023){Garnier}, {Simon}, {Ross}, {Noam}, {Rudis},
  {Robert}, {Camargo}, Pedro, {Sciaini}, {Marco}, {Scherer}, \&
  {Cédric}}]{viridis}
{Garnier}, {Simon}, {Ross}, {et~al.} 2023, {viridis(Lite)} -
  Colorblind-Friendly Color Maps for R, \dodoi{10.5281/zenodo.4679424}

\bibitem[{{Ge} {et~al.}(2022){Ge}, {Zhang}, {Zang}, {Deng}, {Mao}, {Xie},
  {Liu}, {Zhou}, {Willis}, {Huang}, {Howell}, {Feng}, {Zhu}, {Yao}, {Liu},
  {Aizawa}, {Zhu}, {Li}, {Ma}, {Ye}, {Yu}, {Xiang}, {Yu}, {Liu}, {Yang},
  {Wang}, {Shi}, {Fang}, {Zong}, {Liu}, {Zhang}, {Zhang}, {El-Badry}, {Shen},
  {Tam}, {Hu}, {Yang}, {Zou}, {Wu}, {Lei}, {Wei}, {Wu}, {Sun}, {Wang}, {Zhang},
  {Xu}, {Yang}, {Li}, {Xiang}, {Wang}, {Wang}, {Zhang}, {Jia}, {Yuan}, {Zhang},
  {Xuesong Wang}, {Gan}, {Wang}, {Zhao}, {Liu}, {Wei}, {Kang}, {Yang}, {Qi},
  {Liu}, {Zhang}, {Zhu}, {Zhou}, {Zhang}, {Yu}, {Zhang}, {Li}, {Tang}, {Wang},
  {Wang}, {Li}, {Cheng}, {Shen}, {Li}, {Pan}, {Yang}, {Gao}, {Song}, {Wang},
  {Zhang}, {Chen}, {Wang}, {Zhang}, {Wang}, {Zeng}, {Zheng}, {Zhu}, {Guo},
  {Zhang}, {Li}, {Wen}, {Feng}, {Chen}, {Chen}, {Han}, {Yang}, {Wang}, {Duan},
  {Huang}, {Liang}, {Bi}, {Gai}, {Ge}, {Guo}, {Huang}, {Li}, {Li}, {Li},
  {Yuxi}, {Lu}, {Rix}, {Shi}, {Song}, {Tang}, {Ting}, {Wu}, {Wu}, {Yang},
  {Yin}, {Gould}, {Lee}, {Dong}, {Yee}, {Shvartzvald}, {Yang}, {Kuang},
  {Zhang}, {Liao}, {Qi}, {Yang}, {Zhang}, {Jiang}, {Ou}, {Li}, {Beck},
  {Bedding}, {Campante}, {Chaplin}, {Christensen-Dalsgaard}, {Garc{\'\i}a},
  {Gaulme}, {Gizon}, {Hekker}, {Huber}, {Khanna}, {Li}, {Mathur}, {Miglio},
  {Mosser}, {Ong}, {Santos}, {Stello}, {Bowman}, {Lares-Martiz}, {Murphy},
  {Niu}, {Ma}, {Moln{\'a}r}, {Fu}, {De Cat}, {Su}, \& {consortium}}]{ge22}
{Ge}, J., {Zhang}, H., {Zang}, W., {et~al.} 2022, arXiv e-prints,
  arXiv:2206.06693, \dodoi{10.48550/arXiv.2206.06693}

\bibitem[{{Gomes} {et~al.}(2005){Gomes}, {Levison}, {Tsiganis}, \&
  {Morbidelli}}]{gomes05}
{Gomes}, R., {Levison}, H.~F., {Tsiganis}, K., \& {Morbidelli}, A. 2005, \nat,
  435, 466, \dodoi{10.1038/nature03676}

\bibitem[{{Halbwachs} {et~al.}(2023){Halbwachs}, {Pourbaix}, {Arenou},
  {Galluccio}, {Guillout}, {Bauchet}, {Marchal}, {Sadowski}, \&
  {Teyssier}}]{halbwachs23}
{Halbwachs}, J.-L., {Pourbaix}, D., {Arenou}, F., {et~al.} 2023, \aap, 674, A9,
  \dodoi{10.1051/0004-6361/202243969}

\bibitem[{{Hall} {et~al.}(2018){Hall}, {Thompson}, {Handley}, \&
  {Queloz}}]{hall18}
{Hall}, R.~D., {Thompson}, S.~J., {Handley}, W., \& {Queloz}, D. 2018, \mnras,
  479, 2968, \dodoi{10.1093/mnras/sty1464}

\bibitem[{{Heger} {et~al.}(2003){Heger}, {Fryer}, {Woosley}, {Langer}, \&
  {Hartmann}}]{heger03}
{Heger}, A., {Fryer}, C.~L., {Woosley}, S.~E., {Langer}, N., \& {Hartmann},
  D.~H. 2003, \apj, 591, 288, \dodoi{10.1086/375341}

\bibitem[{{H{\o}g} {et~al.}(2000){H{\o}g}, {Fabricius}, {Makarov}, {Urban},
  {Corbin}, {Wycoff}, {Bastian}, {Schwekendiek}, \& {Wicenec}}]{hog00}
{H{\o}g}, E., {Fabricius}, C., {Makarov}, V.~V., {et~al.} 2000, \aap, 355, L27

\bibitem[{{Holl} {et~al.}(2023){Holl}, {Sozzetti}, {Sahlmann}, {Giacobbe},
  {S{\'e}gransan}, {Unger}, {Delisle}, {Barbato}, {Lattanzi}, {Morbidelli}, \&
  {Sosnowska}}]{holl23}
{Holl}, B., {Sozzetti}, A., {Sahlmann}, J., {et~al.} 2023, \aap, 674, A10,
  \dodoi{10.1051/0004-6361/202244161}

\bibitem[{{Horner} {et~al.}(2010){Horner}, {Jones}, \& {Chambers}}]{horner10}
{Horner}, J., {Jones}, B.~W., \& {Chambers}, J. 2010, International Journal of
  Astrobiology, 9, 1, \dodoi{10.1017/S1473550409990346}

\bibitem[{{Horner} {et~al.}(2020){Horner}, {Vervoort}, {Kane}, {Ceja},
  {Waltham}, {Gilmore}, \& {Kirtland Turner}}]{horner20}
{Horner}, J., {Vervoort}, P., {Kane}, S.~R., {et~al.} 2020, \aj, 159, 10,
  \dodoi{10.3847/1538-3881/ab5365}

\bibitem[{Kass \& Raftery(1995)}]{kass95}
Kass, R.~E., \& Raftery, A.~E. 1995, Journal of the american statistical
  association, 90, 773

\bibitem[{{Kervella} {et~al.}(2019){Kervella}, {Arenou}, {Mignard}, \&
  {Th{\'e}venin}}]{kervella19}
{Kervella}, P., {Arenou}, F., {Mignard}, F., \& {Th{\'e}venin}, F. 2019, \aap,
  623, A72, \dodoi{10.1051/0004-6361/201834371}

\bibitem[{{Kreidberg} {et~al.}(2012){Kreidberg}, {Bailyn}, {Farr}, \&
  {Kalogera}}]{kreidberg12}
{Kreidberg}, L., {Bailyn}, C.~D., {Farr}, W.~M., \& {Kalogera}, V. 2012, \apj,
  757, 36, \dodoi{10.1088/0004-637X/757/1/36}

\bibitem[{{Laliotis} {et~al.}(2023){Laliotis}, {Burt}, {Mamajek}, {Li},
  {Perdelwitz}, {Zhao}, {Butler}, {Holden}, {Rosenthal}, {Fulton}, {Feng},
  {Kane}, {Bailey}, {Carter}, {Crane}, {Furlan}, {Gnilka}, {Howell},
  {Laughlin}, {Shectman}, {Teske}, {Tinney}, {Vogt}, {Wang}, \&
  {Wittenmyer}}]{laliotis23}
{Laliotis}, K., {Burt}, J.~A., {Mamajek}, E.~E., {et~al.} 2023, \aj, 165, 176,
  \dodoi{10.3847/1538-3881/acc067}

\bibitem[{{Lam} {et~al.}(2022){Lam}, {Lu}, {Udalski}, {Bond}, {Bennett},
  {Skowron}, {Mr{\'o}z}, {Poleski}, {Sumi}, {Szyma{\'n}ski}, {Koz{\l}owski},
  {Pietrukowicz}, {Soszy{\'n}ski}, {Ulaczyk}, {Wyrzykowski}, {Miyazaki},
  {Suzuki}, {Koshimoto}, {Rattenbury}, {Hosek}, {Abe}, {Barry}, {Bhattacharya},
  {Fukui}, {Fujii}, {Hirao}, {Itow}, {Kirikawa}, {Kondo}, {Matsubara},
  {Matsumoto}, {Muraki}, {Olmschenk}, {Ranc}, {Okamura}, {Satoh}, {Silva},
  {Toda}, {Tristram}, {Vandorou}, {Yama}, {Abrams}, {Agarwal}, {Rose}, \&
  {Terry}}]{lam22}
{Lam}, C.~Y., {Lu}, J.~R., {Udalski}, A., {et~al.} 2022, \apjl, 933, L23,
  \dodoi{10.3847/2041-8213/ac7442}

\bibitem[{{Leclerc} {et~al.}(2023){Leclerc}, {Babusiaux}, {Arenou}, {van
  Leeuwen}, {Bonnefoy}, {Delfosse}, {Forveille}, {Le Bouquin}, \&
  {Rodet}}]{leclerc23}
{Leclerc}, A., {Babusiaux}, C., {Arenou}, F., {et~al.} 2023, \aap, 672, A82,
  \dodoi{10.1051/0004-6361/202244144}

\bibitem[{Levenberg(1944)}]{levenberg44}
Levenberg, K. 1944, Quarterly of applied mathematics, 2, 164

\bibitem[{{Li} {et~al.}(2021){Li}, {Brandt}, {Brandt}, {Dupuy}, {Michalik},
  {Jensen-Clem}, {Zeng}, {Faherty}, \& {Mitra}}]{li21}
{Li}, Y., {Brandt}, T.~D., {Brandt}, G.~M., {et~al.} 2021, \aj, 162, 266,
  \dodoi{10.3847/1538-3881/ac27ab}

\bibitem[{{Lindegren}(2020)}]{lindegren20}
{Lindegren}, L. 2020, \aap, 633, A1, \dodoi{10.1051/0004-6361/201936161}

\bibitem[{{Lindegren} {et~al.}(2016){Lindegren}, {Lammers}, {Bastian},
  {Hern{\'a}ndez}, {Klioner}, {Hobbs}, {Bombrun}, {Michalik}, {Ramos-Lerate},
  {Butkevich}, {Comoretto}, {Joliet}, {Holl}, {Hutton}, {Parsons},
  {Steidelm{\"u}ller}, {Abbas}, {Altmann}, {Andrei}, {Anton}, {Bach},
  {Barache}, {Becciani}, {Berthier}, {Bianchi}, {Biermann}, {Bouquillon},
  {Bourda}, {Br{\"u}semeister}, {Bucciarelli}, {Busonero}, {Carlucci},
  {Casta{\~n}eda}, {Charlot}, {Clotet}, {Crosta}, {Davidson}, {de Felice},
  {Drimmel}, {Fabricius}, {Fienga}, {Figueras}, {Fraile}, {Gai}, {Garralda},
  {Geyer}, {Gonz{\'a}lez-Vidal}, {Guerra}, {Hambly}, {Hauser}, {Jordan},
  {Lattanzi}, {Lenhardt}, {Liao}, {L{\"o}ffler}, {McMillan}, {Mignard}, {Mora},
  {Morbidelli}, {Portell}, {Riva}, {Sarasso}, {Serraller}, {Siddiqui}, {Smart},
  {Spagna}, {Stampa}, {Steele}, {Taris}, {Torra}, {van Reeven}, {Vecchiato},
  {Zschocke}, {de Bruijne}, {Gracia}, {Raison}, {Lister}, {Marchant},
  {Messineo}, {Soffel}, {Osorio}, {de Torres}, \& {O'Mullane}}]{lindegren16}
{Lindegren}, L., {Lammers}, U., {Bastian}, U., {et~al.} 2016, \aap, 595, A4,
  \dodoi{10.1051/0004-6361/201628714}

\bibitem[{{Lindegren} {et~al.}(2018){Lindegren}, {Hern{\'a}ndez}, {Bombrun},
  {Klioner}, {Bastian}, {Ramos-Lerate}, {de Torres}, {Steidelm{\"u}ller},
  {Stephenson}, {Hobbs}, {Lammers}, {Biermann}, {Geyer}, {Hilger}, {Michalik},
  {Stampa}, {McMillan}, {Casta{\~n}eda}, {Clotet}, {Comoretto}, {Davidson},
  {Fabricius}, {Gracia}, {Hambly}, {Hutton}, {Mora}, {Portell}, {van Leeuwen},
  {Abbas}, {Abreu}, {Altmann}, {Andrei}, {Anglada}, {Balaguer-N{\'u}{\~n}ez},
  {Barache}, {Becciani}, {Bertone}, {Bianchi}, {Bouquillon}, {Bourda},
  {Br{\"u}semeister}, {Bucciarelli}, {Busonero}, {Buzzi}, {Cancelliere},
  {Carlucci}, {Charlot}, {Cheek}, {Crosta}, {Crowley}, {de Bruijne}, {de
  Felice}, {Drimmel}, {Esquej}, {Fienga}, {Fraile}, {Gai}, {Garralda},
  {Gonz{\'a}lez-Vidal}, {Guerra}, {Hauser}, {Hofmann}, {Holl}, {Jordan},
  {Lattanzi}, {Lenhardt}, {Liao}, {Licata}, {Lister}, {L{\"o}ffler},
  {Marchant}, {Martin-Fleitas}, {Messineo}, {Mignard}, {Morbidelli}, {Poggio},
  {Riva}, {Rowell}, {Salguero}, {Sarasso}, {Sciacca}, {Siddiqui}, {Smart},
  {Spagna}, {Steele}, {Taris}, {Torra}, {van Elteren}, {van Reeven}, \&
  {Vecchiato}}]{lindegren18}
{Lindegren}, L., {Hern{\'a}ndez}, J., {Bombrun}, A., {et~al.} 2018, \aap, 616,
  A2, \dodoi{10.1051/0004-6361/201832727}

\bibitem[{{Lindegren} {et~al.}(2021{\natexlab{a}}){Lindegren}, {Bastian},
  {Biermann}, {Bombrun}, {de Torres}, {Gerlach}, {Geyer}, {Hern{\'a}ndez},
  {Hilger}, {Hobbs}, {Klioner}, {Lammers}, {McMillan}, {Ramos-Lerate},
  {Steidelm{\"u}ller}, {Stephenson}, \& {van Leeuwen}}]{lindegren21a}
{Lindegren}, L., {Bastian}, U., {Biermann}, M., {et~al.} 2021{\natexlab{a}},
  \aap, 649, A4, \dodoi{10.1051/0004-6361/202039653}

\bibitem[{{Lindegren} {et~al.}(2021{\natexlab{b}}){Lindegren}, {Klioner},
  {Hern{\'a}ndez}, {Bombrun}, {Ramos-Lerate}, {Steidelm{\"u}ller}, {Bastian},
  {Biermann}, {de Torres}, {Gerlach}, {Geyer}, {Hilger}, {Hobbs}, {Lammers},
  {McMillan}, {Stephenson}, {Casta{\~n}eda}, {Davidson}, {Fabricius},
  {Gracia-Abril}, {Portell}, {Rowell}, {Teyssier}, {Torra}, {Bartolom{\'e}},
  {Clotet}, {Garralda}, {Gonz{\'a}lez-Vidal}, {Torra}, {Abbas}, {Altmann},
  {Anglada Varela}, {Balaguer-N{\'u}{\~n}ez}, {Balog}, {Barache}, {Becciani},
  {Bernet}, {Bertone}, {Bianchi}, {Bouquillon}, {Brown}, {Bucciarelli},
  {Busonero}, {Butkevich}, {Buzzi}, {Cancelliere}, {Carlucci}, {Charlot},
  {Cioni}, {Crosta}, {Crowley}, {del Peloso}, {del Pozo}, {Drimmel}, {Esquej},
  {Fienga}, {Fraile}, {Gai}, {Garcia-Reinaldos}, {Guerra}, {Hambly}, {Hauser},
  {Jan{\ss}en}, {Jordan}, {Kostrzewa-Rutkowska}, {Lattanzi}, {Liao}, {Licata},
  {Lister}, {L{\"o}ffler}, {Marchant}, {Masip}, {Mignard}, {Mints}, {Molina},
  {Mora}, {Morbidelli}, {Murphy}, {Pagani}, {Panuzzo}, {Pe{\~n}alosa Esteller},
  {Poggio}, {Re Fiorentin}, {Riva}, {Sagrist{\`a} Sell{\'e}s}, {Sanchez
  Gimenez}, {Sarasso}, {Sciacca}, {Siddiqui}, {Smart}, {Souami}, {Spagna},
  {Steele}, {Taris}, {Utrilla}, {van Reeven}, \& {Vecchiato}}]{lindegren21b}
{Lindegren}, L., {Klioner}, S.~A., {Hern{\'a}ndez}, J., {et~al.}
  2021{\natexlab{b}}, \aap, 649, A2, \dodoi{10.1051/0004-6361/202039709}

\bibitem[{{Lunine}(2001)}]{lunine01}
{Lunine}, J.~I. 2001, Proceedings of the National Academy of Science, 98, 809,
  \dodoi{10.1073/pnas.98.3.809}

\bibitem[{{Lunz} {et~al.}(2023){Lunz}, {Anderson}, {Xu}, {Titov},
  {Heinkelmann}, {Johnson}, \& {Schuh}}]{lunz23}
{Lunz}, S., {Anderson}, J.~M., {Xu}, M.~H., {et~al.} 2023, \aap, 676, A11,
  \dodoi{10.1051/0004-6361/202040266}

\bibitem[{Marquardt(1963)}]{marquardt63}
Marquardt, D.~W. 1963, Journal of the society for Industrial and Applied
  Mathematics, 11, 431

\bibitem[{{Michalik} {et~al.}(2015){Michalik}, {Lindegren}, \&
  {Hobbs}}]{michalik15}
{Michalik}, D., {Lindegren}, L., \& {Hobbs}, D. 2015, \aap, 574, A115,
  \dodoi{10.1051/0004-6361/201425310}

\bibitem[{{Perryman} {et~al.}(1997){Perryman}, {Lindegren}, {Kovalevsky},
  {Hoeg}, {Bastian}, {Bernacca}, {Cr{\'e}z{\'e}}, {Donati}, {Grenon},
  {Grewing}, {van Leeuwen}, {van der Marel}, {Mignard}, {Murray}, {Le Poole},
  {Schrijver}, {Turon}, {Arenou}, {Froeschl{\'e}}, \& {Petersen}}]{perryman97}
{Perryman}, M.~A.~C., {Lindegren}, L., {Kovalevsky}, J., {et~al.} 1997, \aap,
  323, L49

\bibitem[{{R Core Team}(2023)}]{r23}
{R Core Team}. 2023, R: A Language and Environment for Statistical Computing, R
  Foundation for Statistical Computing, Vienna, Austria

\bibitem[{Schloerke {et~al.}(2021)Schloerke, Cook, Larmarange, Briatte,
  Marbach, Thoen, Elberg, \& Crowley}]{ggally}
Schloerke, B., Cook, D., Larmarange, J., {et~al.} 2021, GGally: Extension to
  'ggplot2'

\bibitem[{Schwarz {et~al.}(1978)}]{schwarz78}
Schwarz, G., {et~al.} 1978, The annals of statistics, 6, 461

\bibitem[{{Shahaf} {et~al.}(2023){Shahaf}, {Bashi}, {Mazeh}, {Faigler},
  {Arenou}, {El-Badry}, \& {Rix}}]{shahaf23}
{Shahaf}, S., {Bashi}, D., {Mazeh}, T., {et~al.} 2023, \mnras, 518, 2991,
  \dodoi{10.1093/mnras/stac3290}

\bibitem[{{Snellen} \& {Brown}(2018)}]{snellen18}
{Snellen}, I.~A.~G., \& {Brown}, A.~G.~A. 2018, Nature Astronomy, 2, 883,
  \dodoi{10.1038/s41550-018-0561-6}

\bibitem[{{Tsiganis} {et~al.}(2005){Tsiganis}, {Gomes}, {Morbidelli}, \&
  {Levison}}]{tsiganis05}
{Tsiganis}, K., {Gomes}, R., {Morbidelli}, A., \& {Levison}, H.~F. 2005, \nat,
  435, 459, \dodoi{10.1038/nature03539}

\bibitem[{{van Leeuwen}(2007)}]{leeuwen07}
{van Leeuwen}, F. 2007, \aap, 474, 653, \dodoi{10.1051/0004-6361:20078357}

\bibitem[{Wickham(2016)}]{ggplot2}
Wickham, H. 2016, ggplot2: Elegant Graphics for Data Analysis (Springer-Verlag
  New York).
\newblock \url{https://ggplot2.tidyverse.org}

\bibitem[{{Wittenmyer} {et~al.}(2020){Wittenmyer}, {Wang}, {Horner}, {Butler},
  {Tinney}, {Carter}, {Wright}, {Jones}, {Bailey}, {O'Toole}, \&
  {Johns}}]{wittenmyer20}
{Wittenmyer}, R.~A., {Wang}, S., {Horner}, J., {et~al.} 2020, \mnras, 492, 377,
  \dodoi{10.1093/mnras/stz3436}

\bibitem[{{Ye} \& {Fishbach}(2022)}]{ye22}
{Ye}, C., \& {Fishbach}, M. 2022, \apj, 937, 73,
  \dodoi{10.3847/1538-4357/ac7f99}

\end{thebibliography}

\appendix
\section{Error inflation and jitter optimization}
\begin{figure*}
    \centering
    \includegraphics[scale=0.35]{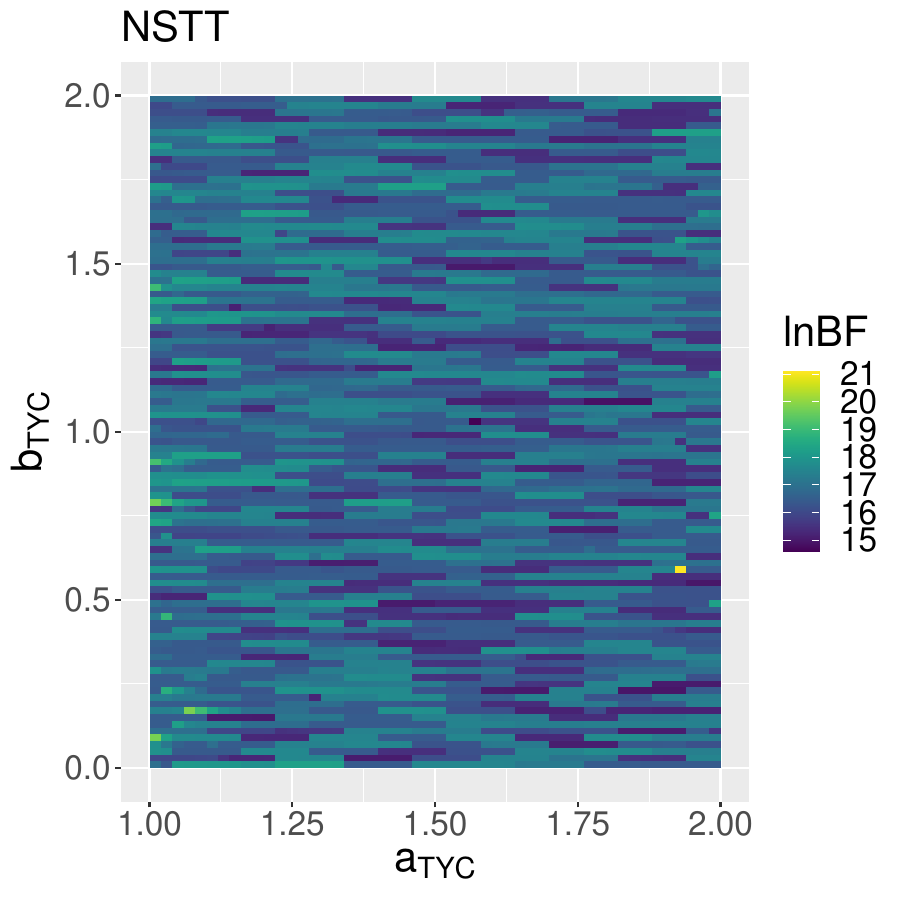}
    \includegraphics[scale=0.35]{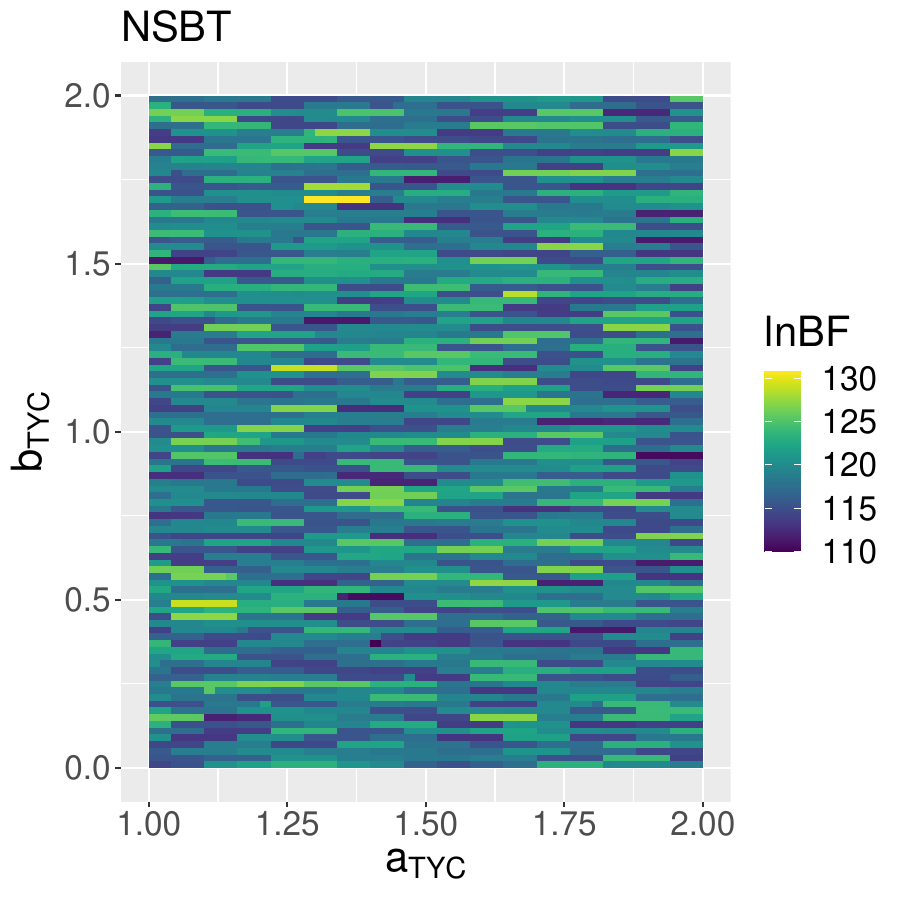}
    \includegraphics[scale=0.35]{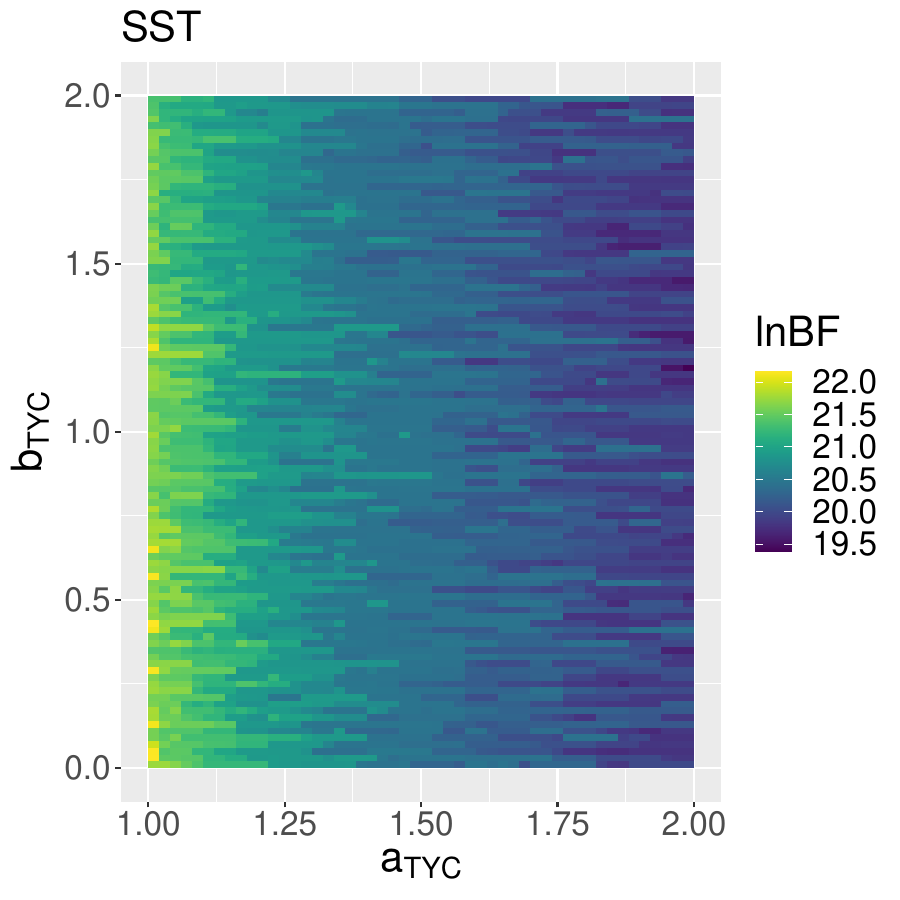}
    
    \includegraphics[scale=0.35]{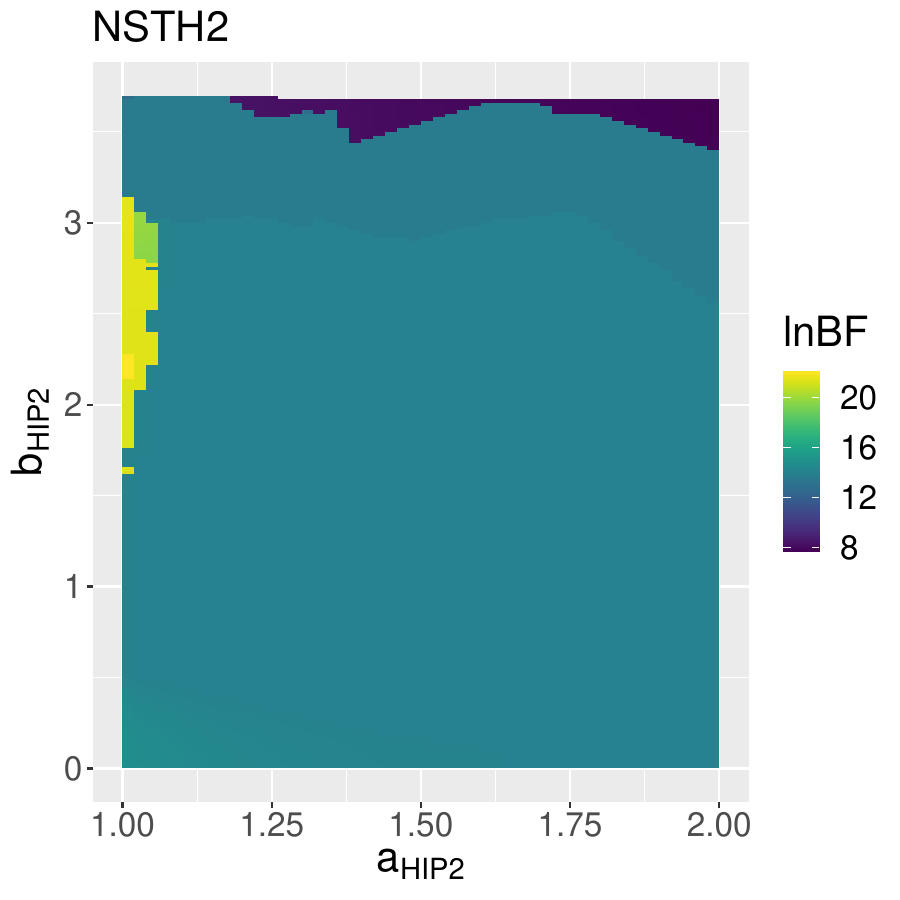}
    \includegraphics[scale=0.35]{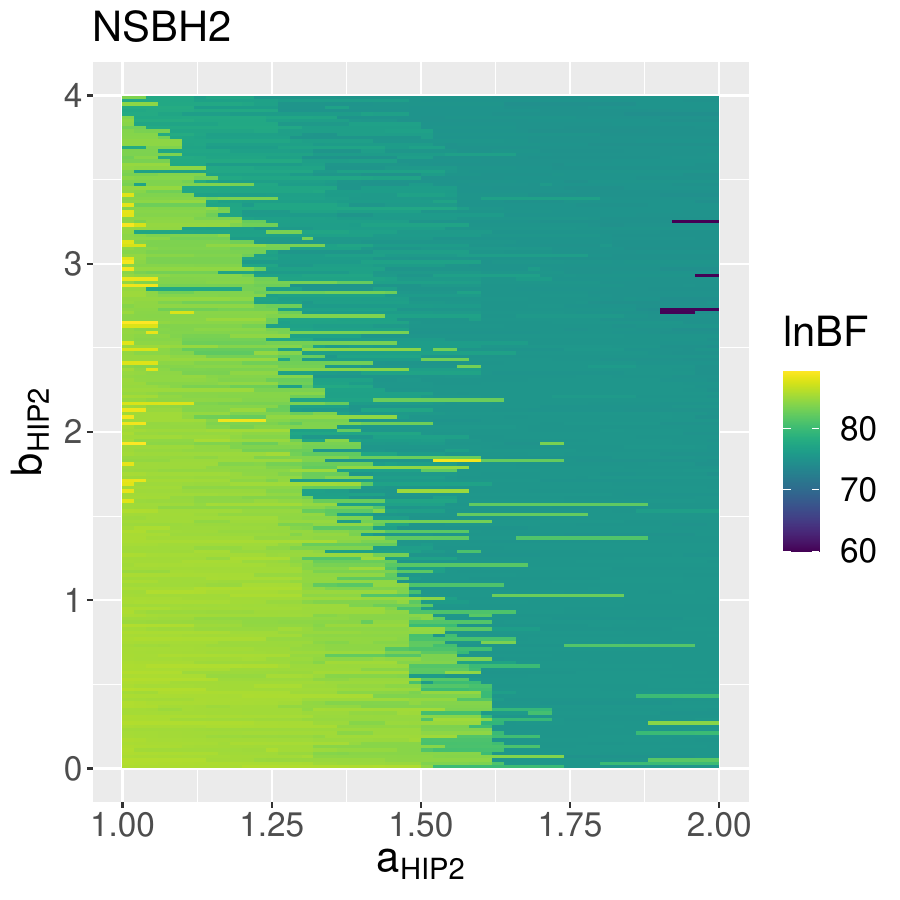}
    \includegraphics[scale=0.35]{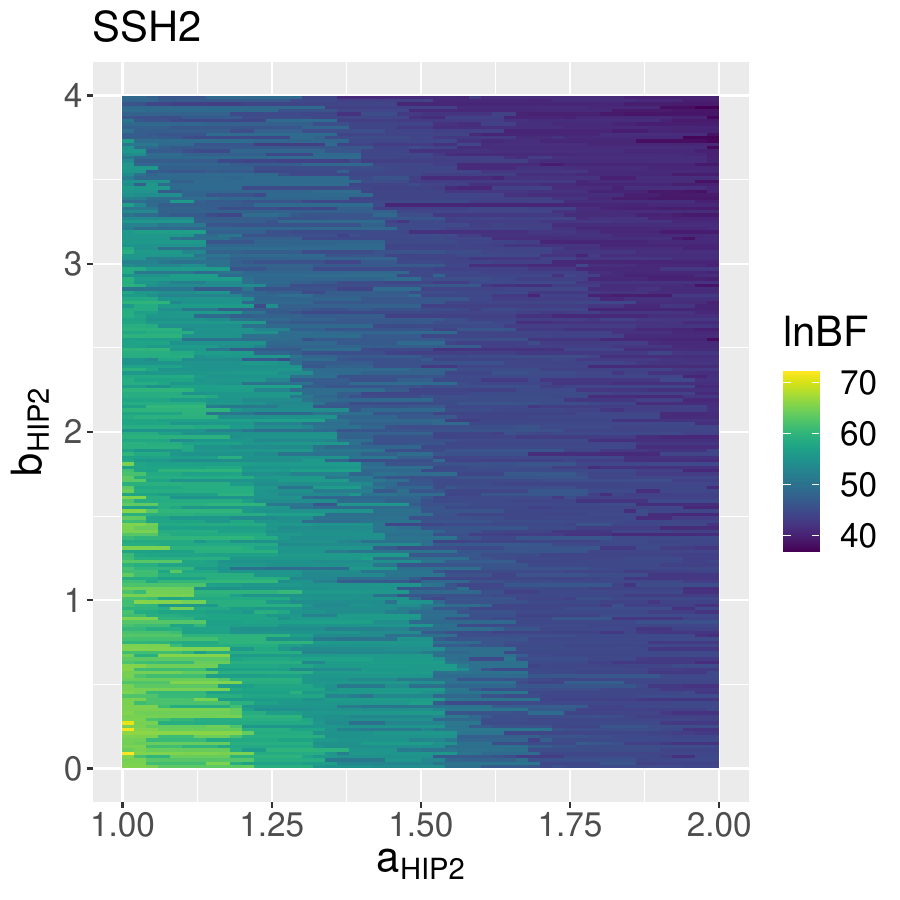}

    \includegraphics[scale=0.35]{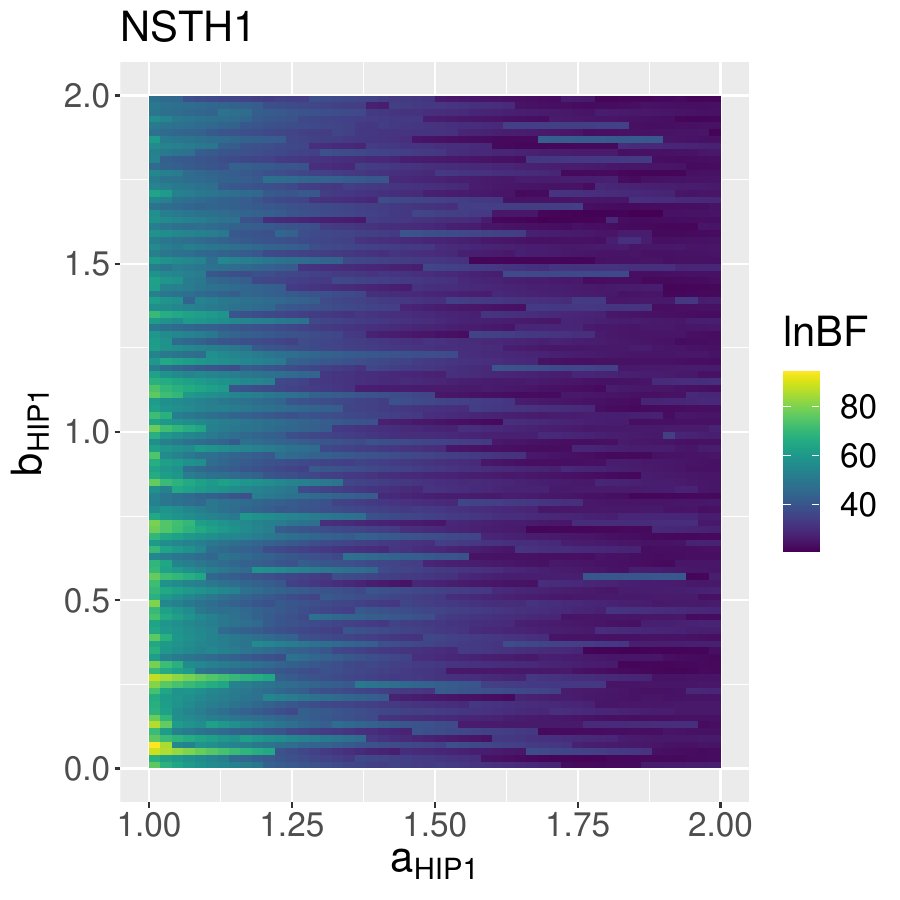}
    \includegraphics[scale=0.35]{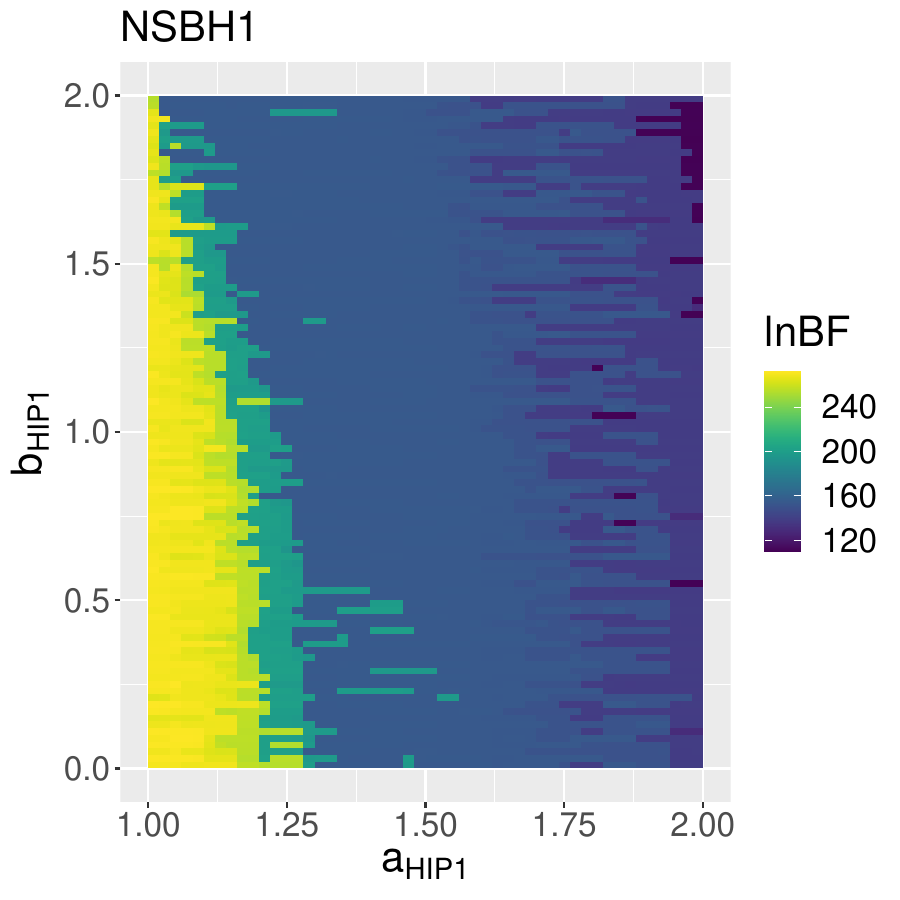}
    \includegraphics[scale=0.35]{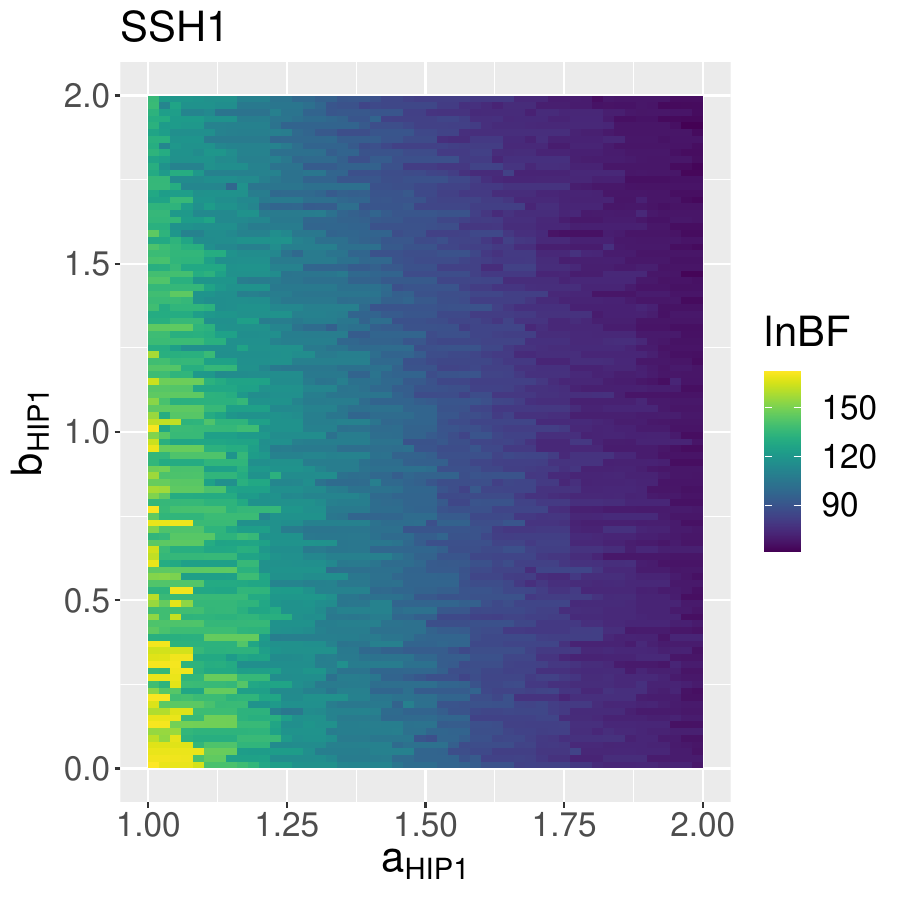}
    
    \caption{Similar to Fig. \ref{fig:ab_g123}, but for TYC (first row), HIP1 (second row) and HIP2 (third row).}
    \label{fig:ab_th12}
  \end{figure*}

Fig. \ref{fig:ab_th12} shows the distribution of lnBF with error
inflation and jitter for various catalogs based on analyses of
different calibration sources.
  
\begin{figure}
  \centering
  \includegraphics[scale=0.3]{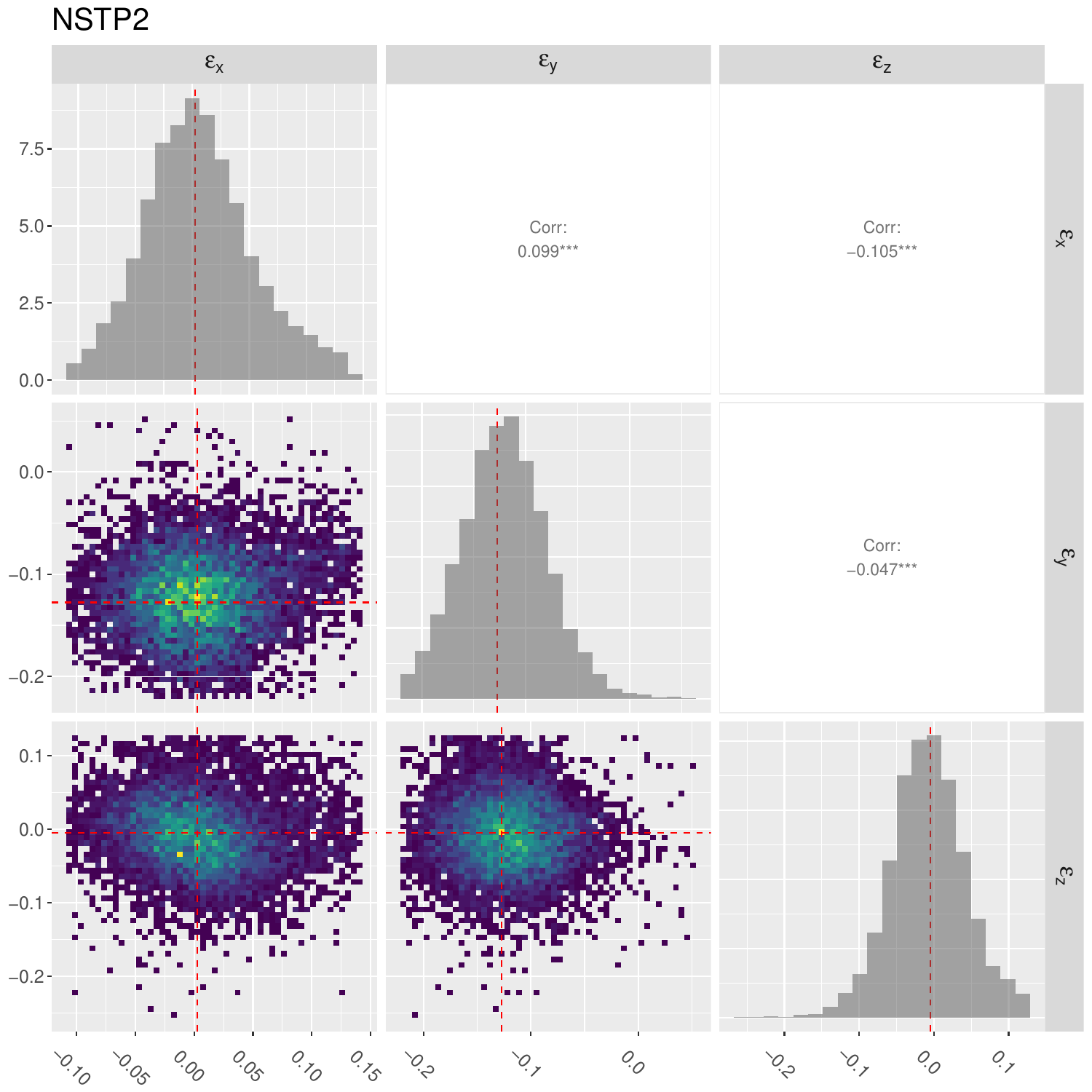}
  \includegraphics[scale=0.3]{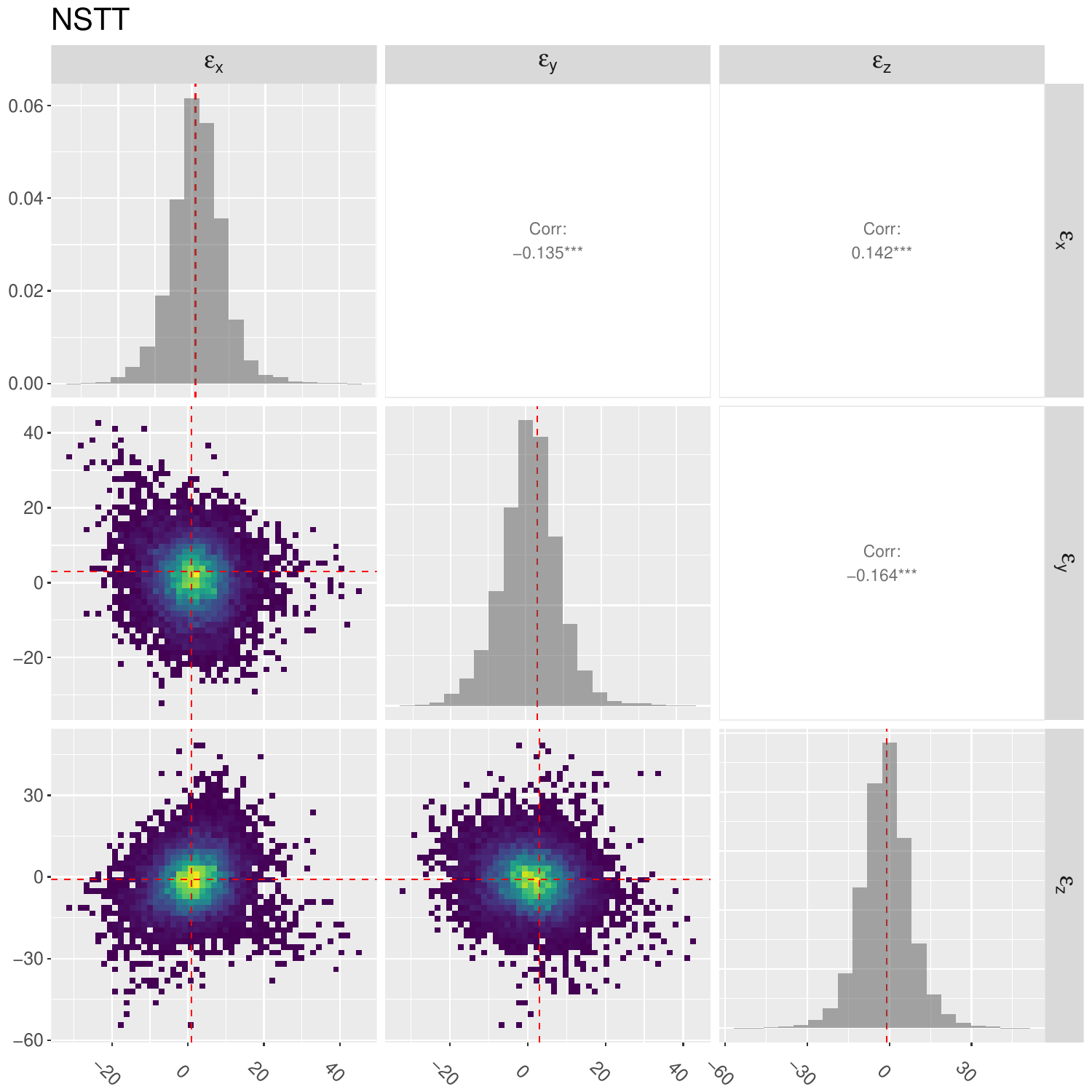}
  
  \includegraphics[scale=0.3]{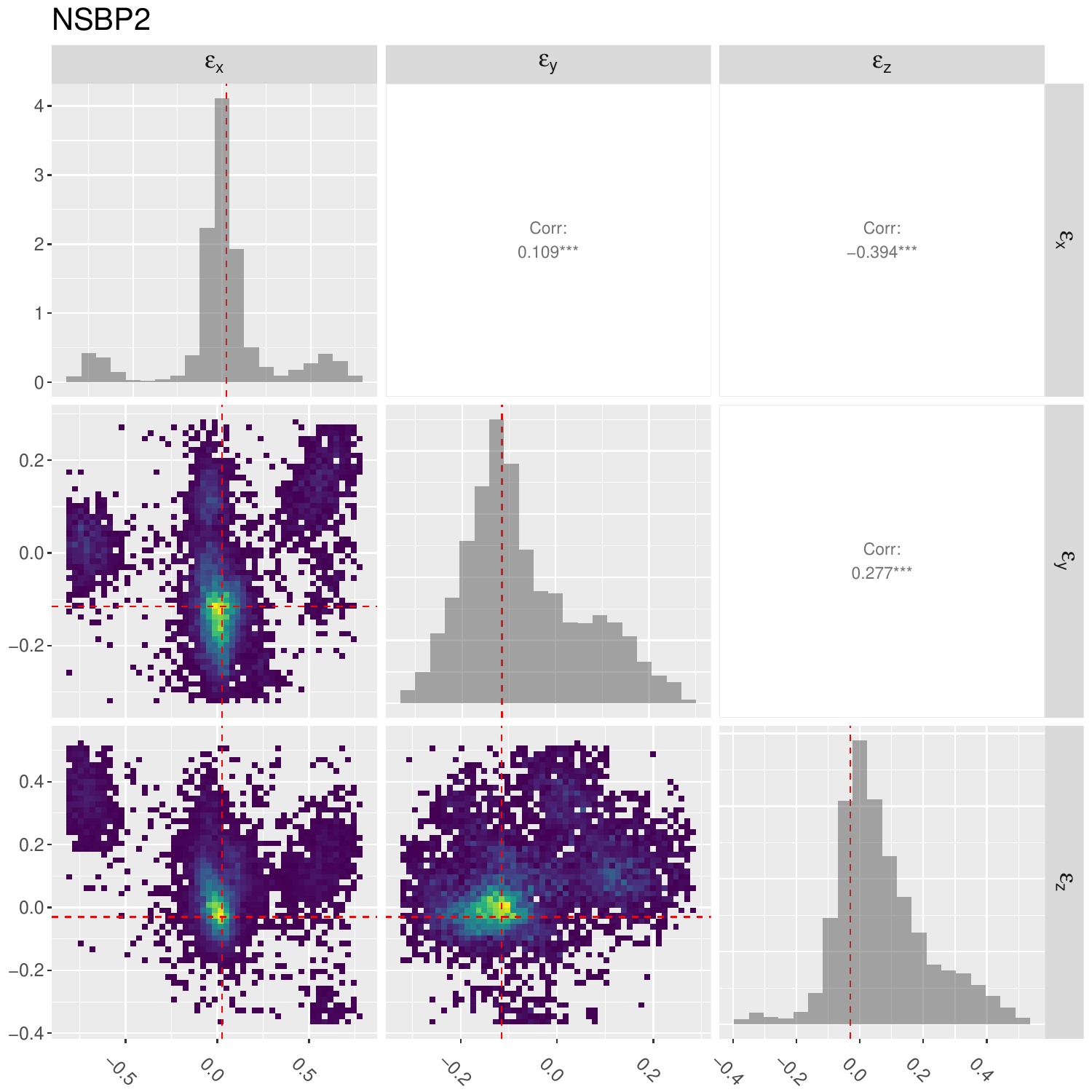}
  \includegraphics[scale=0.3]{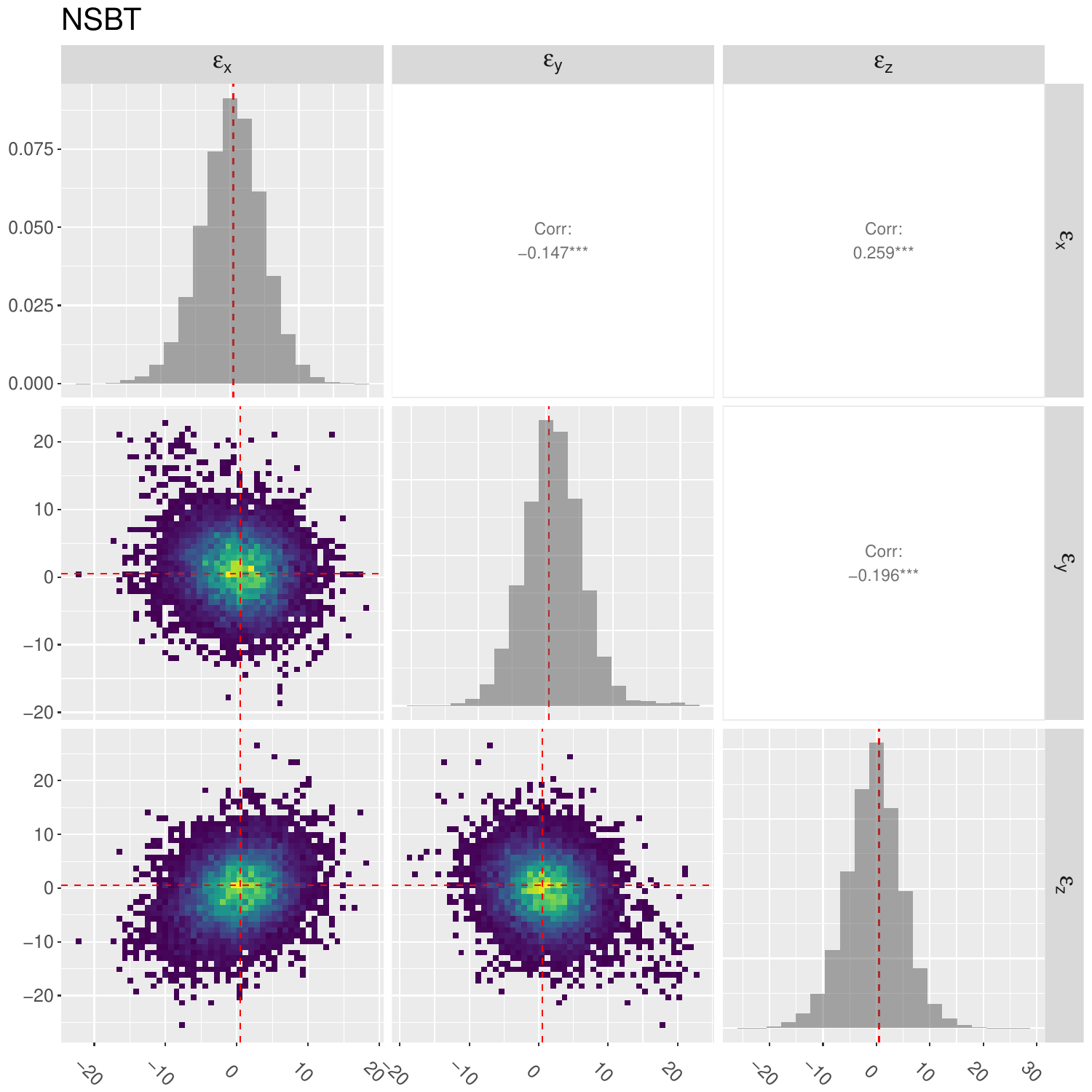}
  
  \includegraphics[scale=0.3]{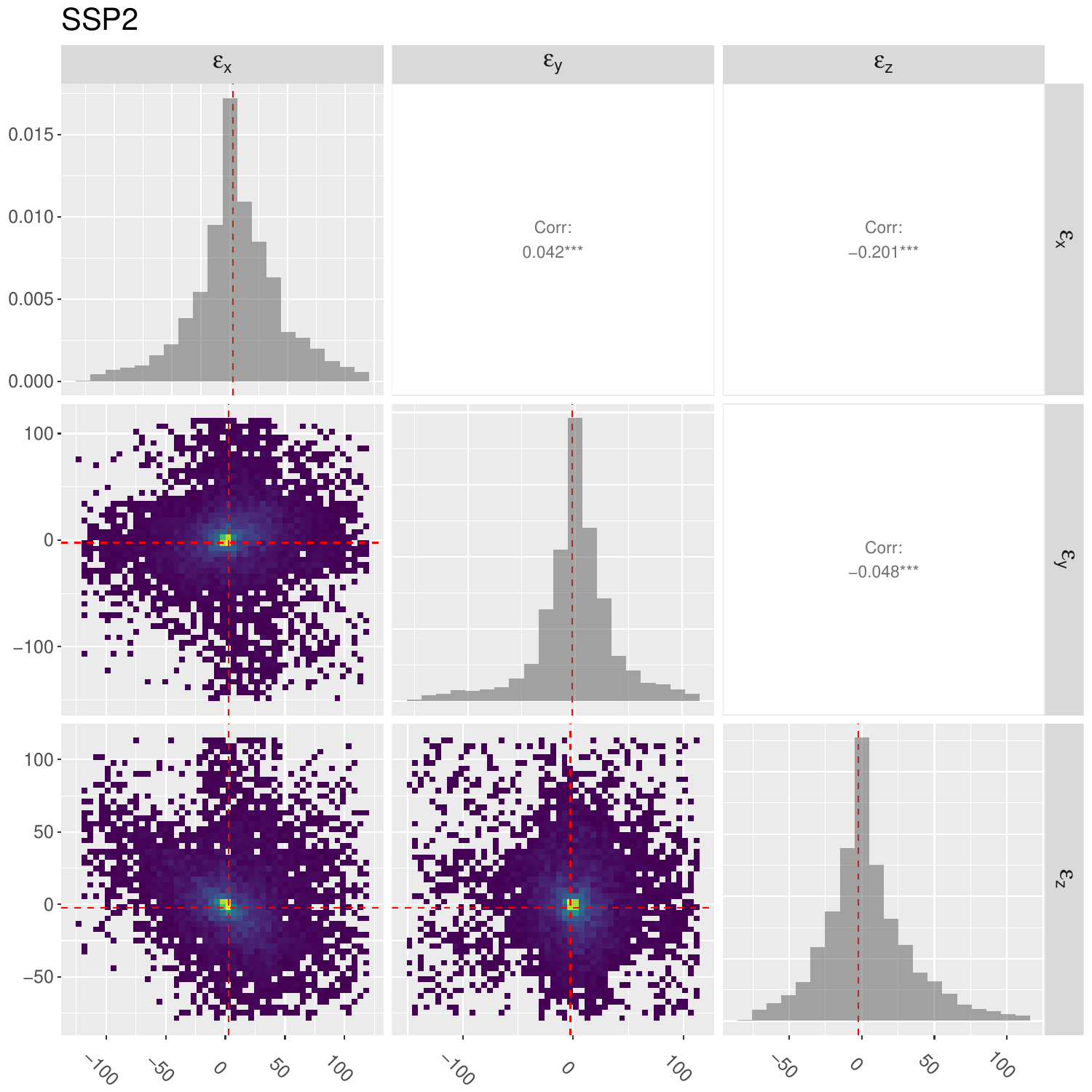}
  \includegraphics[scale=0.3]{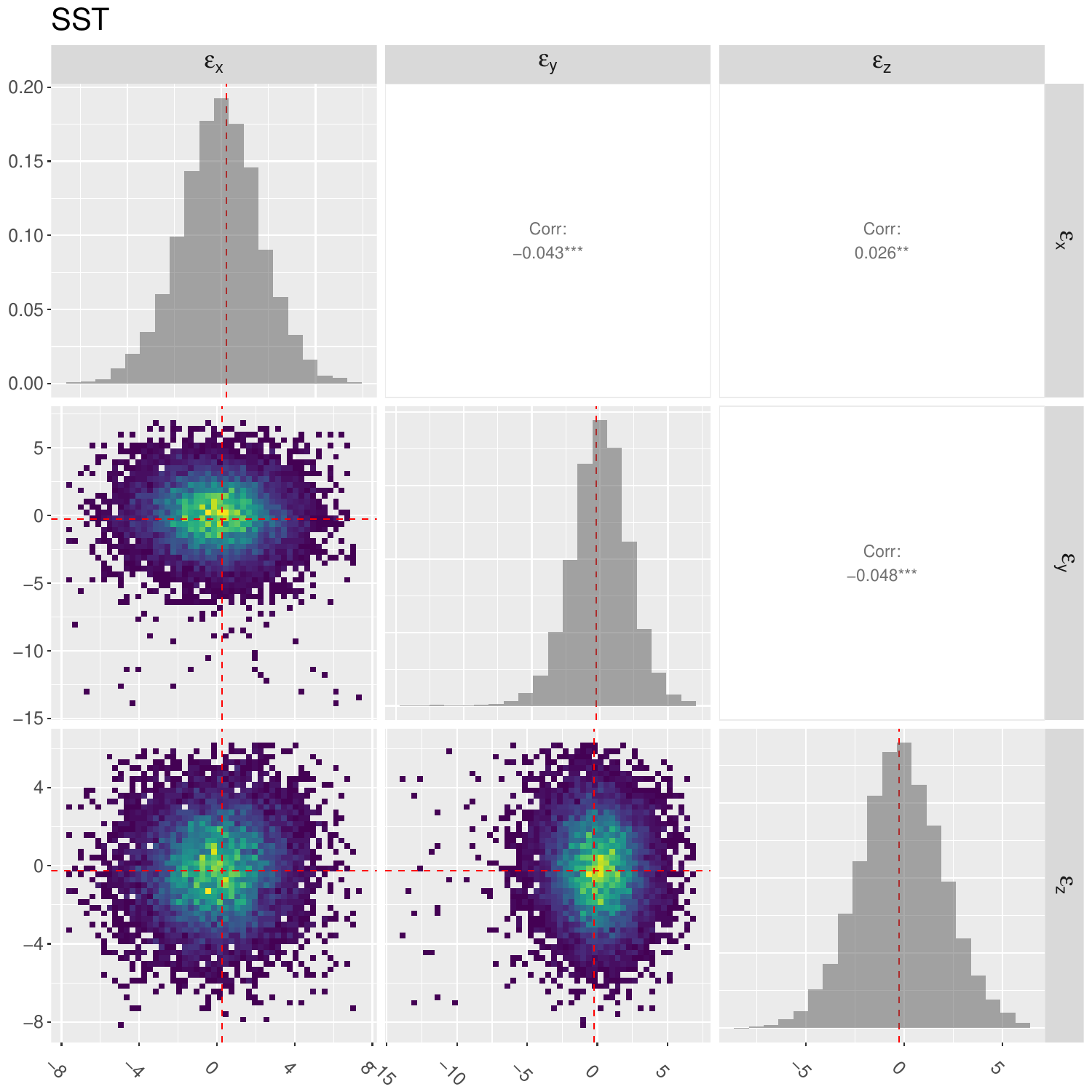}
    \caption{Similar to Fig. \ref{fig:pair_none_g1p5}, but for distribution of optimized calibration parameters for G1P2 (left) and TYC (right) based on 10,000 draws of 100 samples from various catalibration
      sources.}
  \label{fig:pair_g1p2_tyc}
\end{figure}

\begin{figure}
  \centering
  \includegraphics[scale=0.3]{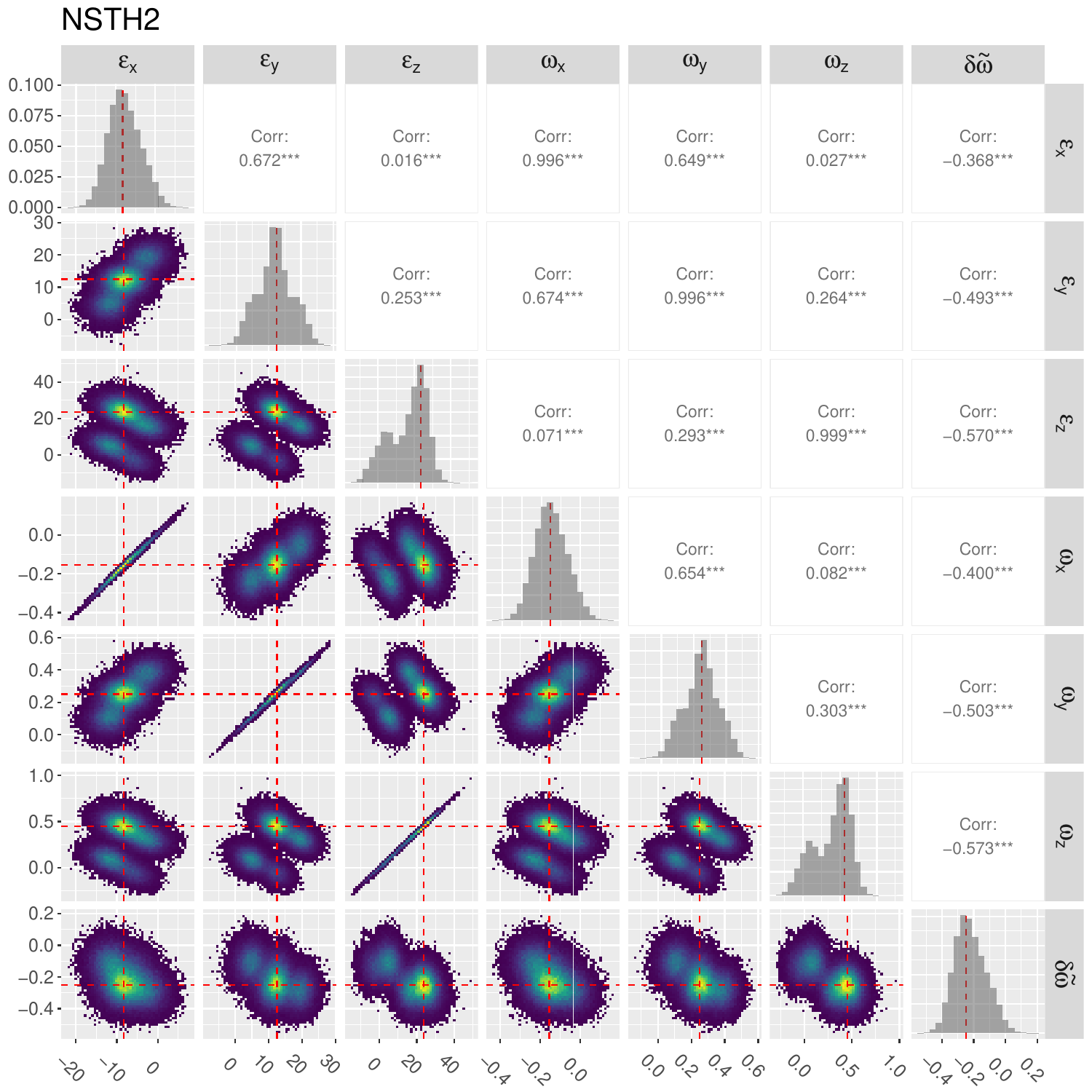}
  \includegraphics[scale=0.3]{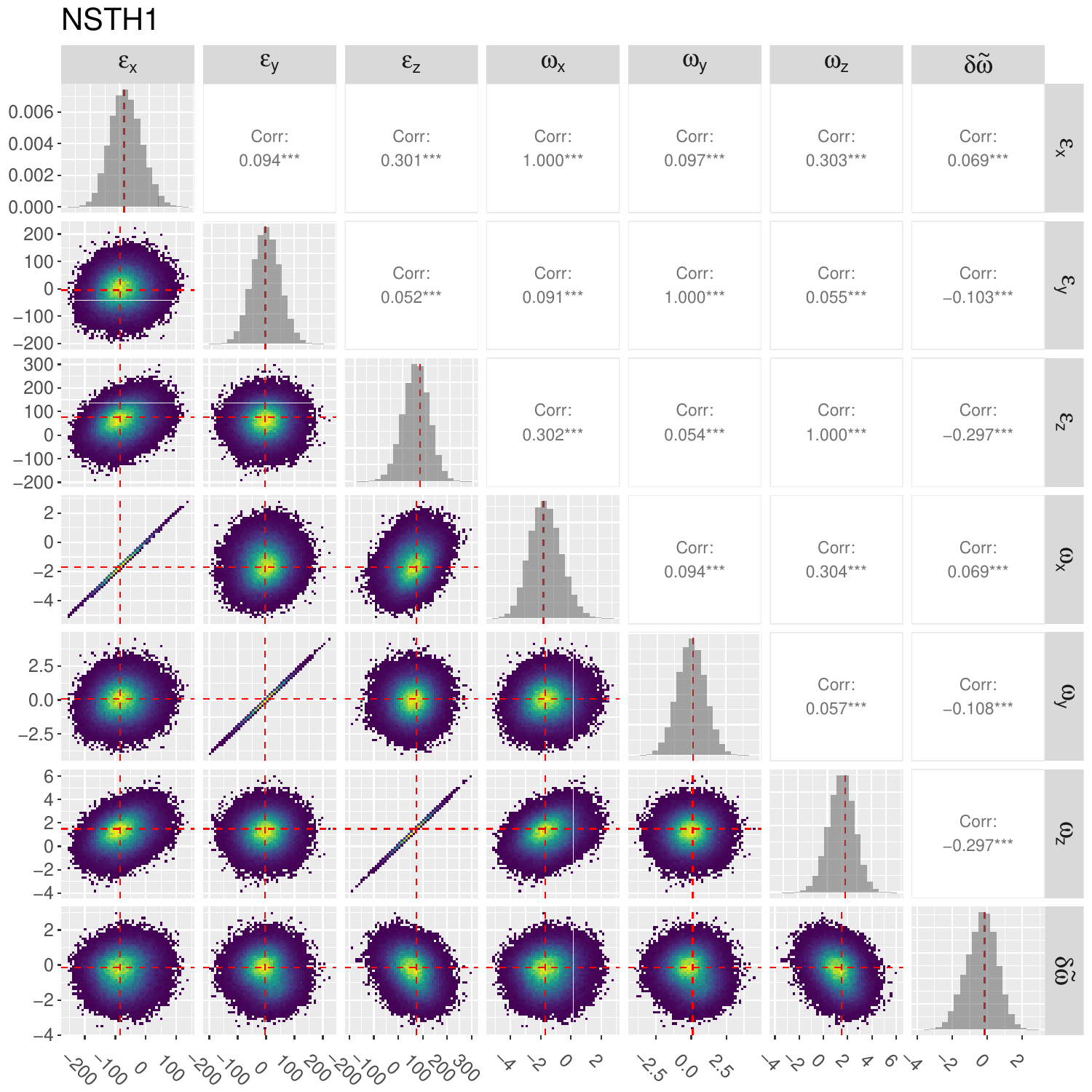}
  
  \includegraphics[scale=0.3]{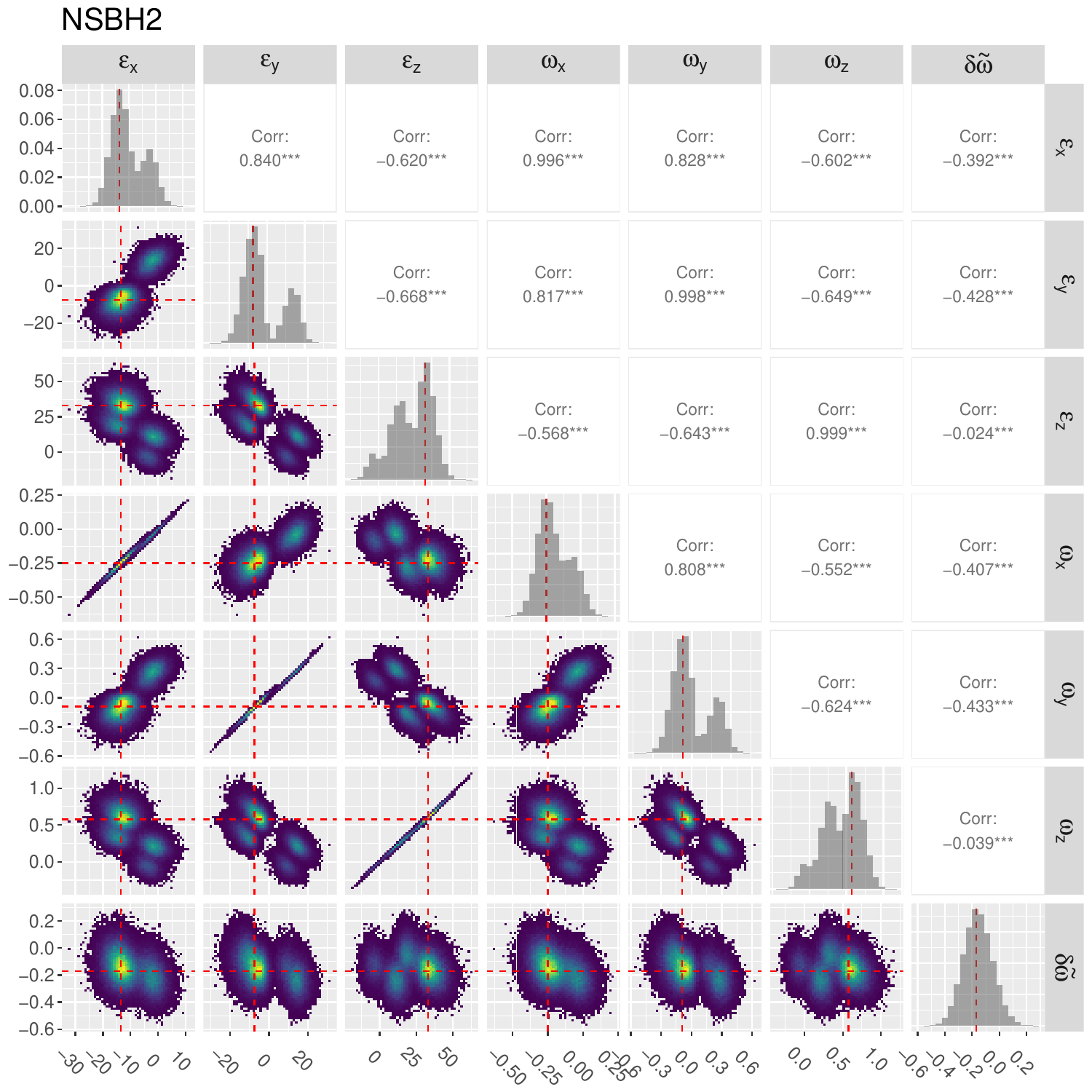}
  \includegraphics[scale=0.3]{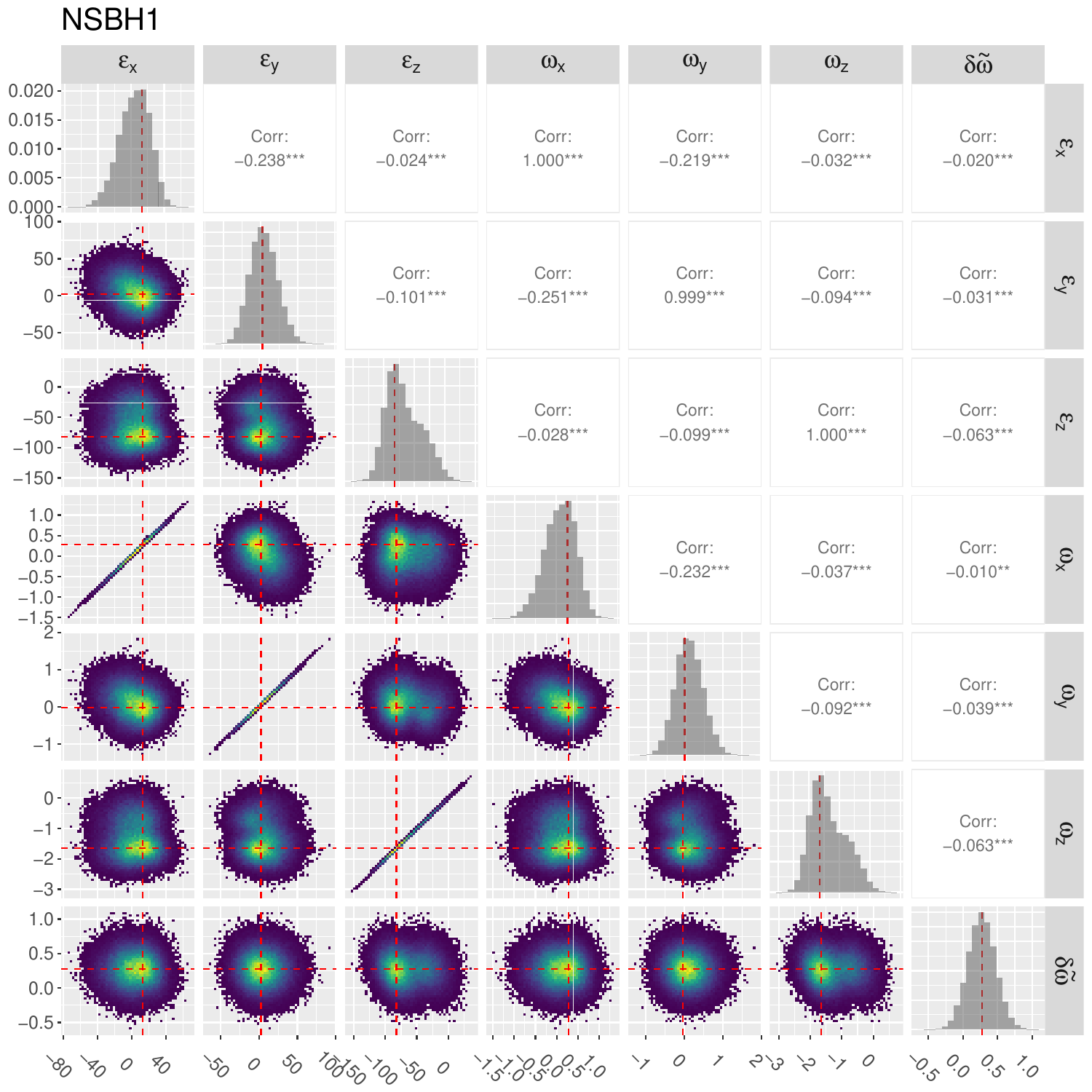}
  
  \includegraphics[scale=0.3]{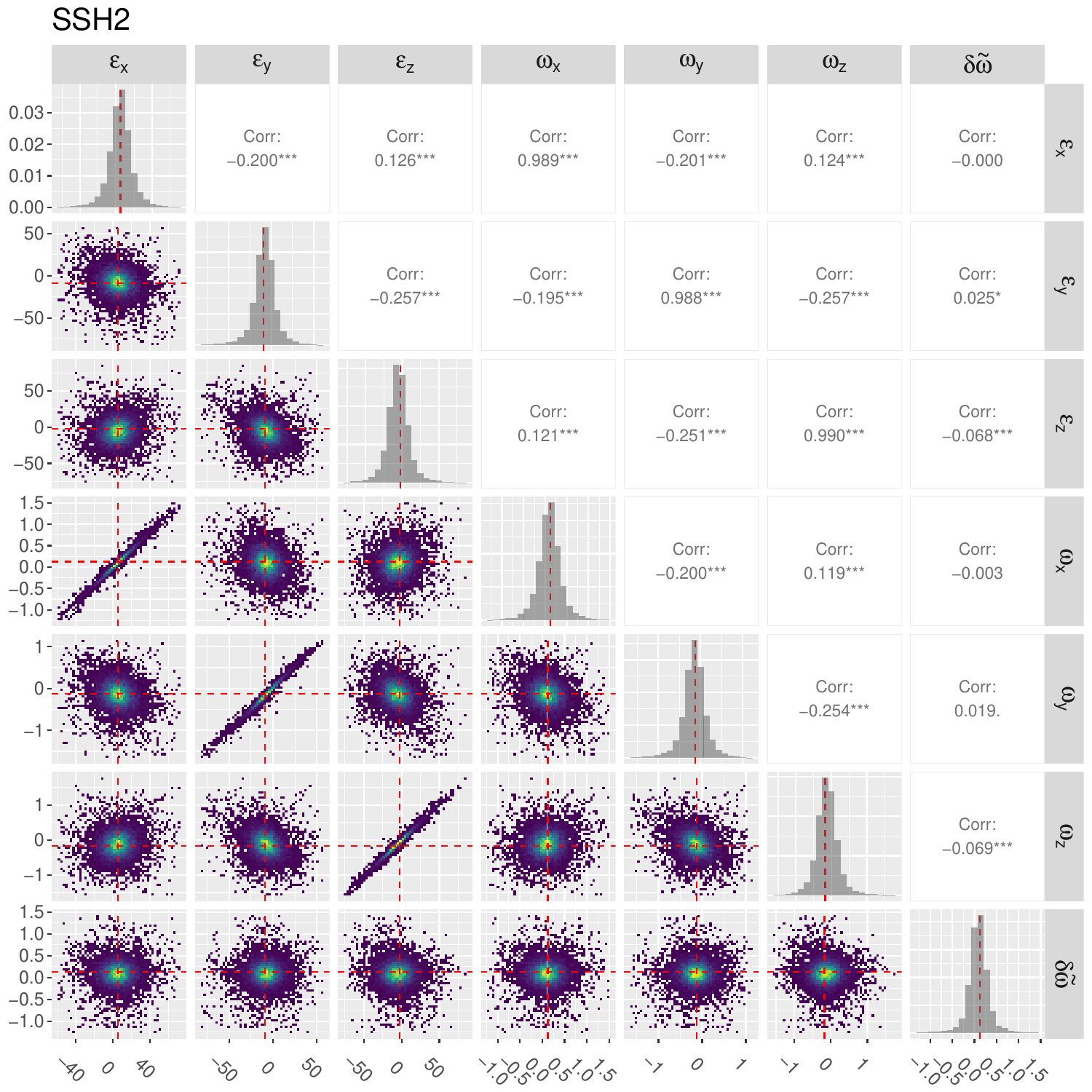}
  \includegraphics[scale=0.3]{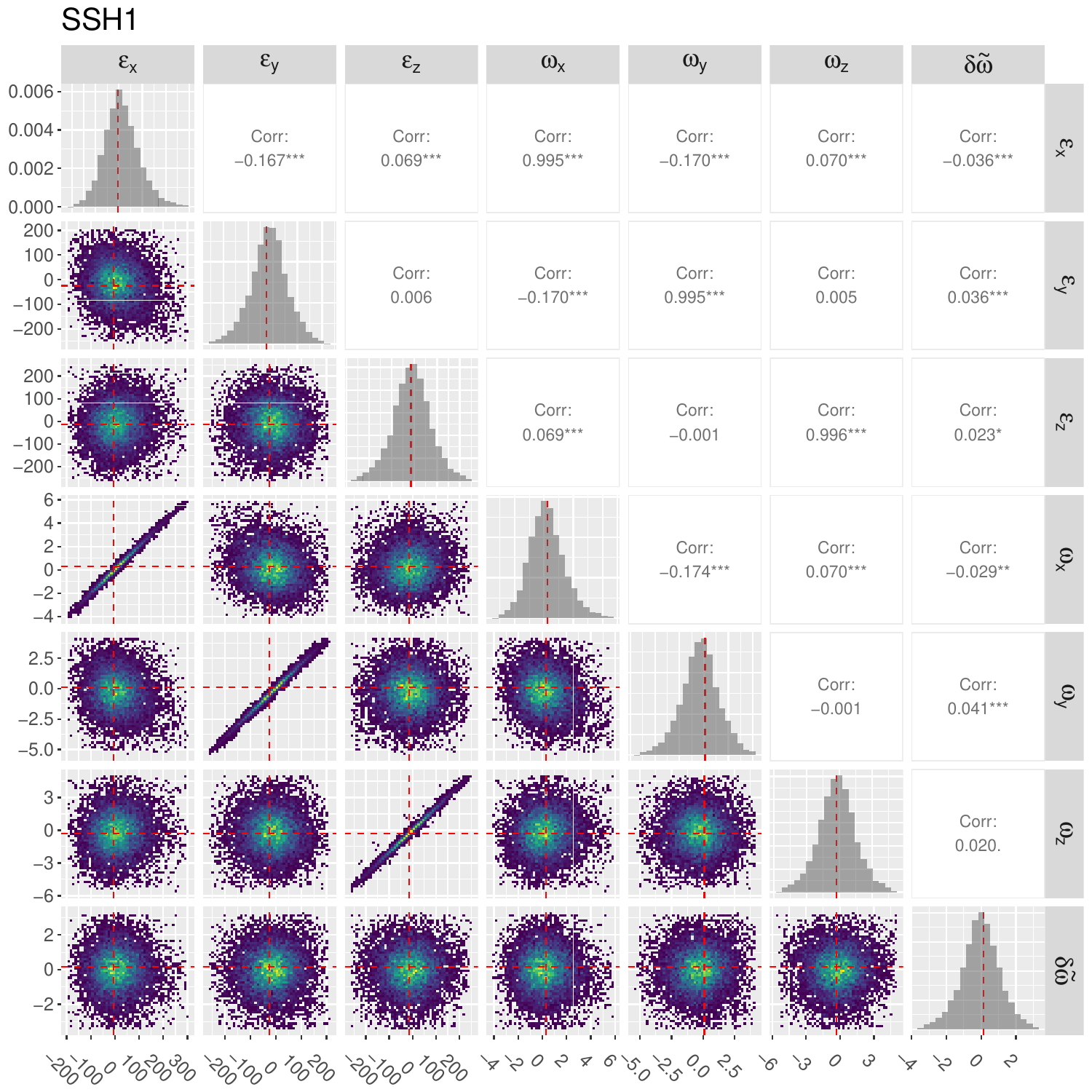}
    \caption{Similar to Fig. \ref{fig:pair_none_g1p5}, but for distribution of optimized calibration parameters for HIP2 (left) and HIP1 (right) based on 10,000 draws of 100 samples from various catalibration sources.}
  \label{fig:pair_hip2_hip1}
\end{figure}

Fig. \ref{fig:pair_g1p2_tyc} and Fig. \ref{fig:pair_hip2_hip1} show
the corner plots of $\vec\beta^d$ for various calibration sources.

\section{Pipeline for downloading data}
The data used in this work is downloaded using several scripts, which are available in the GitHub repository: \url{https://github.com/ruiyicheng/Download_HIP_Gaia_GOST}. The description of these scripts is as follows:

\begin{itemize}
    \item {\it get\_hipIAD1997.py}\quad This code is dedicated to
      downloading IAD of the Hipparcos
      1997 reductions. There are a total of 118,204 entries which have
      non-empty Hipparcos IAD information in the 1997 version. Using
      the \verb|get_hipIAD1997| function, the Hipparcos IAD of the
      corresponded source can be downloaded as a csv file. If the
      input HIP entry is non-empty, this csv file will contain columns
      of orbit number, source of abscissa (FAST or NDAC), partial
      derivatives of the abscissa with respect to five astrometric
      parameters ($\alpha* = \alpha \cos\delta, \delta, \pi,
      \mu_{\alpha*}, \mu_{\delta}$), abscissa residual in mas,
      standard error of the abscissa in mas, correlation coefficient
      between abscissae, reference great circle mid-epoch in years,
      reference great circle mid-epoch in days, RA of the great circle
      pole in degrees, decl. of the great circle pole in degrees. Otherwise, \verb|The Hipparcos IAD of this HIP entry cannot be found| will be printed.
    \item {\it nssGDR123HIPTYC.sql}\quad This script is used to queue the Gaia data via ADQL (\url{https://gea.esac.esa.int/archive/}). It retrieves the five-parameter astrometric solution of Gaia DR1, DR2, and DR3 for non-single stars with two-body orbital solutions. Additionally, it obtains the cross-match results of these stars with the Hipparcos and Tycho catalogues.
    \item {\it Obtain\_GOST.py}\quad This script is used to queue the Gaia GOST data using the interface in \url{https://gaia.esac.esa.int/gost/GostServlet}. It employs a similar pre-processing method as described in \cite{brandt21} to convert the server's returned results into a human-friendly pandas dataframe. The data is then filtered to match the GOST results obtained through the web page interface at \url{https://gaia.esac.esa.int/gost/}. The script requires the path of the csv file obtained using {\it nssGDR123HIPTYC.sql} as an argument. The results are recorded in separate files named using the Gaia DR3 IDs.
    \item {\it frame\_rotation\_correction.py}\quad This script is used to align the astrometry of different cataloges with Gaia DR3. The script requires the catalog astrometry, together with source and CAT1 as shown in Table 1. The results are recorded in a dictionary. An exception value would be given if there is no input proper motion or parallax.
    \item {\it download\_single\_target.py}\quad This script offers a user-friendly interface for downloading and collaborating astrometric data. Users can queue the data using HIP id, TYC id, or Gaia DR1, DR2, DR3 source id. The Gaia astrometric data is acquired using an ADQL script similar to \textit{nssGDR123HIPTYC.sql}. The HIP and TYC astrometry is gathered through cross-matching with the bulk data obtained from VizieR\footnote{\url{https://vizier.cds.unistra.fr/}}. The collaborated astrometry is obtained using \textit{frame\_rotation\_correction.py} with the recommended startegy in section 4.2. Users have the option to download either the Hipparcos epoch data or Gaia GOST data for their target. The results are stored in the \textit{/results} folder.
\end{itemize}

    \section{Symbols and acronyms}
    The following table provides abbreviations used in the paper along
    with the meaning of acronyms and symbols used. 

    \clearpage
\startlongtable
\begin{deluxetable*}{ll}
\tablecaption{Symbols used in this work.\label{tab:symbol}}
\tablehead{
  \colhead{Symbols or Acronyms}&\colhead{Meaning}\\
}
\startdata
$a$&Error inflation\\
$b$&Astrometric jitter\\
$a_p$&Semi-major axis of the photocenter relative to the mass center\\
$a_b$& Binary semi-major axis\\
$a_r$&Semi-major axis of the reflex motion of the target\\
$A_u$& 1 au\\
$A$, $B$, $F$, $G$& Thiele Innes constants\\
$A'$, $B'$, $F'$, $G'$& Scaled Thiele Innes constants\\
$\vec{b}$&Vector of dependent variable in the normal equation $\bm{\eta}\vec\beta=\vec{b}$\\
$C^p_i$&$\cos\theta_i$\\
$d$& Heliocentric distance\\
$e$&Eccentricity\\
$E$&Eccentric anomaly\\
$f^{\rm AL}_i$& Along-scan  parallax factor \\
$F_1$& Luminosity of the primary\\
$F_2$& Luminosity of the secondary\\
$I$& Inclination\\
$m_1$& Mass of the primary in a binary system\\
$m_2$& Mass of the secondary in a binary system\\
$n_d$&Number of differential calibration parameters $\vec\beta^d$\\
$n_b$&Number of barycentric astrometry parameters $\vec\beta^b$\\
$n_r$&Number of reflex motion parameters $\vec\beta^r$\\
$M_0$&Mean anomaly at the reference epoch\\
$P$&Orbital period\\
$S^p_i$&$\sin\theta_i$\\
$X_i$& Reflex motion in the X direction of the orbital plane\\
$\vec{y}$&Data vector consists of catalog data or IAD\\
$Y_i$& Reflex motion in the Y direction of the orbital plane\\
\hline
$\alpha$&R.A.\\
$\alpha_*$&$\alpha\cos\delta$\\
$\vec\beta^b$&Linear parameters of the barycentric motion model\\
$\vec\beta^d$&Parameters of the calibration model\\
$\vec\beta^n$&Nonlinear parameters of the reflex motion model\\
$\vec\beta^r$&Parameters of the reflex motion model\\
$\vec\beta_{\rm sample}$&Parameters of a sample of calibration sources\\
$\bm{\gamma}$&Coefficient matrix for $\vec\beta^r$ and $\vec{y}$\\
$\bm{\eta}$&Design matrix for the normal equation $\bm{\eta}\vec\beta=\vec{b}$\\
$\delta$& Decl.\\
$\delta\varpi$& Parallax offset between two reference frames\\
$\delta\xi$ &Vector of abscissae residual\\
$\Delta \delta^r$&Reflex motion projected onto the increasing Decl. direction\\
$\Delta\alpha^b$&$\alpha^b_*-\alpha^{\rm ref}_*$\\
$\Delta\delta^b$&$\delta^b-\delta^{\rm ref}$\\
$\Delta\mu_\alpha^b$&$\mu_\alpha^b-\mu_\alpha^{\rm ref}$\\
$\Delta\mu_\delta^b$&$\mu_\delta^b-\mu_\delta^{\rm ref}$\\
$\Delta \delta_i^b$&Reflex motion projection in the increasing Decl. direction at $t_i$\\
$\Delta \delta_i^b$&Barycentric motion projection in the increasing Decl. direction\\
$\Delta \alpha$&R.A. offset, equivalent to $\Delta \alpha_*$\\
$\Delta t_i$ &$t_i-t_j$\\
$\Delta t_{ij}$ &Time difference between epoch $t_i$ and epoch $t_j$\\
$\vec\epsilon \equiv (\epsilon_x,\epsilon_y,\epsilon_z)$&Offset between two reference frames at a reference epoch\\
$\theta$& Gaia scan angle\\
$\bm{\kappa^b}$&Coefficient matrix for $\vec\beta^b$ and $\vec\xi$\\
$\bm{\kappa^d}$&Coefficient matrix for $\vec\beta^d$ and $\vec\xi$\\
$\bm{\kappa^r}$&Coefficient matrix for $\vec\beta^r$ and $\vec\xi$\\
$\bm{\lambda}$&Coefficient matrix for $\vec\beta^b$ and $\vec{y}$\\
$\mu_\alpha$& Proper motion in R.A.\\
$\mu_\delta$& Proper motion in Decl.\\
$\mu_\alpha^{\rm ref}$& $\mu_\alpha$ at the reference epoch\\
$\mu_\delta^{\rm ref}$& $\mu_\delta$ at the reference epoch\\
$\bm{\mu^b}$&Combination of $\bm{\kappa^b}$ of multiple catalogs\\
$\bm{\mu^d}$&Combination of $\bm{\kappa^d}$ of multiple catalogs\\
$\bm{\mu^r}$&Combination of $\bm{\kappa^r}$ of multiple catalogs\\
$\nu$&Number of extra parameters of model 2 compared with model 1\\
$\varpi$& Parallax\\
$\phi$& Orbital phase\\
$\psi$& Hipparcos scan angle and $\psi=\pi/2-\theta$\\
$\xi_i$ &Raw or synthetic abscissae\\
$\vec\xi$&$\bm{\kappa^b}\vec\beta^b+\bm{\kappa^r}\vec\beta^r$\\
$\vec\xi$ &Abscissae vector\\
$\omega_T$&Argument of periastron for target reflex motion\\
$\Omega$&Longitude of ascending node\\
$\bm\Sigma$&Covariance matrix of data $\vec{y}$\\
$\chi_\xi^2$&$\chi^2$ for abscissae modeling\\
$\vec\omega \equiv (\omega_x,\omega_y,\omega_z)$&Rotation between two reference frames \\
\hline
AL & Along scan\\
CAT1 & First catalog in a crossmatch of three catalogs\\
G1P2 & GDR1 targets with two-parameter solutions\\
G1P5 & GDR1 targets with five-parameter solutions\\
G3NS & GDR3 NSS sources of the {\tt Orbital} type\\
G3NSB &G3NS sources with barycentric astrometry\\
G3NSP &G3NS sources with target photocentric astrometry\\
G3SS & GDR3 non-NSS sources that have Hipparcos data\\
GDR1 & Gaia Data Release 1\\
GDR2 & Gaia Data Release 2\\
GDR3 & Gaia Data Release 3\\
GEDR3 & Gaia Early Data Release 3\\
GOST & Gaia Observation Forecast Tool\\
HIP1 & Hipparcos catalog released in 1997\\
HIP2 & Revised Hipparcos catalog released in 2007\\
IAD & Intermediate Astrometric Data\\
lnBF & Logarithmic Bayes Factor\\
NSB &  GDR2+G3NSB\\
NSBH1 &  HIP1+GDR2+G3NSB\\
NSBH2 &  HIP2+GDR2+G3NSB\\
NSBP2 &  G1P2+GDR2+G3NSB\\
NSBP5 &  G1P5+GDR2+G3NSB\\
NSBT &  TYC+GDR2+G3NSB\\
NSS & GDR3 non-single-star sources\\
NST &  GDR2+G3NST\\
NSTP2 &  G1P2+GDR2+G3NST\\
NSTP5 &  G1P5+GDR2+G3NST\\
NSTT &  TYC+GDR2+G3NST\\
NSTH1 &  HIP1+GDR2+G3NST\\
NSTH2 &  HIP2+GDR2+G3NST\\
SS &  GDR2+G3SS\\
SSH1 &  HIP1+GDR2+G3SS\\
SSH2 &  HIP2+GDR2+G3SS\\
SSP2 &  G1P2+GDR2+G3SS\\
SSP5 &  G1P5+GDR2+G3SS\\
SST &  TYC+GDR2+G3SS\\
TYC & Tycho-2\\
\enddata
\tablecomments{Some of the variations of symbols are not
  shown and the meaning of symbol variations could be found in the corresponding text.}
\end{deluxetable*}
\end{document}